\documentclass[10pt,twocolumn,letterpaper]{article}

\usepackage{cvpr}              % To produce the CAMERA-READY version
\usepackage[dvipsnames]{xcolor}
\usepackage{array}
\usepackage[export]{adjustbox} % enables valign=c on \includegraphics
\usepackage{multirow}
\usepackage{float}
\usepackage{wrapfig}
\usepackage{tikz} % for the arrow
\usepackage{placeins}
\usepackage{graphicx,subcaption}
\usepackage{booktabs,siunitx}
\usepackage{booktabs,multirow}
\usepackage[table]{xcolor}
\newcommand{\best}[1]{\cellcolor{green!25}\textbf{#1}}
\newcommand{\second}[1]{\cellcolor{yellow!25}#1}
\newcommand{\worst}[1]{\cellcolor{red!20}#1}

\usepackage{booktabs} % For formal tables

\usepackage{graphicx}
\usepackage{tikz}
\usetikzlibrary{shapes.geometric, arrows, positioning}
\usepackage{subcaption}

\usepackage[ruled]{algorithm2e} % For algorithms

\SetAlFnt{\small}
\SetAlCapFnt{\small}
\SetAlCapNameFnt{\small}
\SetAlCapHSkip{0pt}

\usepackage{tikz}
\usetikzlibrary{arrows.meta}        % nicer arrow heads
\definecolor{ellRed}{HTML}{D64045}  % keep colours consistent
\definecolor{ellBlue}{HTML}{1E88C8}
\definecolor{rayClr}{HTML}{0D47A1}
\definecolor{normClr}{HTML}{388E3C}

\definecolor{cvprblue}{rgb}{0.21,0.49,0.74}
\usepackage[pagebackref,breaklinks,colorlinks,citecolor=cvprblue]{hyperref}

\def\paperID{271} % *** Enter the Paper ID here
\def\confName{3DV\xspace}
\def\confYear{2026\xspace}

\title{SH-SAS: An Implicit Neural Representation for Complex Spherical‑Harmonic Scattering Fields for 3D Synthetic Aperture Sonar}

\author{Omkar Shailendra Vengurlekar\\
Arizona State University\\
\and
Adithya Pediredla \\
Dartmouth College \\
\and
Suren Jayasuriya\\
Arizona State University\\
}

\begin{document}
\maketitle

\graphicspath{{./images/}}

\begin{abstract}
Synthetic aperture sonar (SAS) reconstruction requires recovering both the spatial distribution of acoustic scatterers and their direction-dependent response. Time-domain backprojection is the most common 3D SAS reconstruction algorithm, but it does not model directionality and can suffer from sampling limitations, aliasing and occlusion. Prior neural volumetric methods applied to synthetic aperture sonar, e.g. Reed et al.~\cite{reed2023neural}, treat each voxel as an isotropic scattering density, not modeling anisotropic returns. We introduce SH-SAS, an implicit neural representation that expresses the complex acoustic scattering field as a set of spherical harmonic (SH) coefficients. A multi-resolution hash encoder feeds a lightweight MLP that outputs complex SH coefficients up to a specified degree L. The zeroth-order coefficient acts as an isotropic scattering field, which also serves as the density term, while higher orders compactly capture directional scattering with minimal parameter overhead. Because the model predicts the complex amplitude for any transmit–receive baseline, training is performed directly from 1-D time-of-flight (ToF) signals without the need to beamform intermediate images for supervision. Across synthetic and real SAS (both in-air and underwater) benchmarks, results show that SH-SAS performs better in terms of 3D reconstruction quality and geometric metrics than previous methods like time-domain backprojection and Reed et al.~\cite{reed2023neural}.

% , while also fitting the ground truth signal better than previous INR methods like \cite{reed2023neural}. On average, our method achieves 40\% lower chamfer distance and 52\% lower MSE on rendered signals compared to \cite{reed2023neural}, despite using an $8\times$ smaller MLP and $\approx$50\% faster convergence.
\end{abstract}    
\section{Introduction}
\label{sec:intro}

Synthetic aperture sonar (SAS) is a powerful modality for obtaining high-resolution 2D/3D reconstructions using acoustic signals~\cite{bellettini2008design, hayes2009synthetic}. A transmitting transducer (e.g. a speaker or hydrophone) emits acoustic signals to insonify the environment, and then a receiving transducer  records the reflected signal to estimate the time-of-flight (ToF) to various objects in the scene. To obtain high resolution scene reconstruction, this transmitter/receiver pair takes multiple measurements at different physical locations, creating a synthetic aperture whose signals can be coherently combined to reconstruct the scene. SAS technology has seen wide-scale deployment for both underwater sensing~\cite{hansen2011challenges} and even in-air sensing~\cite{blanford2024air,jideani2014air,saruwatari1999synthetic}. 

The conventional approach to SAS reconstruction, time-domain backprojection~\cite{Callow2003}, is among the fastest methods for generating volumetric representations~\cite{gerg2020gpu}. However, it is prone to geometric and acoustic inaccuracies, particularly when viewpoints are limited or sparsely distributed, necessitating dense sampling that is often impractical in real-world deployments. Recent approach~\cite{reed2023neural} address these limitations by representing the 3D scene as an implicit surface: an MLP is trained to predict the complex scattering field, with parameters optimized in an analysis-by-synthesis fashion. However, a key limitation of this methods for 3D SAS is their treatment of each voxel as an isotropic density term, implicitly assuming that scattering properties are identical in all directions, thus making a diffuse assumption. In practice, the scattering field often exhibits strong directionality, with the measured response varying as a function of both the incident and received angles due to underlying surface structure and material properties. As a result, isotropic models remain fundamentally limited in their ability to accurately capture anisotropic phenomena, motivating the need for more expressive representations that can efficiently encode these directional dependencies.

In this paper, we perform enhanced 3D scene and acoustic scatterer reconstruction for synthetic aperture sonar. Specifically, we propose an implicit neural representation, SH-SAS, that parameterizes the complex scattering field at each spatial location using spherical harmonic coefficients. This approach enables a more expressive and compact modeling of the underlying scattering phenomena compared to conventional isotropic volumetric neural methods.

Our formulation allows for direct end-to-end training from raw time-of-flight (ToF) measurements leveraging a point-based sonar scattering model originally introduced by Brown et al.~\cite{brown2017point} and made differentiable by Reed et al.~\cite{reed2023neural}, eliminating the need for voxelized supervision and streamlining the reconstruction pipeline. 
% Through comprehensive experiments on standard SAS datasets, we demonstrate that our method achieves improved 3D reconstruction quality, with sharper features and reduced artifacts, while also exhibiting faster convergence during training. While Spherical Harmonics are traditionally used to model the effects of illumination and view-dependent radiance, we use them to model the view-dependent complex acoustic scattering due to inter-reflection. Moreover, we do not model incident wave effects in this study, as a full acoustic BSDF formulation exceeds our scope.

% Notably, while the spherical harmonics basis is capable of expressing directionality in principle, our improvements originate primarily from a better fit to ground truth measurements rather than explicit modeling of highly anisotropic effects.

The key contributions of SH-SAS are as follows:
\begin{itemize}
\item \textbf{Implicit Spherical Harmonic Representation:} We introduce SH-SAS, a neural framework that implicitly models the complex scattering field at each voxel using compact spherical harmonic coefficients, offering improved expressive power over isotropic neural volumetric methods.

\item \textbf{Improved 3D Reconstruction Quality:} Experiments demonstrate that our method produces sharper features and fewer artifacts on simulated and real (both in-air and underwater) SAS data, yielding higher-fidelity 3D scene reconstruction compared to conventional neural methods.

\item \textbf{Code Release:} We will release our implementation as open source after peer review, including the simulation tools, experimental frameworks, and datasets used in our study (see anonymous submission for the first version). This will enable reproducibility and facilitate further research in this domain.

% \item \textbf{Faster Training Convergence:} Leveraging the spherical harmonics basis yields more efficient optimization, resulting in faster convergence during the training process.
\end{itemize}

\begin{figure*}
  \centering
  \includegraphics[width=0.8\textwidth]{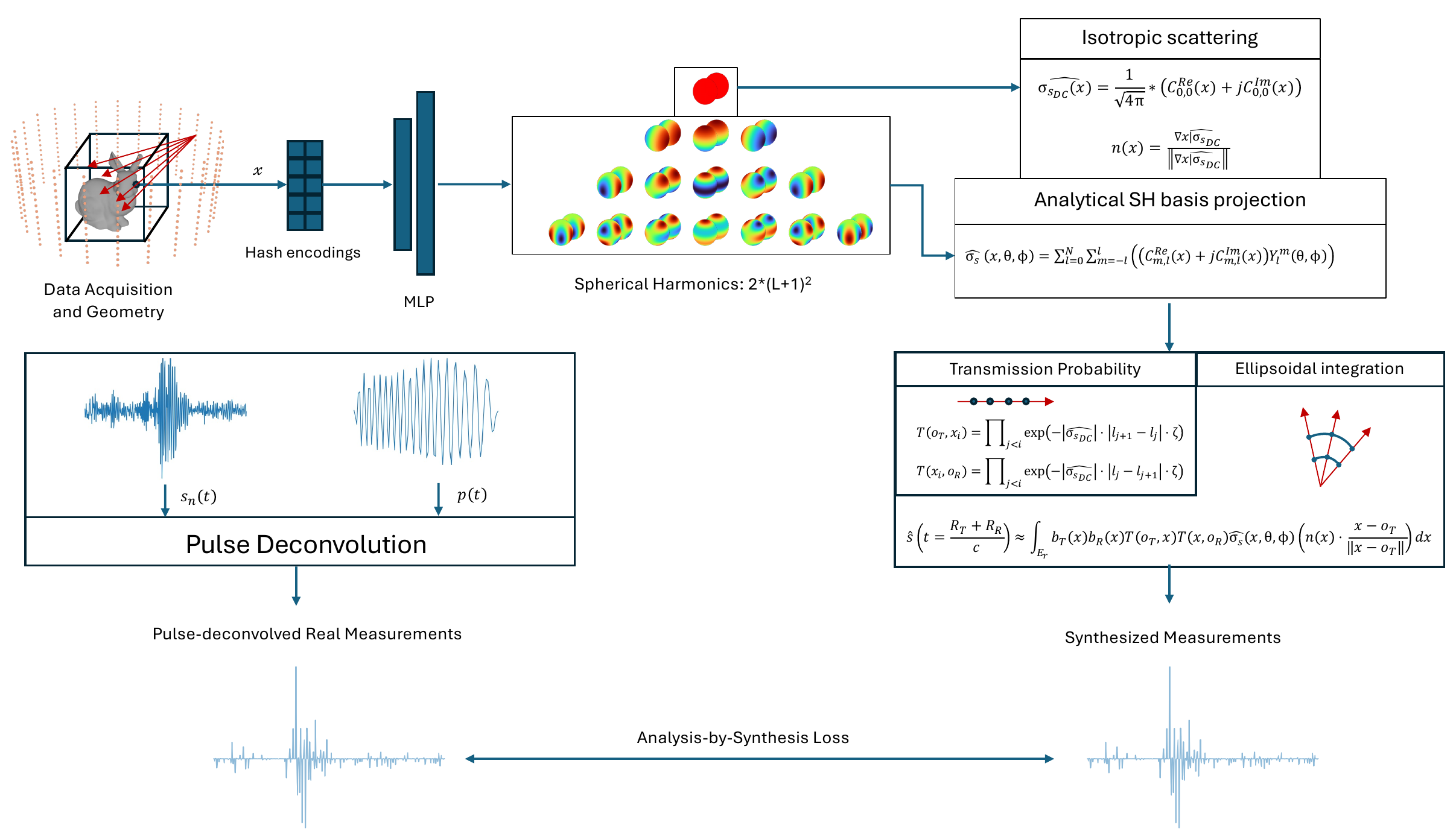}
  % \vspace{-5pt}
  \caption{Pipeline: Neural-field architecture for Synthetic Aperture Sonar Reconstruction. Our approach uses hash encodings and a multi-layer perceptron (MLP) that outputs complex scattering coefficients. The isotropic component $\hat{\sigma}_{s_{DC}}(x)$ is computed from the DC spherical harmonic coefficient, while the directional scattering $\hat{\sigma}_s(x, \theta, \phi)$ is modeled using spherical harmonics (SH) basis functions with analytical projection. Surface normals $n(x)$ are derived from the gradient of the isotropic scattering amplitude. Transmission probabilities $T(o_T,x)$ and $T(x,o_R)$ model acoustic propagation between transmitter/receiver positions and scatterer locations using ellipsoidal integration. The forward model synthesizes time-of-flight signals $\hat{s}(t)$ through volumetric integration over the ellipsoidal region $E_r$, incorporating beam patterns $b_T(x)$ and $b_R(x)$. During training, synthesized signals are compared against pulse-deconvolved measurements $s_{PD}(t)$ through an analysis-by-synthesis optimization framework.}
  \label{fig:pipeline}
\end{figure*}

\section{Related Work}
\label{sec:relatedwork}

\paragraph{Time-Domain Backprojection for Synthetic Aperture Sonar:} Synthetic aperture sonar (SAS) utilizes a distributed set of time-of-flight measurements, typically collected from a moving platform, to reconstruct scene geometry. The most widely used and flexible algorithm is time-domain backprojection (also called delay-and-sum beamforming), which backprojects received measurements to image voxels using their time-of-flight measurements~\cite{hayes2009synthetic, soumekh1999synthetic}. This algorithm can accommodate arbitrary synthetic aperture geometries, and recent work has accelerated it using GPU computing for real-time performance~\cite{baralli2013gpu, song2016processing, gerg2020gpu}. However, backprojection is computationally expensive, especially for high-resolution 3D imaging, and assumes that all voxels are equally visible to the sensor. As a result, traditional implementations do not account for self-occlusion, leading to artifacts when parts of the scene are shadowed or obscured. More specialized algorithms in the wavenumber domain~\cite{eaves2012principles, hayes2009synthetic}, circular scanning geometries~\cite{plotnick2014fast, marston2011coherent, marston2014autofocusing}, and interferometry~\cite{hansen2003signal,griffiths1997interferometric,gens1996review} exist for 2D and 3D reconstruction, but typically require additional constraints on sampling, acquisition, or hardware, which we do not consider in this manuscript.

\paragraph{Neural Fields for ToF Modalities:} Neural Radiance Fields (NeRF) \cite{mildenhall2020nerf} revolutionized 3D scene representation by introducing implicit volumetric representations using multi-layer perceptrons (MLPs), enabling photorealistic view synthesis from sparse image collections. This neural field paradigm has since been successfully extended to diverse sensing modalities, including radar, sonar, and time-resolved imaging systems \cite{malik2023transient, Luo_2025_WACV, reed2023neural, borts2024radar}, with each adaptation carefully tailored to the underlying sensor physics and measurement characteristics. In optical time-of-flight sensing, researchers have integrated NeRF-style architectures with transient-based volume rendering techniques to achieve significant improvements in continuous-wave ToF systems \cite{attal2021torf}, non-line-of-sight imaging using SPAD sensors \cite{NTF}, and superior 3D reconstruction for single-photon LIDAR applications \cite{malik2023transient,Luo_2025_WACV}. However, these optical ToF implementations share a common characteristic: their hardware can precisely focus light collection along individual rays through optical elements such as lenses or scanning mirrors. In contrast, our work addresses synthetic aperture measurements where time-of-flight data can only be determined for rays within a broad field-of-view, necessitating sophisticated backprojection or computational focusing algorithms to recover accurate 3D geometry. 

These advances have also been successfully translated to radar applications, including synthetic aperture radar (SAR) \cite{barbierrenard2025multiview3dsurfacereconstruction}, inverse SAR \cite{oshim2024nerf,deng2024isar}, FMCW radar systems \cite{borts2024radar,takawale2025spinr}, and Doppler-based measurements \cite{huang2024dart}, with neural fields also proving effective for multimodal fusion, particularly in combining imaging sonar with optical imagery for enhanced performance \cite{qadri2024aoneus}. The acoustic sensing community has witnessed remarkable progress in applying neural fields to various applications, with advances extending to medical ultrasound imaging \cite{chen2024neural}, while imaging sonars with beam-steering capabilities have been enhanced through differentiable rendering techniques \cite{qadri2023neural,NVRFLS}. For synthetic aperture sonar specifically, researchers have formulated coherent imaging as analysis-by-synthesis optimization problems, leading to improved 2D circular SAS reconstructions \cite{SINR} and sophisticated 3D volumetric SAS imaging \cite{reed2023neural}. Our work builds most directly upon \cite{reed2023neural}, but introduces key innovations including directional scattering modeling via Spherical Harmonics representation and significantly improved reconstruction quality with faster convergence times.

\section{Background}
\label{sec:background}

In this section, we describe the forward measurement model based on point-based sonar scattering~\cite{brown2017modeling,brown2017point} and outline a version of time-domain backprojection~\cite{Callow2003} for 3D reconstruction. For our processing pipeline in particular, we avoid traditional matched-filtering approaches in favor of the recently proposed pulse deconvolution technique~\cite{reed2023neural} to improve coherent backprojection results.

\subsection{Pulse Deconvolution}

To obtain deconvolved, temporally compact signals from raw synthetic aperture sonar (SAS) measurements, we follow the neural pulse deconvolution approach introduced by Reed et al.~\cite{reed2023neural}. In this method, an implicit neural representation (INR), denoted $\mathcal{N}_{PD}$, is optimized for each batch of measurements by minimizing the loss
\begin{align}
&\mathcal{L}_{\mathrm{PD}} = \left\|\ \mathcal{N}_{\mathrm{PD}}(t;\, \theta_{\mathrm{PD}}) * p(-t) - s(t) \,\right\|_2^2 \nonumber \\
&+ \lambda_1 \sum_t \left\| \mathcal{N}_{\mathrm{PD}}(t;\, \theta_{\mathrm{PD}}) \right\|_1 + \lambda_2 \sum_t \left\| \nabla \angle \mathcal{N}_{\mathrm{PD}}(t;\, \theta_{\mathrm{PD}}) \right\|_1,
\end{align}
where $s(t)$ is the measured SAS signal, $p(t)$ is the transmit pulse, $\theta_{PD}$ are the network parameters, and $\lambda_1$ , $\lambda_2$ are regularization weights. The first term enforces data fidelity, the second encourages sparsity in the deconvolved waveform, and the third promotes phase smoothness. Following Reed et al., the network is trained using the ADAM optimizer and a multi-resolution hash encoding. The resulting deconvolved signal, $s_{PD}(t) = \mathcal{N}_{PD}(t;\theta_{PD})$, is used for all subsequent processing steps.

\subsection{Ellipsoidal Sampling}
\label{sec:ellipsoidal-sampling}

We adopt \emph{ellipsoidal sampling} following Reed et al.~\cite{reed2023neural}: points with a constant time-of-flight (ToF) lie on an ellipsoid whose foci are the transmitter (TX) and receiver (RX). Rays are traced from the TX and we keep only ray–ellipsoid intersections consistent with the measured ToF trace. This physically grounded scheme matches SAS geometry, concentrates computation where the acoustic energy is highest, and improves both forward modeling and inversion efficiency. We use it as a core component of our pipeline; equations for the ellipsoidal sampling are provided in the supplemental material.

\section{Method}
\label{sec:method}
Our full pipeline is outlined in~\cref{fig:pipeline}. The main reconstruction goal is to estimate $\hat{\sigma}_s$, the complex acoustic scattering field at a position $(x,y,z)$. In this section, we discuss the novel contributions of our pipeline block-by-block and how they contribute to the final analysis-by-synthesis optimization, which yields high-quality 3D reconstruction.

\subsection{Network Design}
Our network architecture leverages hash encodings as introduced in Instant Neural Graphics Primitives \cite{mueller2022instantngp}, which enable efficient, high-resolution feature representation through multi-resolution grids. We configure the hash encoding with a base grid resolution of 16, scaling up to a maximum grid resolution of 4096 across 16 levels. This setup allows the network to capture fine-grained spatial details while maintaining computational efficiency, which is crucial for modeling complex acoustic fields.

At the core of the network is a compact multilayer perceptron (MLP) consisting of two hidden layers, each with a hidden dimension of 32 with hidden ReLU activations. This lightweight design minimizes parameter count and training overhead, making it suitable for scenarios with limited computational resources. The MLP processes the encoded features and outputs a vector of dimension $2*(L+1)^2$, corresponding to two-channel spherical harmonic coefficients (real and imaginary components) up to degree $L$. By keeping the architecture simple yet effective, we ensure rapid convergence during training, which is particularly beneficial for sparse sonar data. This configuration forms the backbone of our method for reconstructing directional scattering fields from 1D signals.

\subsection{Spherical Harmonic Modeling}

Traditional backprojection aggregates directional energy across all sensor measurements at a 3D point. Although sufficient for localization, this isotropic accumulation limits the ability to resolve view-dependent scattering.

To better learn the view-dependent scattering of $\hat{\sigma}_{s}$, we utilize spherical harmonics (SH)~\cite{seeley1966spherical,muller2006spherical}, a compact, continuous basis for angular functions. SH have been used often for modeling view-dependent reflectance in inverse rendering~\cite{ramamoorthi2001signal}. Each sampled point $x$ is parameterized by a set of learnable SH coefficients. The complex scattering function at point $x$ with incident ray direction $(\theta,\phi)$ is given by the following equation:
\begin{equation}
\hat{\sigma}_{s}(x, \theta, \phi) = \sum_{l=0}^{L} \sum_{m=-l}^{l} c_{l,m}(x) Y_{l,m}(\theta, \phi),
\end{equation}
$$
c_{l,m}(x) = c^{\mathrm{Re}}_{l,m}(x) + j c^{\mathrm{Im}}_{l,m}(x),
$$
where $c_{l,m}(x)$ are 2-channel SH coefficients defined at the point $x$, and $Y_{l,m}$ are SH basis functions evaluated at $(\theta, \phi)$. This enables learnable, continuous directional scattering.

\subsection{Geometry}

In optical neural rendering, geometry is usually separated from appearance. NeRF \cite{mildenhall2020nerf} learns a continuous density field $\sigma(x)$ that governs opacity for volumetric rendering, but it neither predicts nor uses surface normals during training; surfaces are typically recovered \emph{post hoc} by thresholding $\sigma$ (or opacity) and extracting an iso-surface with marching cubes. By contrast, several INR variants estimate normals as gradients of a learned scalar field e.g., the SDF in NeuS \cite{wang2021neus} via $\nabla \Phi$, or, in density/occupancy formulations, via $\nabla \sigma$, yet the surface is still obtained by iso-surface extraction. In all cases, density primarily encodes geometric occupancy, while view-dependent color/reflectance is modeled separately.

Acoustic scattering more directly ties geometry to the signal: the scattering coefficient itself often acts as a proxy for density. For example, Reed et al.~\cite{reed2023neural} use isotropic scatterers $\hat{\sigma}_s$ whose strength forms the density field, with surface normals given by the negative gradient of $|\hat{\sigma}_s|$.

We adopt this acoustic viewpoint but with directional, complex-valued scattering via spherical harmonics (SH). Let $c_{0,0}(x)$ be the complex DC SH coefficient of the scattering field. We define
\begin{align}
\hat{\sigma}_{s,\mathrm{DC}}(x) &= \frac{1}{\sqrt{4\pi}}\,c_{0,0}(x),\\
\hat{\rho}(x) &= \big|\hat{\sigma}_{s,\mathrm{DC}}(x)\big|\,\zeta,
\end{align}
where $\zeta$ is an occlusion scaling factor. Surface normals follow as
\begin{align}
\mathbf{n}(x)= -\frac{\nabla_{\mathbf{x}}\big|\hat{\sigma}_{s,\mathrm{DC}}(x)\big|}{\left\|\nabla_{\mathbf{x}}\big|\hat{\sigma}_{s,\mathrm{DC}}(x)\big|\right\|}.
\end{align}
This yields a physically grounded isotropic density from the DC term while preserving directional (anisotropic) scattering in higher-order SH coefficients. Unlike purely isotropic models \cite{reed2023neural}, our learnable SH representation captures geometric detail \emph{and} complex view-dependent behavior, improving robustness under occlusions and varying scattering directions.

\subsection{Time-Resolved Volume Rendering}
\label{sec:synthesis}

 The final step in our volumetric rendering pipeline is to synthesize the analytic transient measurements $\hat{s}_n(t)$. We use the modeling trick of treating $\hat{\sigma}_s$ as a complex scattering field and thus utilize the complex forward model for analytic signals derived by \citet{reed2023neural}:

% \begin{equation}
%    \hat{s}_n\left(t=\frac{R_T+R_R}{c}\right) \approx \int_{\mathbf{E}_{r}} b_T(\mathbf{x})b_R(\mathbf{x})T(\mathbf{o}_T, \mathbf{x})T(\mathbf{x}, \mathbf{o}_R)\hat{\sigma}_S(\mathbf{x},\theta,\phi)g(\omega_{\textbf{x}})\mathrm{d}\mathbf{x},
%    \label{eq:approximateforwardmodel}
% \end{equation}

\small{
\begin{equation}
\begin{gathered}
    K(x, o_T, o_R) = b_T(\mathbf{x})b_R(\mathbf{x})T(\mathbf{o}_T, \mathbf{x})T(\mathbf{x}, \mathbf{o}_R), \\
   \hat{s}_n\left(t=\frac{R_T+R_R}{c}\right) \approx \int_{\mathbf{E}_{r}} K(x, o_T, o_R)\hat{\sigma}_s(\mathbf{x},\theta,\phi)g(\omega_{\textbf{x}})\mathrm{d}\mathbf{x},
\end{gathered}
\label{eq:approximateforwardmodel}
\end{equation}
}
where $\mathbf{E}_{r}$ is the ellipsoid defined by the time-of-flight $\frac{R_T+R_R}{c}$, and foci $o_T$, $o_R$. ${\hat\sigma}_s$ is now a complex function with outgoing ray from receiver origin($\textbf{o}_R$) to spatial coordinate $\textbf{x}$ parametrized by ($\theta,\phi)$, $b_T(x)$ and $b_R(x)$ are transmitter and receiver directivity functions and $g(\omega_{\textbf{x}}) = \max\left(0, \mathbf{n}(\mathbf{x}) \cdot \frac{o_T - \mathbf{x}}{\| o_T - \mathbf{x} \|}\right)$ is the Lambertian cosine term. Similar to \cite{reed2023neural}, we use $b_R(x)=1$ for all our experiments.
% Note that this equation is derived under the assumption of \textit{ideal pulse deconvolution}, and thus is only an approximation for Equation~\ref{eq:pointbasedspectrum}. 

The transmission probability along a ray to point $\mathbf{x}$ is accumulated volumetrically (similar to~\citet{mildenhall2020nerf}):
\begin{equation}
T(o_T, \mathbf{x_i}) = \prod_{j < i} \exp\left(-\hat{\rho}(x_j) \cdot |l_{j+1} - l_j|\right),
\end{equation}
where $x_j$ are sampled points along the ray leading up to $\mathbf{x}$, and $l_j$ are the bin endpoints for this ray marching. 

Combining these components, we synthesize the analytic signal $\hat{s}_n(t)$ using ~\cref{eq:approximateforwardmodel} and optimize the neural field parameters through an analysis-by-synthesis framework. The synthesized signals are compared against the pulse-deconvolved ground truth measurements $s_{PD}(t)$ using a composite loss function:
\begin{equation}
    \mathcal{L} = \lambda_1\mathcal{L}_{\text{ToF}} + \lambda_2\mathcal{L}_{\text{Sparse}} + \lambda_3\mathcal{L}_{\text{TV}}^{\text{Density}} + \lambda_4\mathcal{L}_{\text{TV}}^{\text{Scatter}} + \lambda_5\mathcal{L}_{\text{TV}}^{\text{Phase}},
\end{equation}
where the individual loss terms enforce data fidelity and regularization constraints:
\begin{align*}
\mathcal{L}_{\text{ToF}} &= \sum_n||s_n(t) - \hat{s}_n(t)||_2, \quad \mathcal{L}_{\text{Sparse}} = \sum_{n} \Big\|\big|\hat{\rho}\big|\Big\|_1, \\
    \mathcal{L}_{\text{TV}}^{\text{Density}} &= \sum_n ||\nabla_{d_{\text{reg}}} \hat{\rho}||_1, \quad \mathcal{L}_{\text{TV}}^{\text{Scatter}} = \sum_n ||\nabla_{d_{\text{reg}}} \hat{\sigma_s}||_1, \\
    \mathcal{L}_{\text{TV}}^{\text{Phase}} &= \sum_n ||\nabla_{d_{\text{reg}}} \angle \hat{\sigma}_s||_1.
\end{align*}
The time-of-flight term $\mathcal{L}_{\text{ToF}}$ ensures data consistency between synthesized and measured signals, while $\mathcal{L}_{\text{Sparse}}$ promotes sparsity in the density field. The total variation terms enforce spatial smoothness in the density ($\mathcal{L}_{\text{TV}}^{\text{Density}}$), scattering amplitude ($\mathcal{L}_{\text{TV}}^{\text{Scatter}}$), and scattering phase ($\mathcal{L}_{\text{TV}}^{\text{Phase}}$) fields.

% \begin{align}
%     \mathcal{L}_{\text{ToF}} &= \sum_n||s_n(t) - \hat{s}_n(t)||, \\
%     \mathcal{L}_{\text{Sparse}} &= \sum_{n} \Big\|\big|\hat{\rho}\big|\Big\|_1, \\
%     \mathcal{L}_{\text{TV}_{\text{Space}}}^{\text{Density}} &= \sum_n ||\nabla_{d_{reg}} \hat{\rho}||_1, \\
%     \mathcal{L}_{\text{TV}_{\text{Space}}}^{\text{Scatter}} &= \sum_n ||\nabla_{d_{reg}} \hat{\sigma_s}||_1, \\
%     \mathcal{L}_{\text{TV}_{\text{Phase}}}^{\text{Scatter}} &= \sum_n ||\nabla_{d_{reg}} \angle \hat{\sigma}_s||
% \end{align}

\section{Data and Implementation}
\label{sec:data}

This section describes the datasets used in our experiments, comprising both simulated synthetic aperture sonar (SAS) measurements and real-world acquisitions. The combination of synthetic and measured data enables comprehensive validation of our reconstruction methods under controlled conditions and realistic scenarios.

\subsection{Simulated SAS Data}
\label{sec:simulated-data}

We synthesize synthetic aperture sonar (SAS) measurements using a transient time-of-flight (ToF) renderer built on CUDA/OptiX. The renderer aggregates returned radiance into discrete travel-time bins to form per-viewpoint transient histograms. Transmitter (TX) and receiver (RX) follow concentric circular trajectories around the target with 54,000 different synthetic aperture positions. 
% TX positions: radius $R_{\mathrm{tx}} \approx 0.85\,\mathrm{m}$, elevation $z_{\mathrm{tx}} \in [0.11, 0.86]\,\mathrm{m}$; RX positions: radius $R_{\mathrm{rx}} \approx 0.86\,\mathrm{m}$, elevation $z_{\mathrm{rx}} \in [0.07, 0.82]\,\mathrm{m}$. This creates a constant 13mm radial and 38mm vertical offset. 
Each transient has $N_t = 14{,}360$ bins at $f_s = 100\,\mathrm{kHz}$ ($\Delta t = 10\,\mu\mathrm{s}$). For ray path length $\ell$, arrival-time index is $k = \lfloor \ell f_s / c \rfloor$. We transmit a real LFM chirp sweeping 10–30 kHz (bandwidth $B = 20\,\text{kHz}$):
$$p(t) = w(t)\cos\big(2\pi(f_0 t + \frac{1}{2}\alpha t^2)\big),$$
where $f_0 = 10$ kHz and $\alpha = B/T_{\text{dur}}$. The received signal is:
$$s_i[k] = (h_{\text{tr},i} * p)[k] + n_i[k].$$
% We use the same simulator implementation as Reed et al.~\cite{reed2023neural}.
\textbf{Implementation:} The transient pass uses CUDA atomicAdd into histogram bins with per-block shared-memory optimization. We model geometric acoustics but exclude diffraction and sub-wavelength effects. The simulator enables controlled experiments over bandwidth, noise (SNR), and object geometry.

\subsection{AirSAS Measurements}
\label{sec:airsas-data}

Our experiments utilize data from the AirSAS platform, a circular synthetic aperture sonar system operating in air within an anechoic environment~\cite{Blanford2019}. This in-air configuration provides experimental advantages and controllability that would be difficult to achieve in underwater settings, while maintaining acoustically analogous physics relevant to our reconstruction methods. The AirSAS dataset has been validated across multiple studies for sonar imaging research \cite{blanford2022leveraging, cowen2021airsas, goehle2023approximate, park2020alternative, reed2023neural}.
The system configuration consists of a transmit element (Peerless OX20SC02-04 tweeter) and receive element (GRAS 46AM microphone) positioned approximately 1 meter from a rotating turntable. Data acquisition involves transmitting 1 ms linear frequency modulated (LFM) chirps with center frequency $f_c = 20$ kHz and 20 kHz  bandwidth. Side-lobe suppression is achieved through Tukey windowing with a 0.1 ratio applied to the transmitted waveform.
The measurement protocol captures data at 1-degree azimuthal increments over full 360° rotations of a 0.2×0.2 meter turntable containing 3D printed test objects. Vertical sampling is performed via 5 mm increments using a linear translation stage after each complete rotation. The resulting spatial sampling satisfies standard SAS criteria where inter-element spacing $D \leq \lambda_{\text{min}}/2$, with $\lambda_{\text{min}}$ representing the minimum wavelength of the transmit signal \cite{Callow2003}. 
% This dense sampling enables flexible post-processing options including helical trajectory reconstruction and sparse-angle analysis through sub-sampling of the complete dataset.

\subsection{Sediment Volume Search Sonar (SVSS)}
\label{sec:svss-data}

We further evaluate our approach on underwater measurements from the Sediment Volume Search Sonar (SVSS) platform~\cite{svss-tech-report}, designed for subsurface imaging through lake-bed sediments. SVSS uses linear frequency-modulated (LFM) transmissions with center frequency $f_c=27.5\,\mathrm{kHz}$, bandwidth $\Delta f=15\,\mathrm{kHz}$, and pulse duration $255\,\mu\mathrm{s}$; a Taylor window is applied for sidelobe control~\cite{svss-tech-report}.

% \begin{wrapfigure}{r}{0.46\columnwidth} % r=right, l=left
%   \vspace{-6pt}                         % pull up a bit (optional)
%   \centering
%   \includegraphics[width=\linewidth]{images/svss-cylinder_gt.jpg}
%   \caption{SVSS cylinder target.}
%   \label{fig:svss_cyl_small}
%   \vspace{-10pt}                        % tighten below (optional)
% \end{wrapfigure}

Data were collected with a pontoon-mounted system and precision navigation at Foster Joseph Sayers Reservoir (PA), where calibrated targets were deployed for controlled experiments~\cite{svss-tech-report}. The array comprises five transmitters (sequential firing) and 80 active receivers. Unlike AirSAS, SVSS operates in a bistatic geometry with spatially separated TX/RX; we handle the TX–RX separation in our forward model following Reed et al.~\cite{reed2023neural}.

Our experiments use a single representative exemplar, shown in~\cref{fig:svss_results}, from the SVSS dataset provided by the program sponsors.

\subsection{Implementation Details}

All experiments are implemented in PyTorch. To enable efficient spherical-harmonics (SH) basis projection, we use a custom CUDA kernel wrapped in a \texttt{torch.autograd.Function} that outputs complex-valued scatterers. We train with Adam~\cite{kingma2014adam} at a fixed learning rate of $10^{-3}$ and use SH up to degree $L{=}3$. Meshes are extracted from point clouds via marching cubes~\cite{Lorensen1987mcubes}.

For AirSAS, we employ the \emph{same regularization scheme as Reed et al.}~\cite{reed2023neural}—total variation (TV) and $\ell_1$ sparsity with matching settings—to keep the comparison fair; these priors are disabled for the Simulated and SVSS datasets. The added priors make AirSAS iterations slightly slower per step, but our approach remains faster overall and converges in fewer iterations than Reed et al.~\cite{reed2023neural}.

% Our method runs with nearly the same per-iteration latency as~\cite{reed2023neural}—and is 10–15 minutes faster for the same number of iterations—while also converging in fewer training steps. On real data, we train both the baseline~\cite{reed2023neural} and our method for 26,000 iterations. For simulated data, we follow the original implementation and train~\cite{reed2023neural} for 100,000 iterations, whereas our method achieves convergence in approximately 50,000 iterations.

\subsection{Baselines and Metrics}
We evaluate our approach against two baselines: time-domain backprojection and Reed et al.~\cite{reed2023neural} across three dataset types (Simulated, AirSAS, and SVSS). Evaluation utilizes four core metrics—Chamfer distance, intersection-over-union (IoU), precision, and F1-score—on both point cloud and mesh representations.

For point cloud metrics, we compare the model-generated point cloud against a reference point cloud derived from the ground truth mesh. For mesh-based metrics, we convert the generated point cloud to a mesh and then back to a point cloud, enabling a consistency check against the ground truth mesh's point cloud. Importantly, metric scores differ between raw point clouds and those produced through mesh conversion.

% Across all four evaluated synthetic scenes, our method demonstrates superior performance to both baselines, resulting in lower average Chamfer distance (indicating better geometric alignment) and higher average IoU, precision, and F1-score in both representations. See Table~\ref{tab:metrics-summary} for a quantitative comparison of averaged results, with individual scene metrics provided in the supplemental material.

\section{Experimental Results}
\label{sec:results}

\subsection{Simulated Results}
\label{sec:simulatedresults}
\begin{figure}[H]
    \centering
    % -------- Row 1: Buddha --------
    \begin{subfigure}[t]{0.24\columnwidth}
        \centering
        \includegraphics[width=\linewidth]{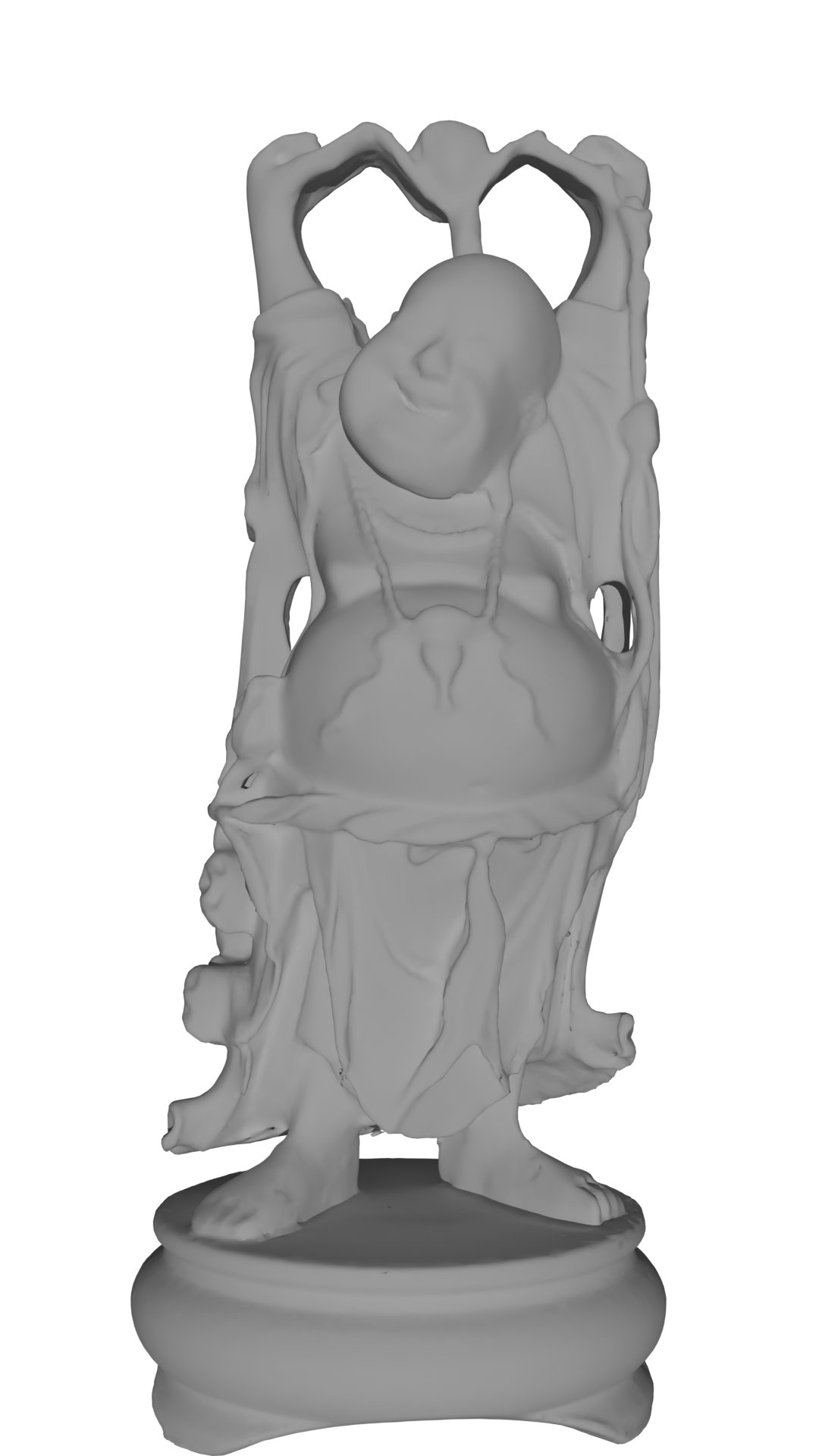}
    \end{subfigure}
    \hfill
    \begin{subfigure}[t]{0.24\columnwidth}
        \centering
        \includegraphics[width=\linewidth]{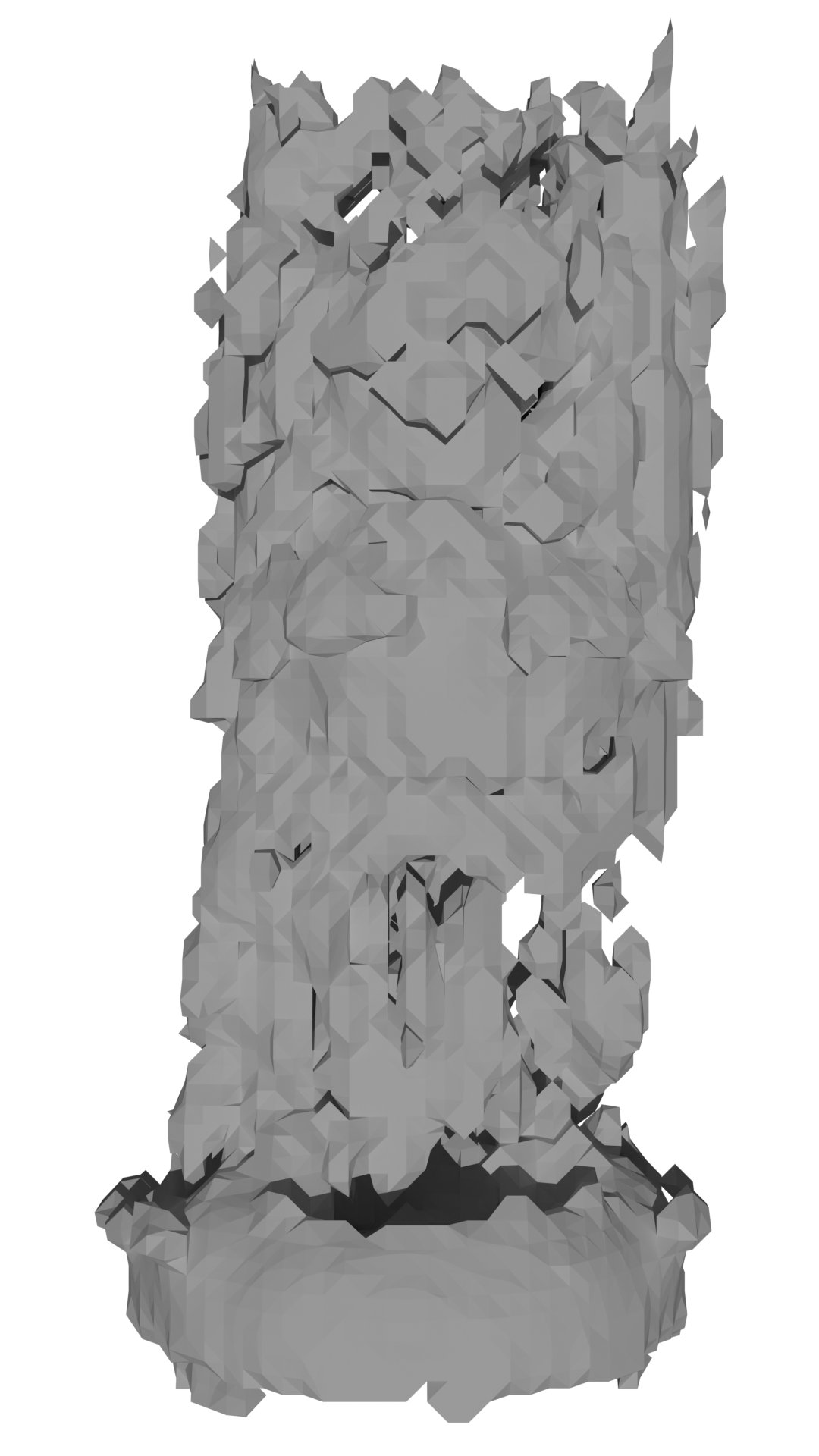}
    \end{subfigure}
    \hfill
    \begin{subfigure}[t]{0.24\columnwidth}
        \centering
        \includegraphics[width=\linewidth]{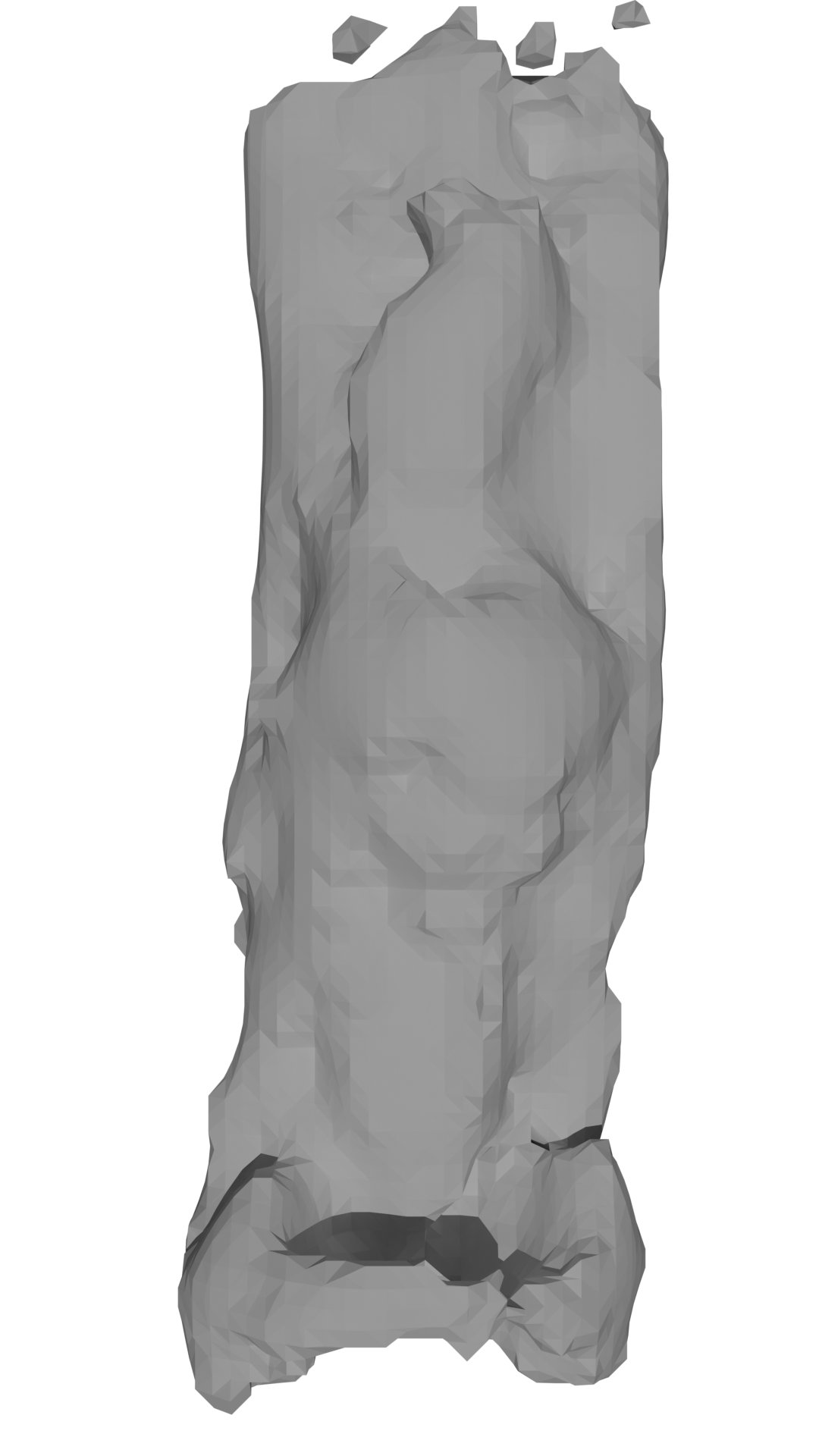}
    \end{subfigure}
    \hfill
    \begin{subfigure}[t]{0.24\columnwidth}
        \centering
        \includegraphics[width=\linewidth]{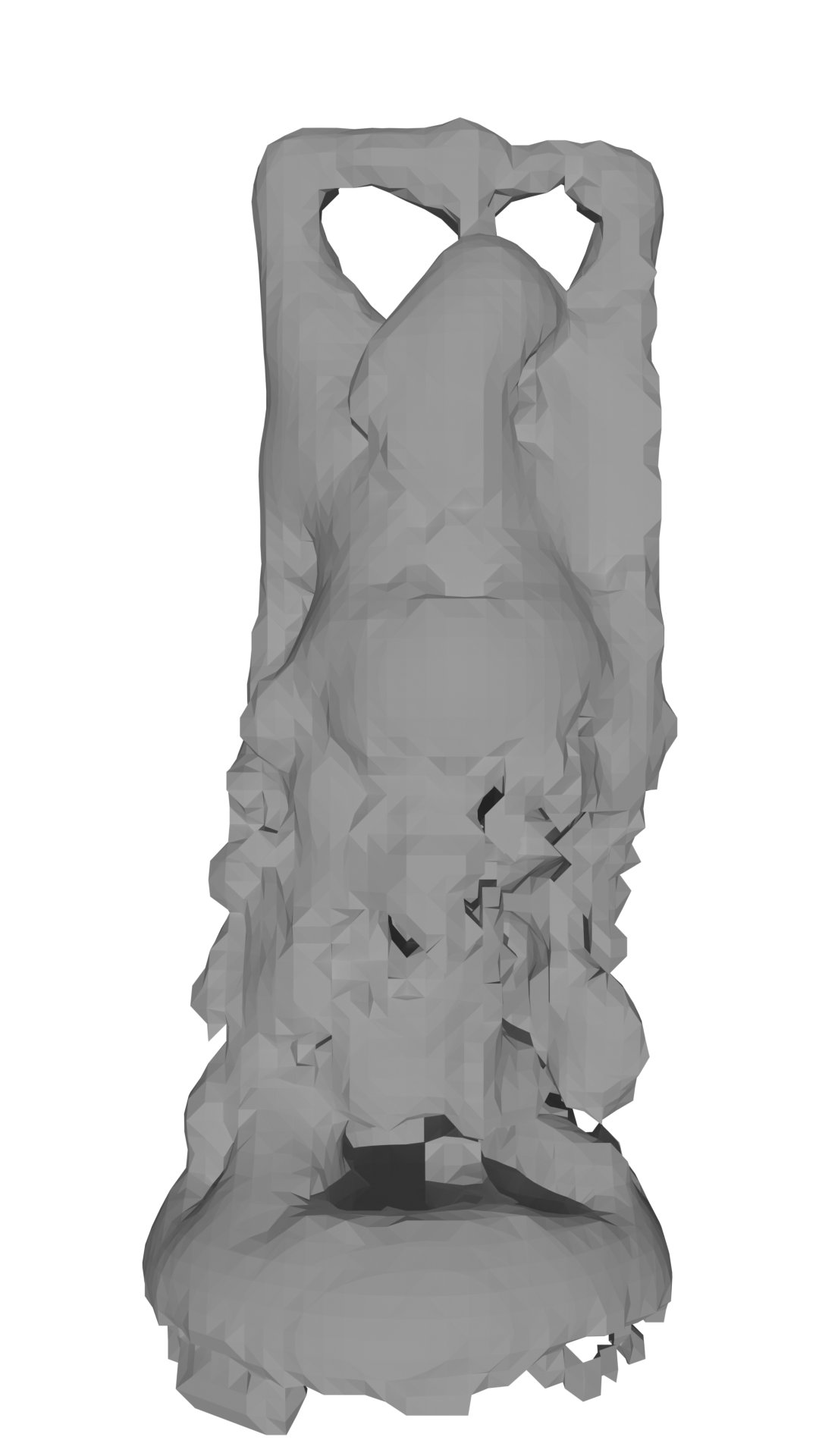}
    \end{subfigure}
    
    \vspace{6pt} % space between rows
    
    % -------- Row 2: Armadillo --------
    \begin{subfigure}[t]{0.24\columnwidth}
        \centering
        \includegraphics[width=\linewidth]{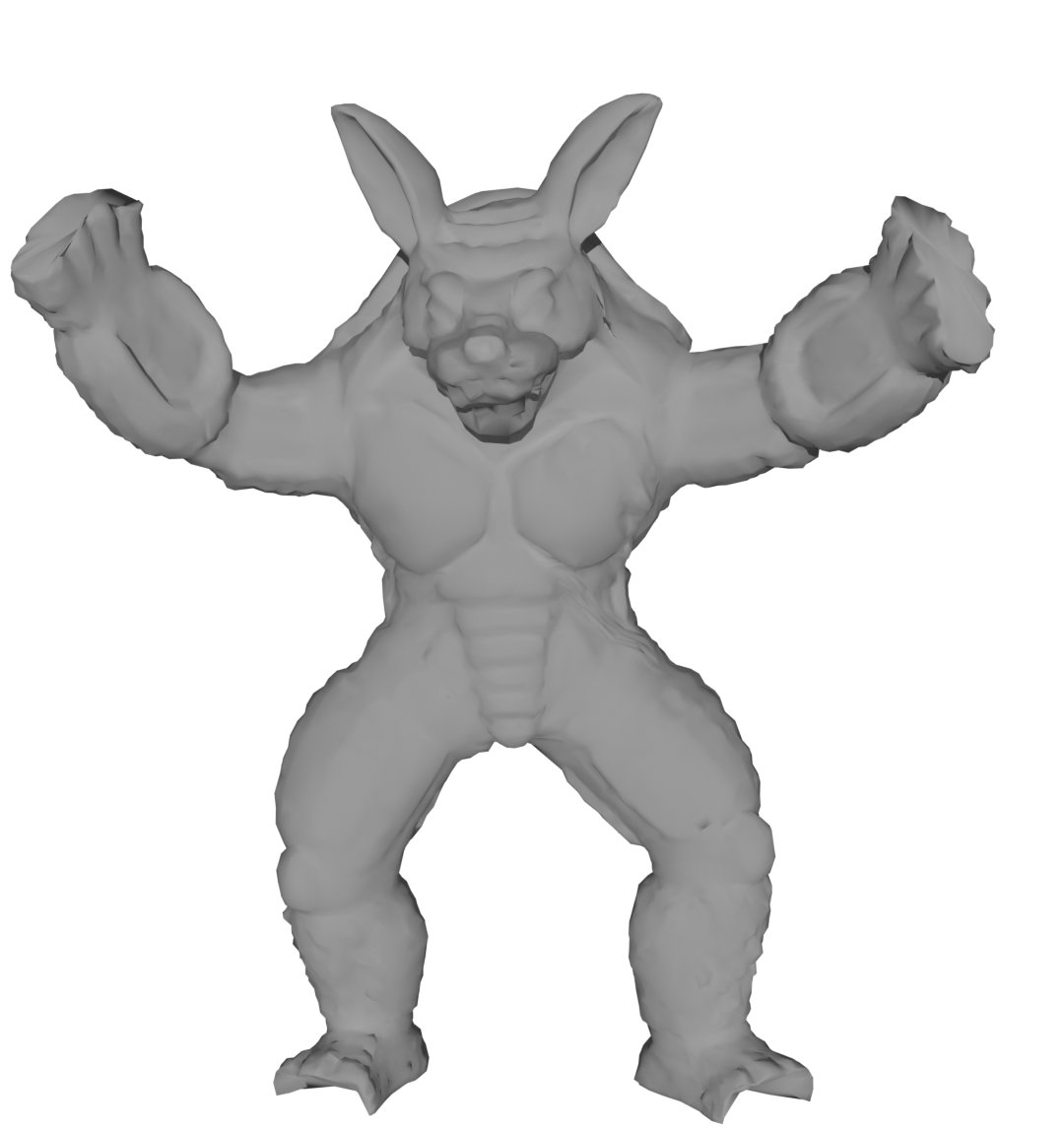}
    \end{subfigure}
    \hfill
    \begin{subfigure}[t]{0.24\columnwidth}
        \centering
        \includegraphics[width=\linewidth]{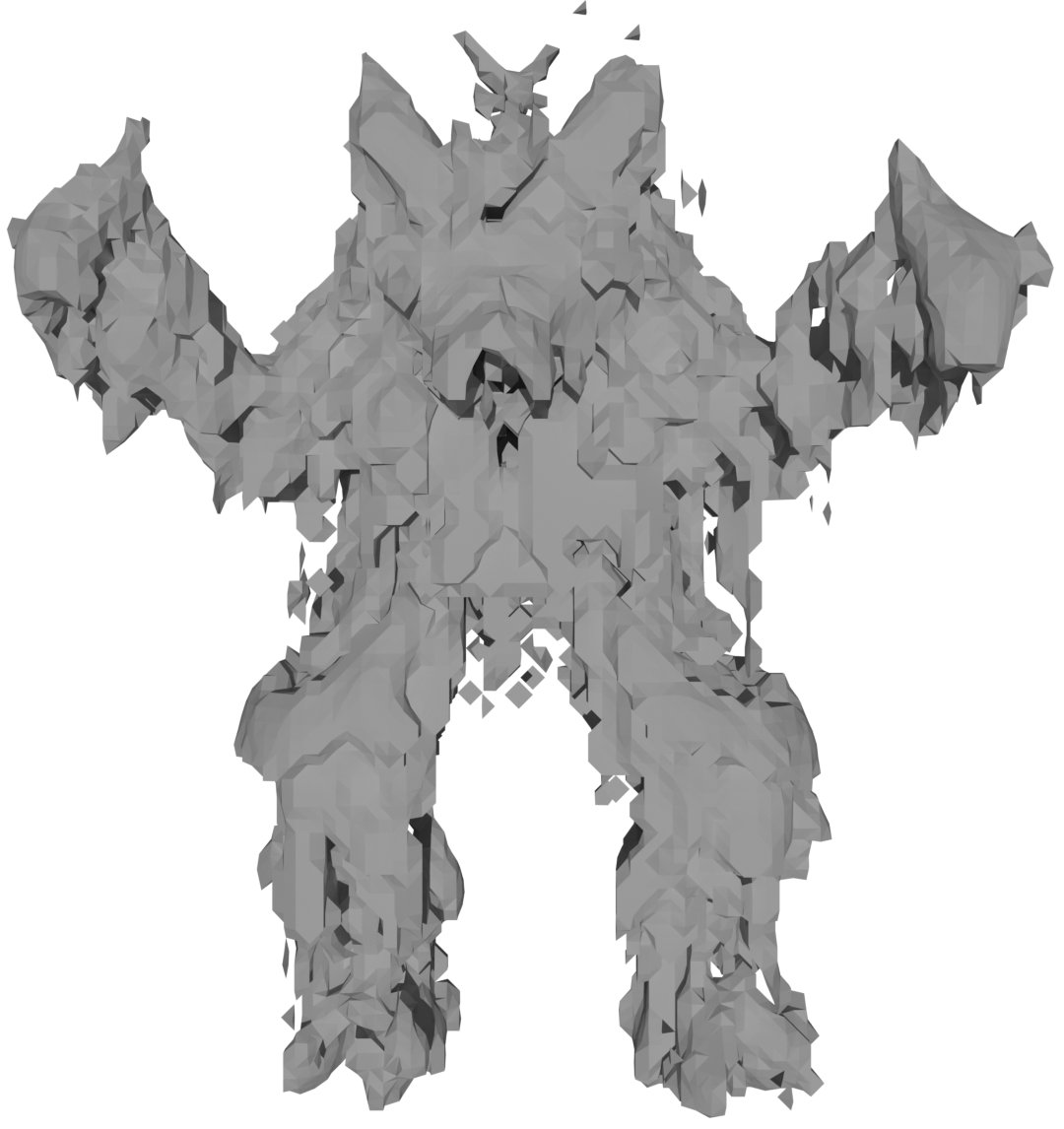}
    \end{subfigure}
    \hfill
    \begin{subfigure}[t]{0.24\columnwidth}
        \centering
        \includegraphics[width=\linewidth]{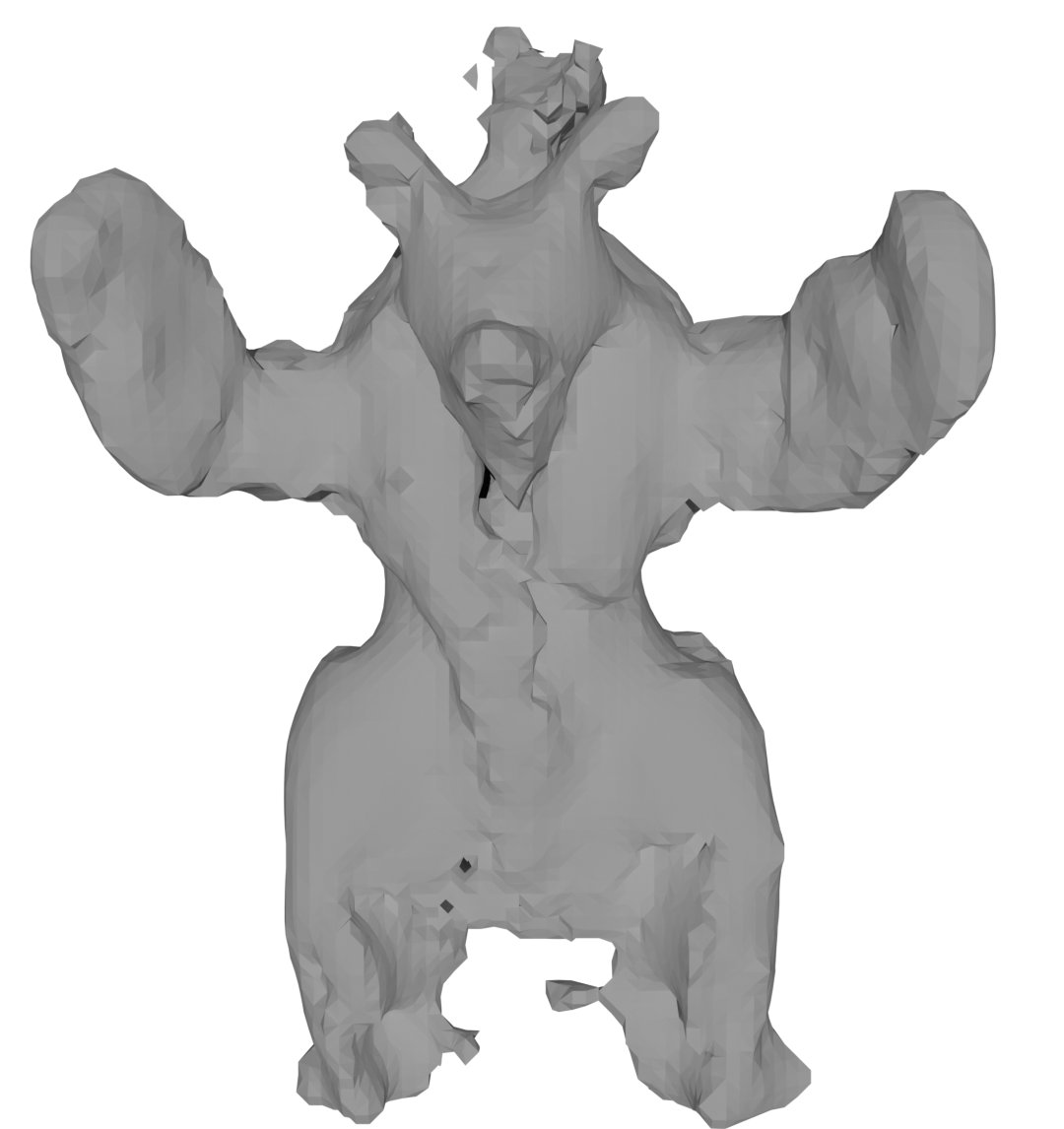}
    \end{subfigure}
    \hfill
    \begin{subfigure}[t]{0.24\columnwidth}
        \centering
        \includegraphics[width=\linewidth]{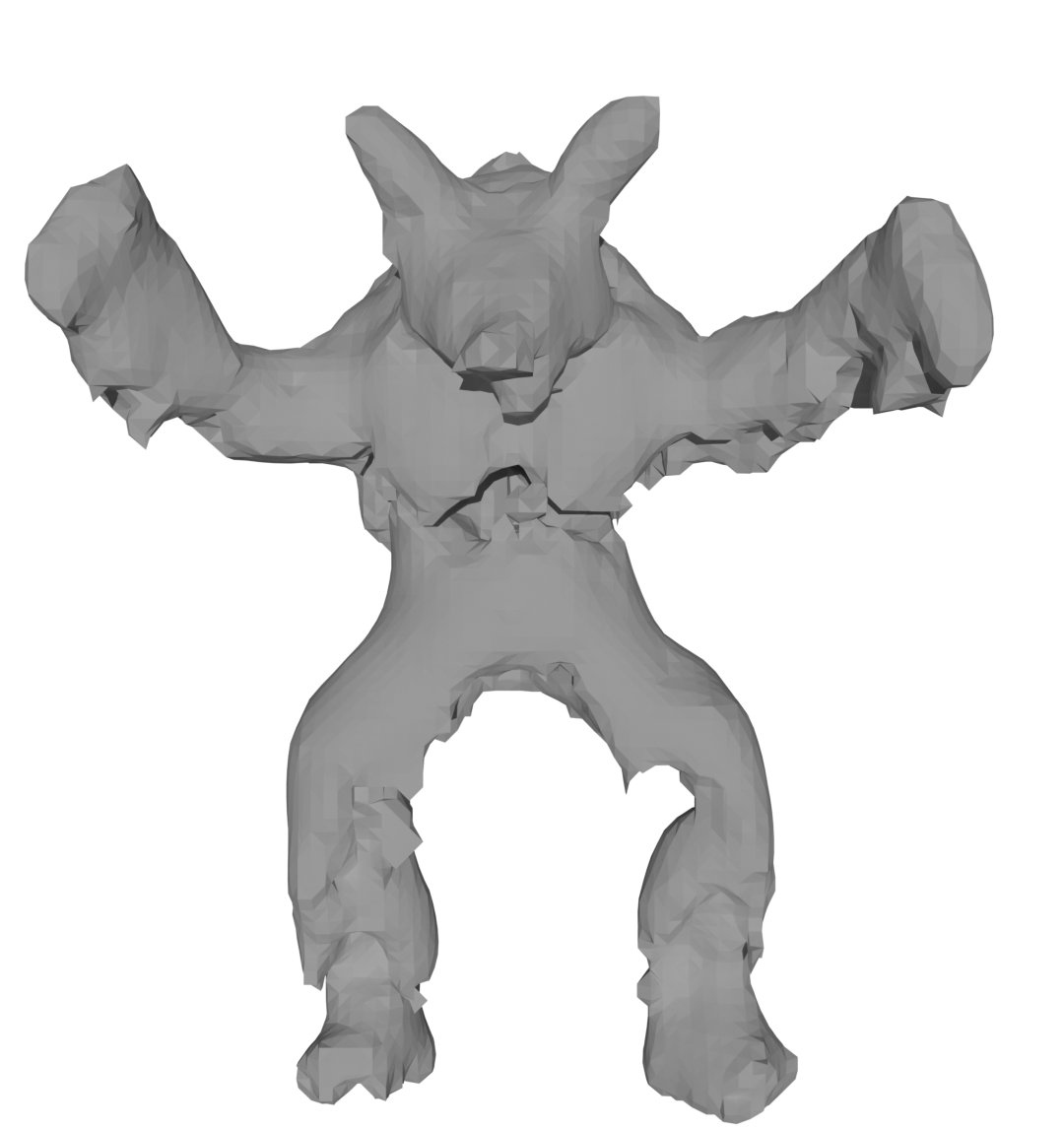}
    \end{subfigure}

    \vspace{6pt} % space between rows
    
    \makebox[0.24\columnwidth]{\footnotesize GT}
    \makebox[0.24\columnwidth]{\footnotesize BP}
    \makebox[0.24\columnwidth]{\footnotesize Reed et al.~\cite{reed2023neural}}
    \makebox[0.24\columnwidth]{\footnotesize Ours}
    
    \caption{Simulated results for Buddha (top row) and Armadillo (bottom row) at 20 dB SNR.}
    \label{fig:simulated_results_buddha_armadillo}
\end{figure}

We evaluate our method on four simulated scenes and compare against backprojection and Reed et al.~\cite{reed2023neural}. All scenes use a linear frequency-modulated (LFM) chirp with center frequency $f_c=20\,\mathrm{kHz}$ and bandwidth $\Delta f=20\,\mathrm{kHz}$ at an SNR of $20\,\mathrm{dB}$. ~\cref{fig:simulated_results_buddha_armadillo} shows qualitative results for Buddha and Armadillo. ~\cref{tab:metrics-summary} reports 3D metrics averaged across all scenes, where our method achieves the best overall performance. Per-scene results and additional metrics are provided in the supplemental material.

\begin{table}[H]
\centering
\scriptsize
\setlength{\tabcolsep}{4pt} % tighten column spacing
\caption{Average 3D metrics across all objects/methods. Chamfer in scientific notation; others rounded to 3 decimals.}
\label{tab:metrics-summary}
\begin{tabular}{llcccc}
\toprule
& Method & Chamfer~($\downarrow$) & IoU~($\uparrow$) & Prec.~($\uparrow$) & F1~($\uparrow$) \\
\midrule
\multirow{3}{*}{Point cloud}
    & Backproj. & 3.04e-4 & 0.201 & 0.244 & 0.333 \\
    & Reed et al. & 1.96e-4 & 0.447 & 0.428 & 0.566 \\
    & Ours & \textbf{8.89e-5} & \textbf{0.497} & \textbf{0.545} & \textbf{0.616} \\
\midrule
\multirow{3}{*}{Mesh}
    & Backproj. & 3.85e-4 & 0.161 & 0.225 & 0.272 \\
    & Reed et al. & 3.97e-4 & 0.133 & 0.171 & 0.232 \\
    & Ours & \textbf{9.70e-5} & \textbf{0.240} & \textbf{0.395} & \textbf{0.384} \\
\bottomrule
\end{tabular}
\end{table}

% \vspace{-10pt}

\subsection{AirSAS Results}
\label{sec:airsasresults}

\begin{figure}[t!]
    \centering
    % -------- Row 1: Armadilo --------
    \begin{subfigure}[t]{0.30\columnwidth}
        \centering
        \includegraphics[width=\linewidth]{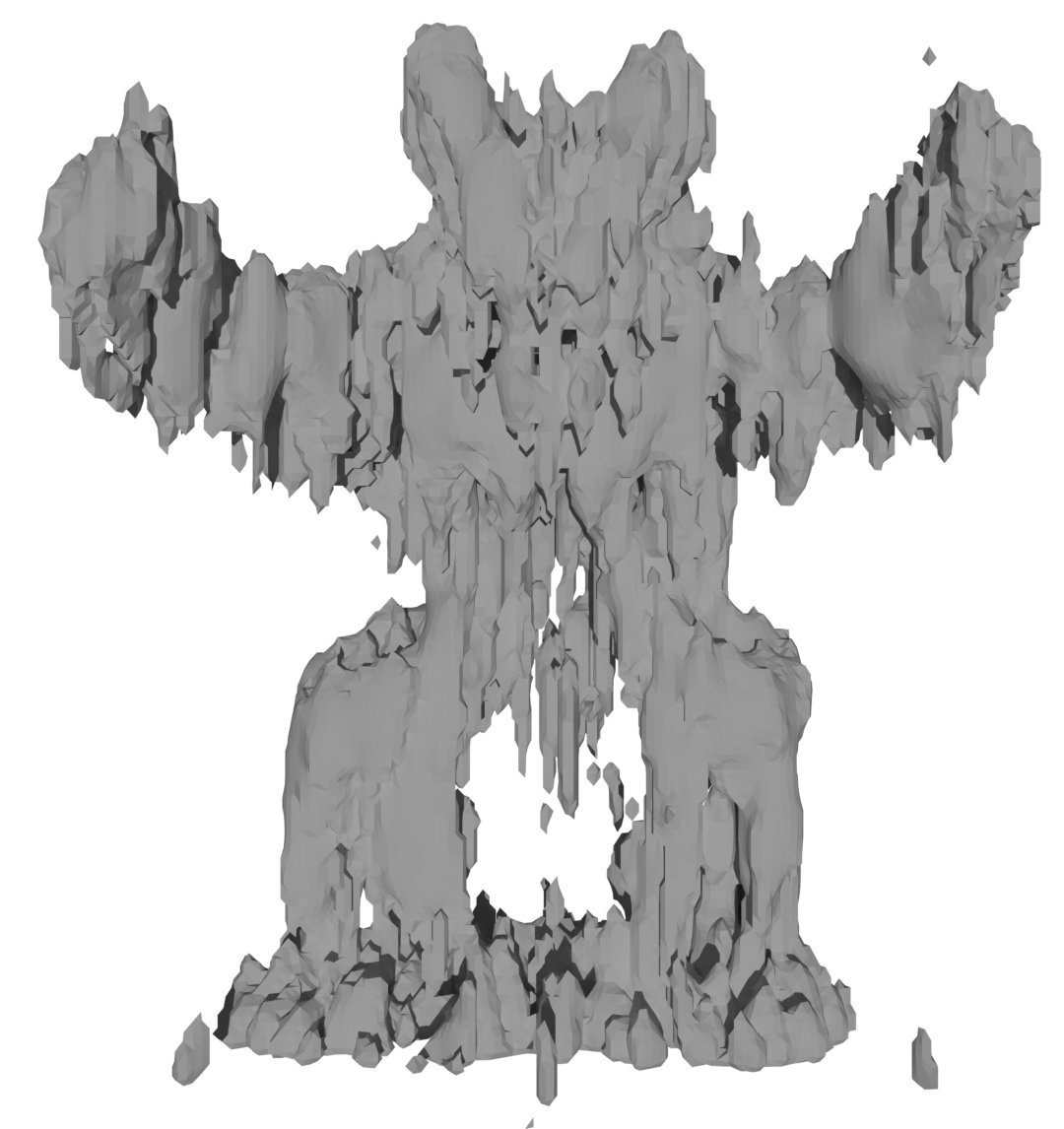}
    \end{subfigure}
    \hfill
    \begin{subfigure}[t]{0.30\columnwidth}
        \centering
        \includegraphics[width=\linewidth]{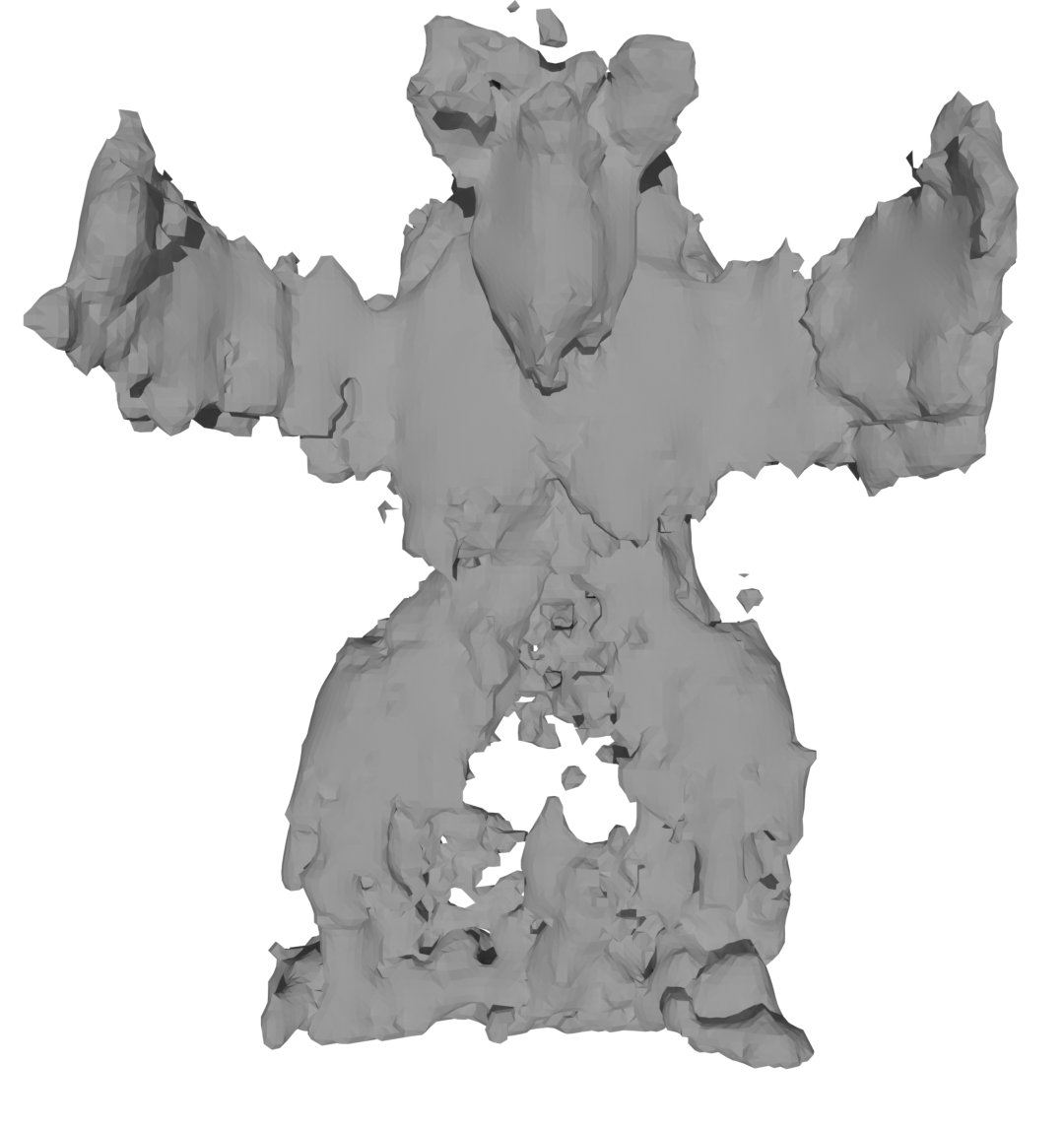}
    \end{subfigure}
    \hfill
    \begin{subfigure}[t]{0.30\columnwidth}
        \centering
        \includegraphics[width=\linewidth]{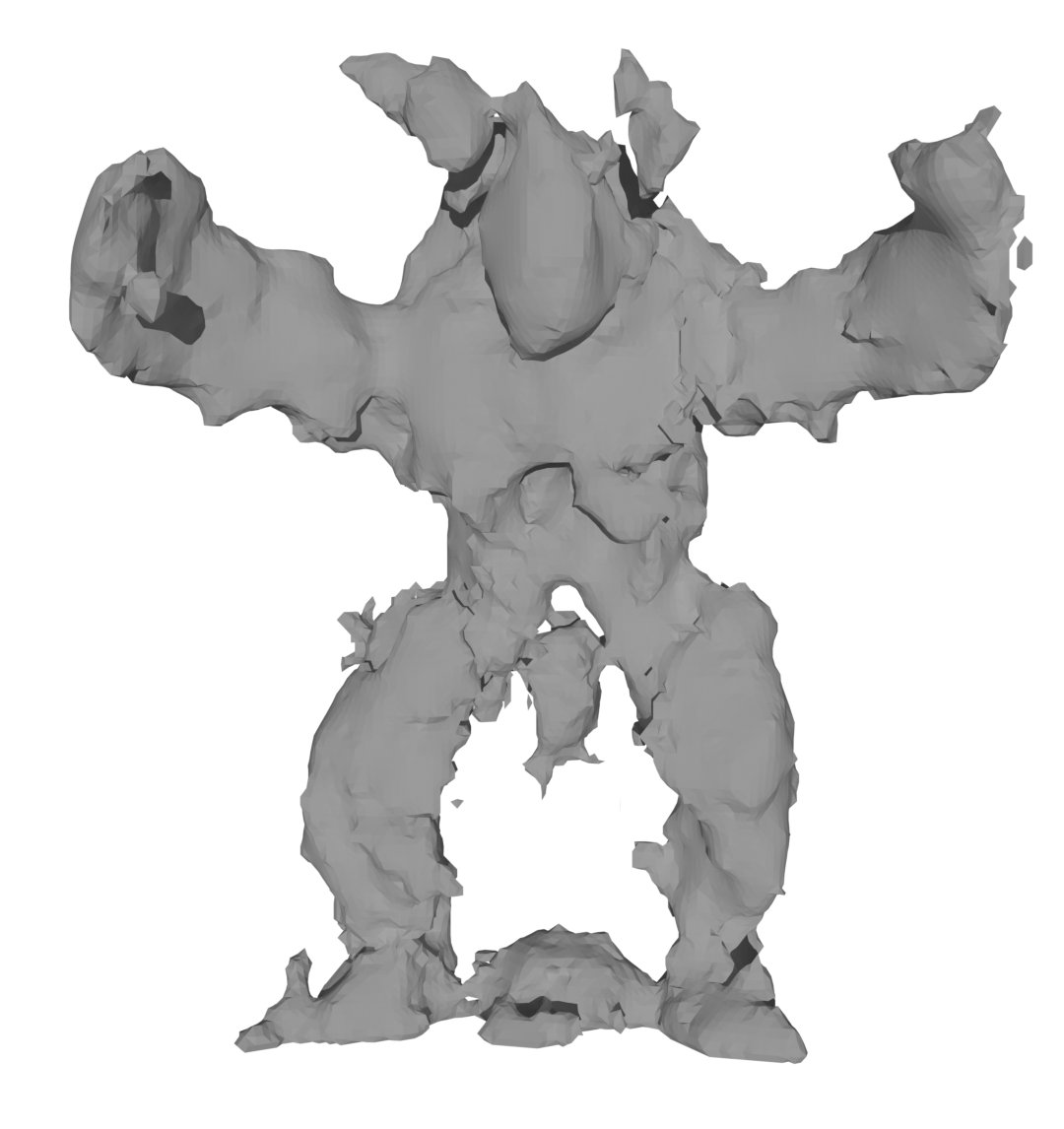}
    \end{subfigure}
    
    % \vspace{6pt} % space between rows
    
    % -------- Row 2: Armadilo side --------
    \begin{subfigure}[t]{0.30\columnwidth}
        \centering
        \includegraphics[width=\linewidth]{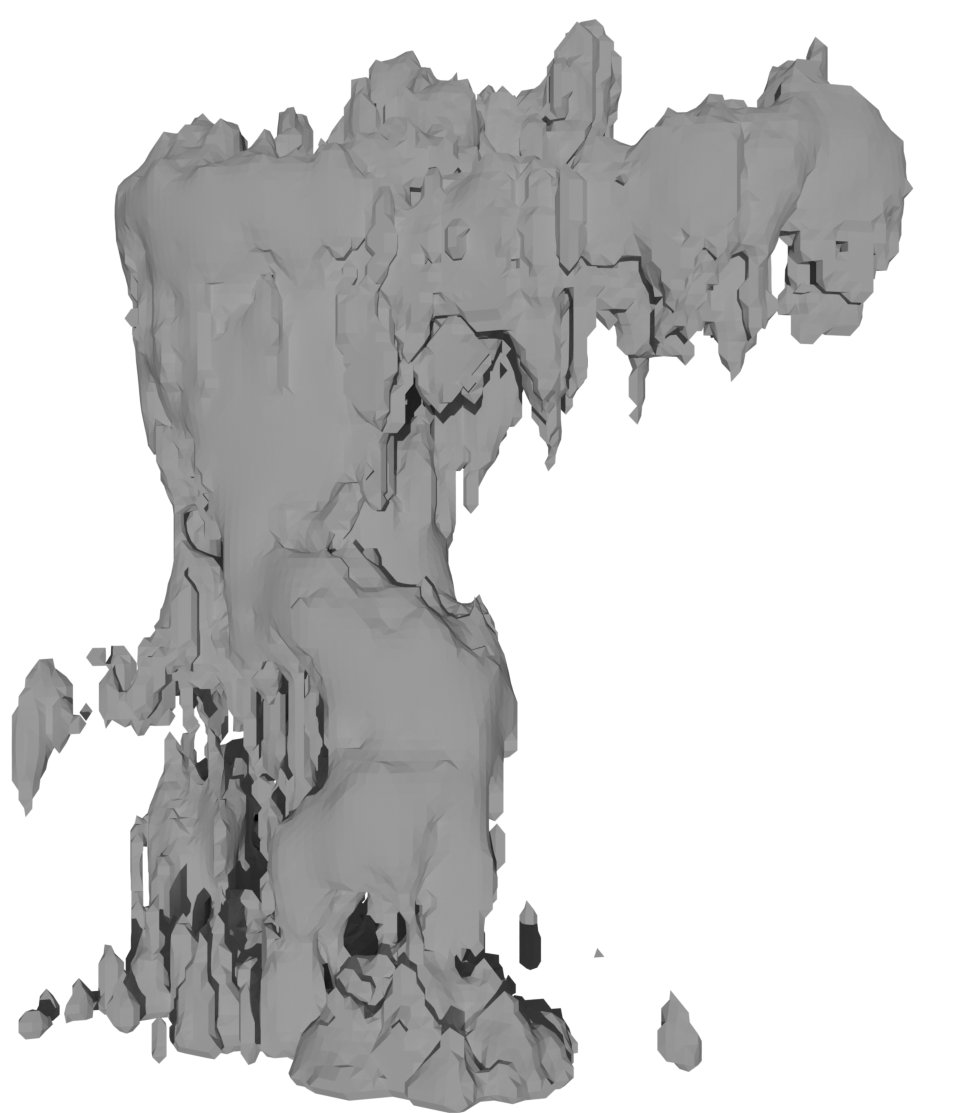}
    \end{subfigure}
    \hfill
    \begin{subfigure}[t]{0.30\columnwidth}
        \centering
        \includegraphics[width=\linewidth]{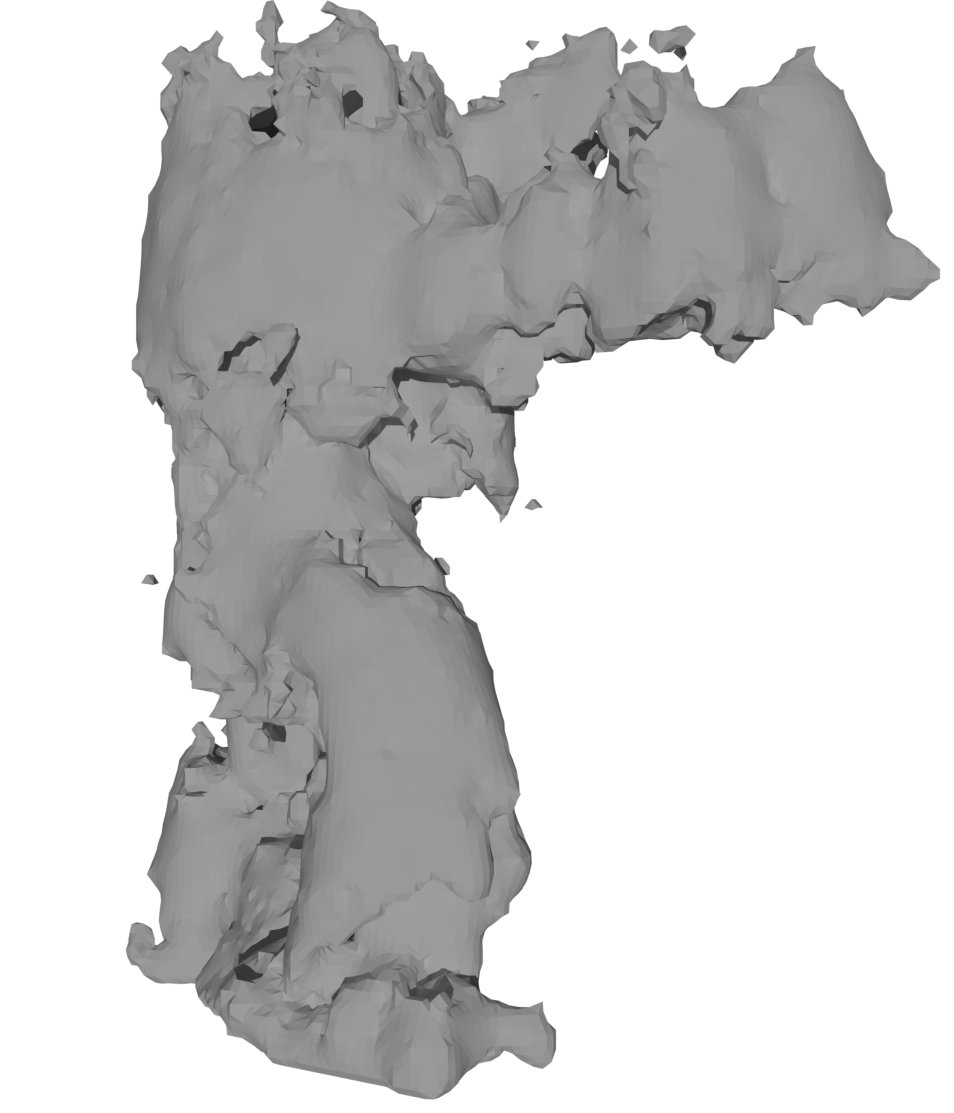}
    \end{subfigure}
    \hfill
    \begin{subfigure}[t]{0.30\columnwidth}
        \centering
        \includegraphics[width=\linewidth]{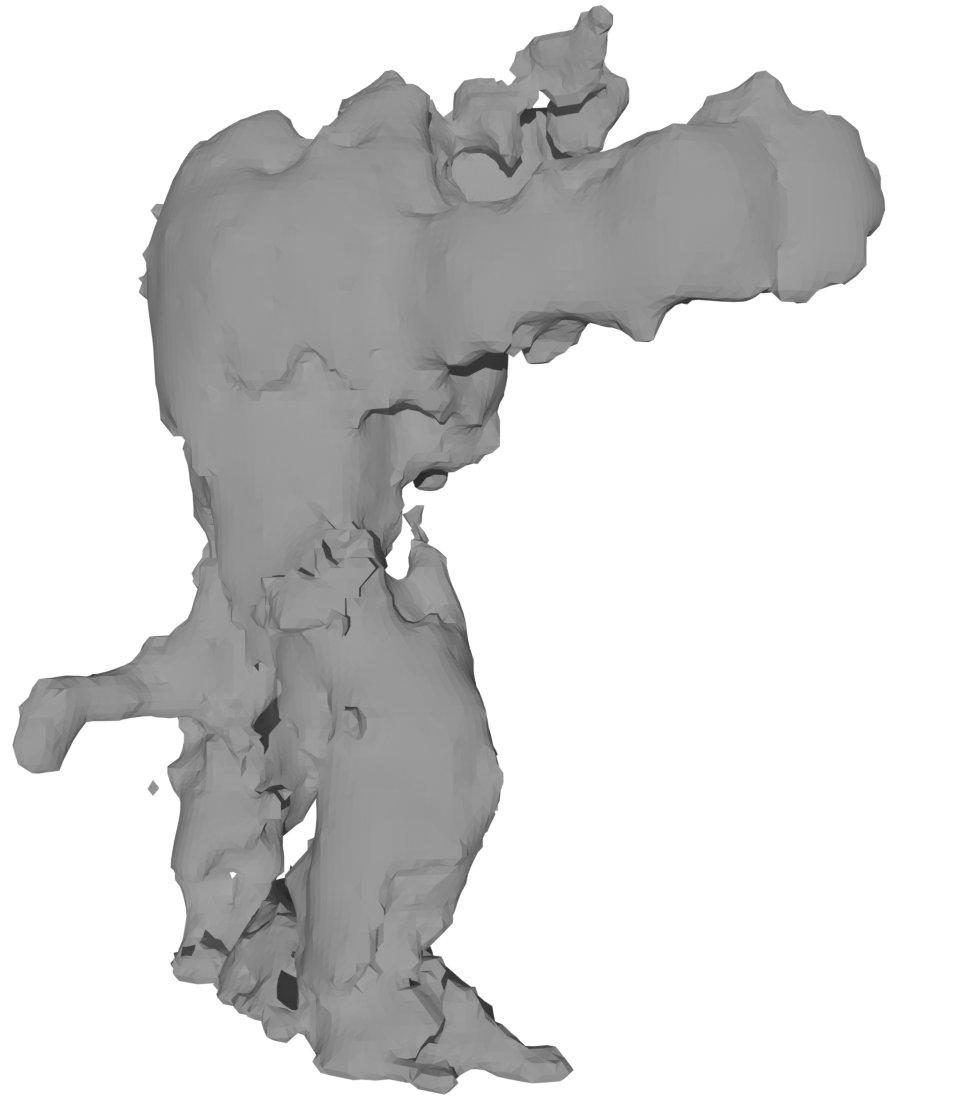}
    \end{subfigure}
    \hfill

%     % \vspace{6pt} % space between rows

%     \makebox[0.30\columnwidth]{\footnotesize Backprojection}
%     \makebox[0.30\columnwidth]{\footnotesize Reed et al.~\cite{reed2023neural}}
%     \makebox[0.30\columnwidth]{\footnotesize Ours}
    
%     \caption{Real AirSAS results for Armadilo at $\Delta f$=20khz.}
%     \label{fig:real_results_arma}
% \end{figure}

% \begin{figure}[t!]
%     \centering
    % -------- Row 3: Bunny --------
    \begin{subfigure}[t]{0.30\columnwidth}
        \centering
        \includegraphics[width=\linewidth]{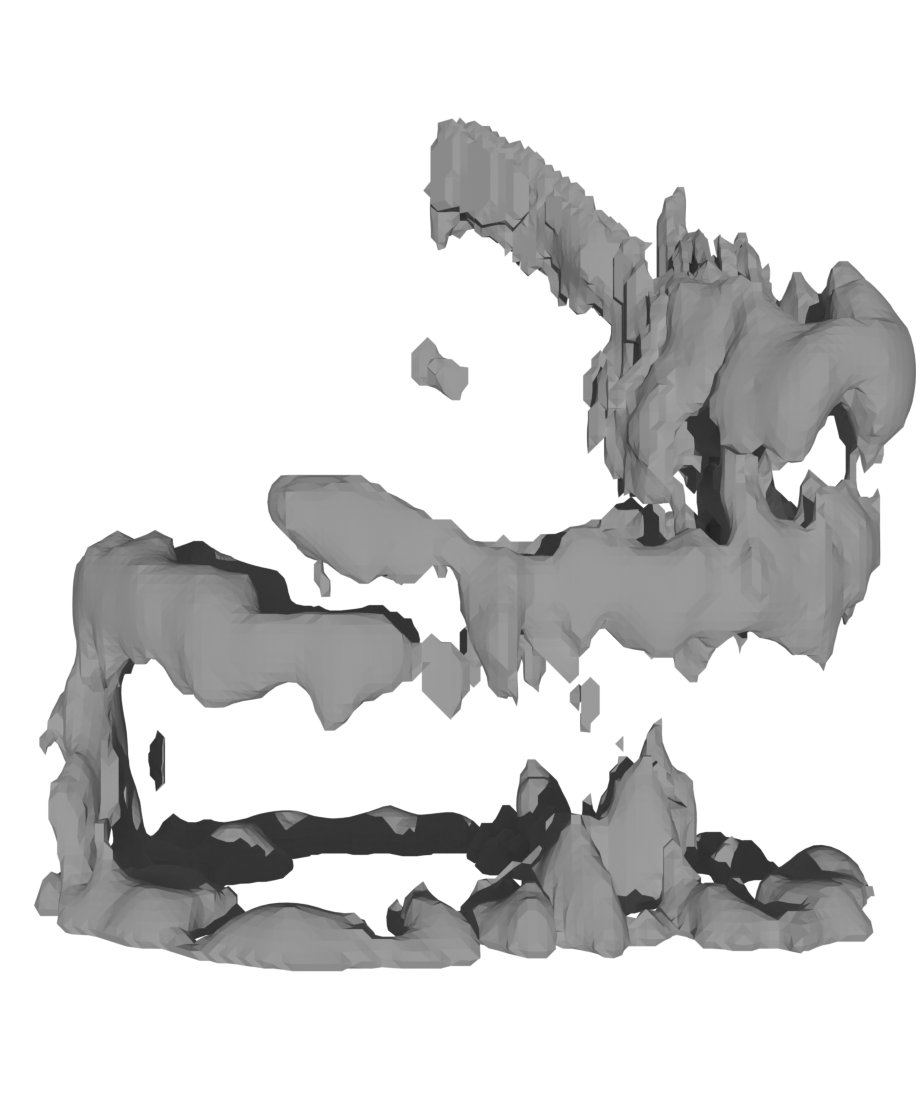}
    \end{subfigure}
    \hfill
    \begin{subfigure}[t]{0.30\columnwidth}
        \centering
        \includegraphics[width=\linewidth]{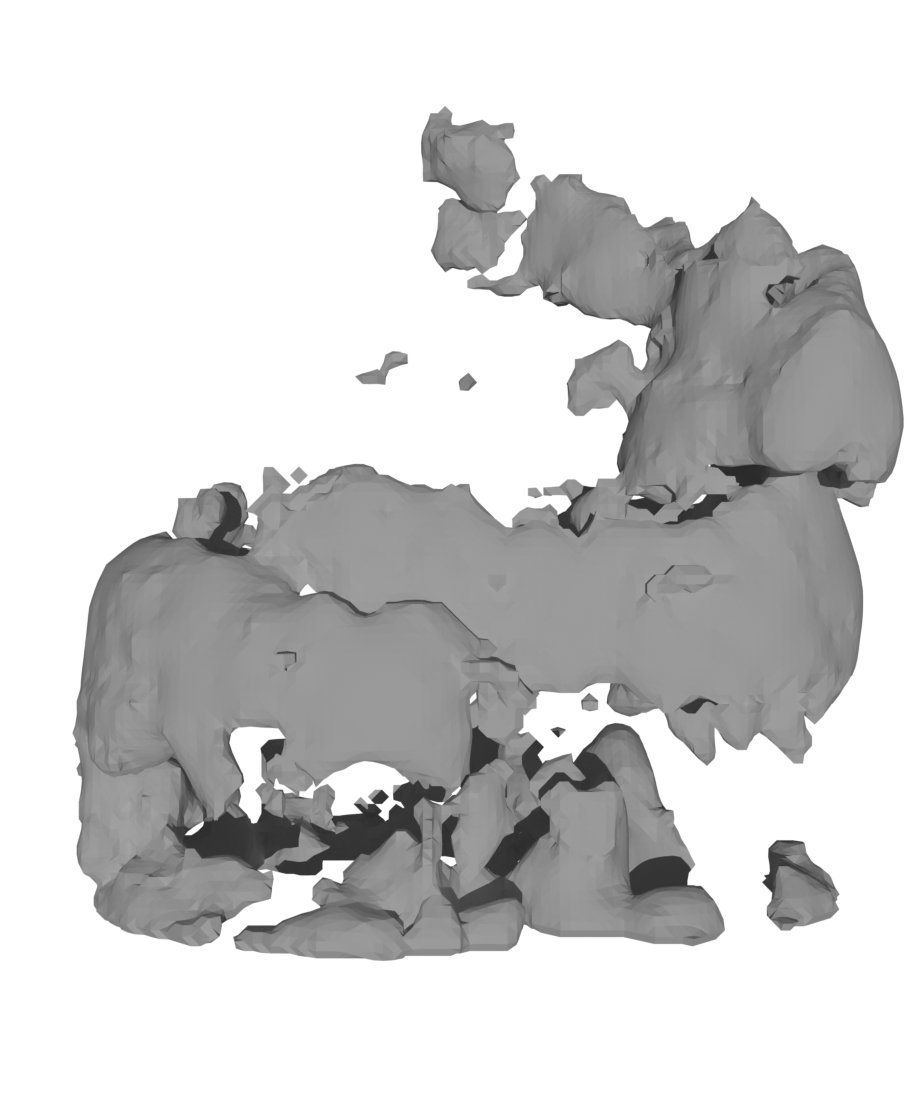}
    \end{subfigure}
    \hfill
    \begin{subfigure}[t]{0.30\columnwidth}
        \centering
        \includegraphics[width=\linewidth]{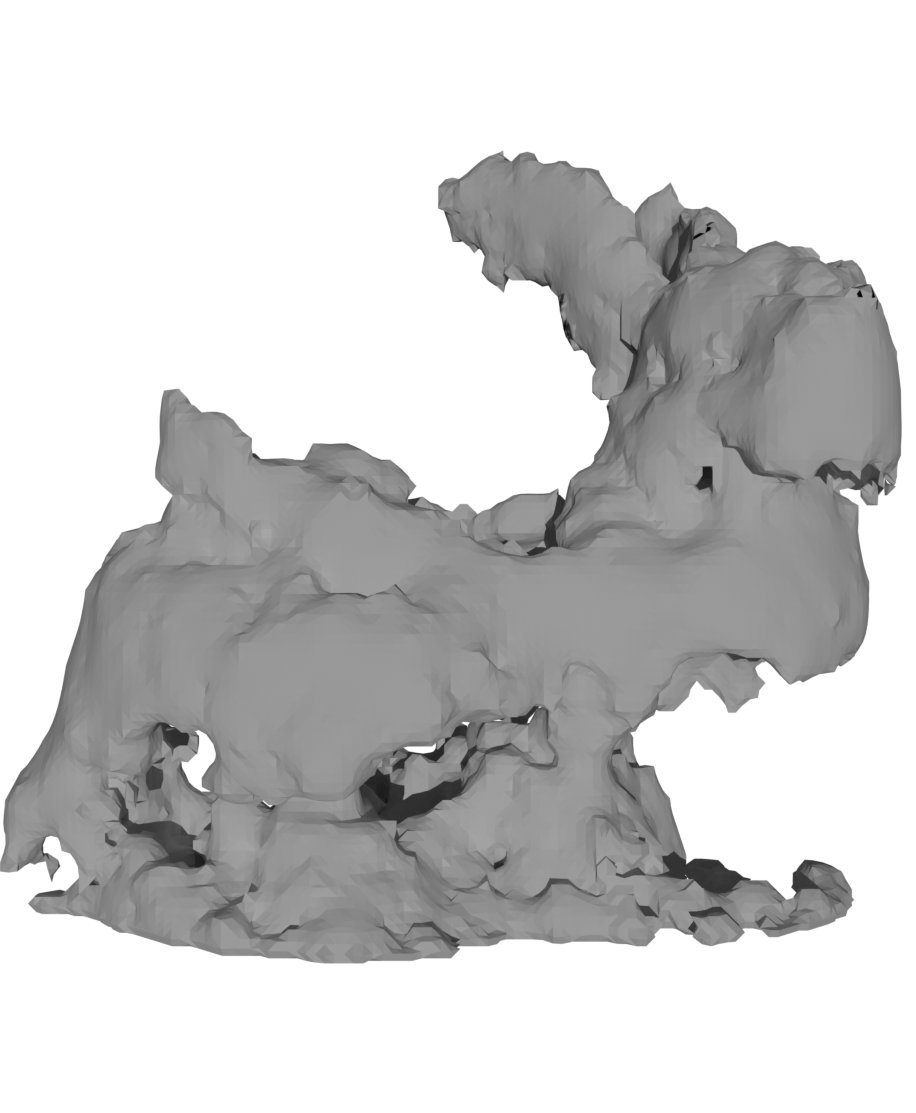}
    \end{subfigure}
    
    % \vspace{6pt} % space between rows
    
    % -------- Row 4: Bunny top --------
    \begin{subfigure}[t]{0.30\columnwidth}
        \centering
        \includegraphics[width=\linewidth]{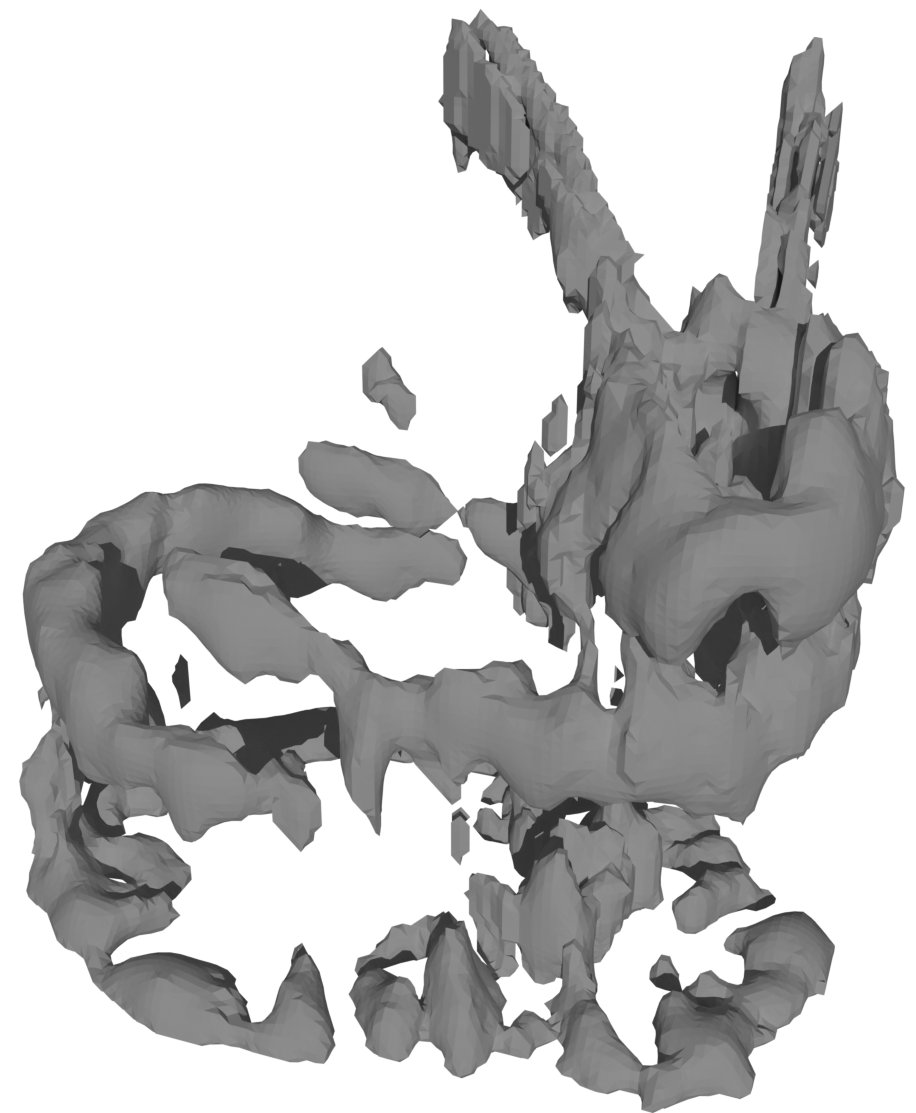}
    \end{subfigure}
    \hfill
    \begin{subfigure}[t]{0.30\columnwidth}
        \centering
        \includegraphics[width=\linewidth]{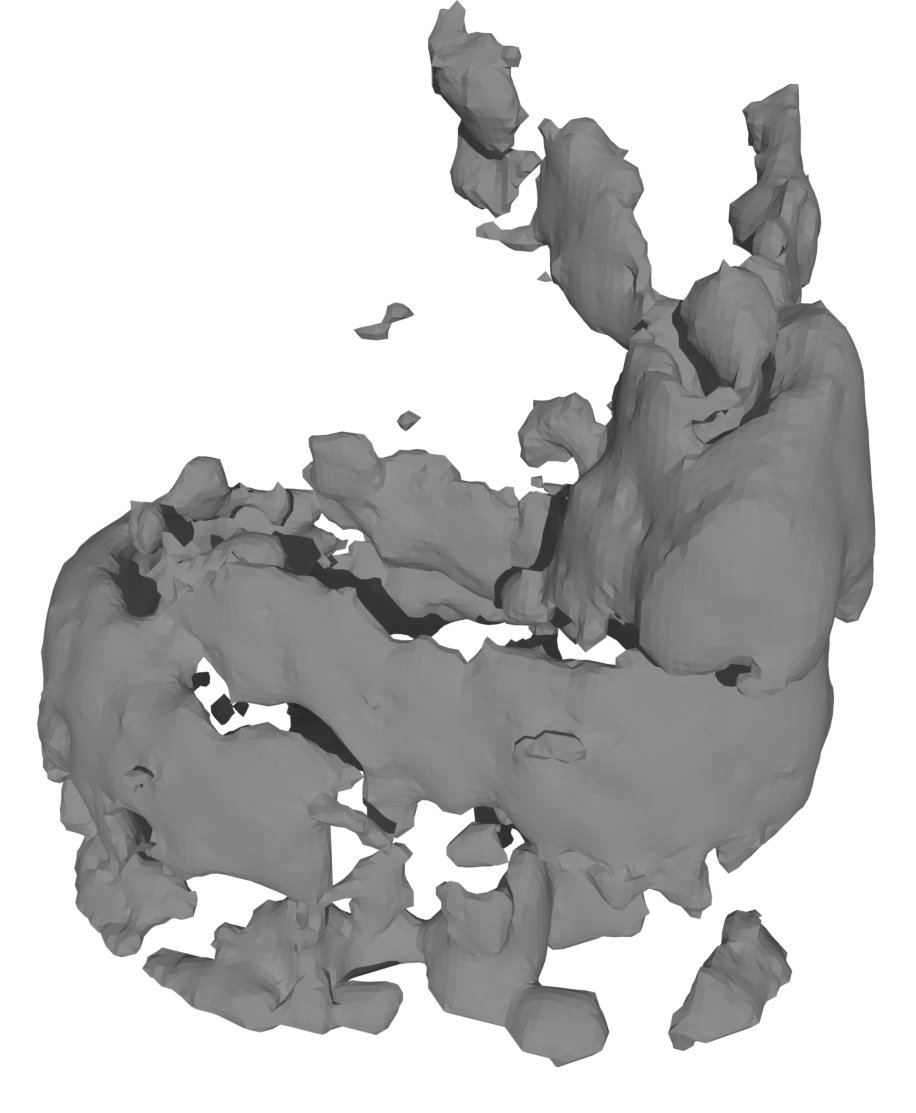}
    \end{subfigure}
    \hfill
    \begin{subfigure}[t]{0.30\columnwidth}
        \centering
        \includegraphics[width=\linewidth]{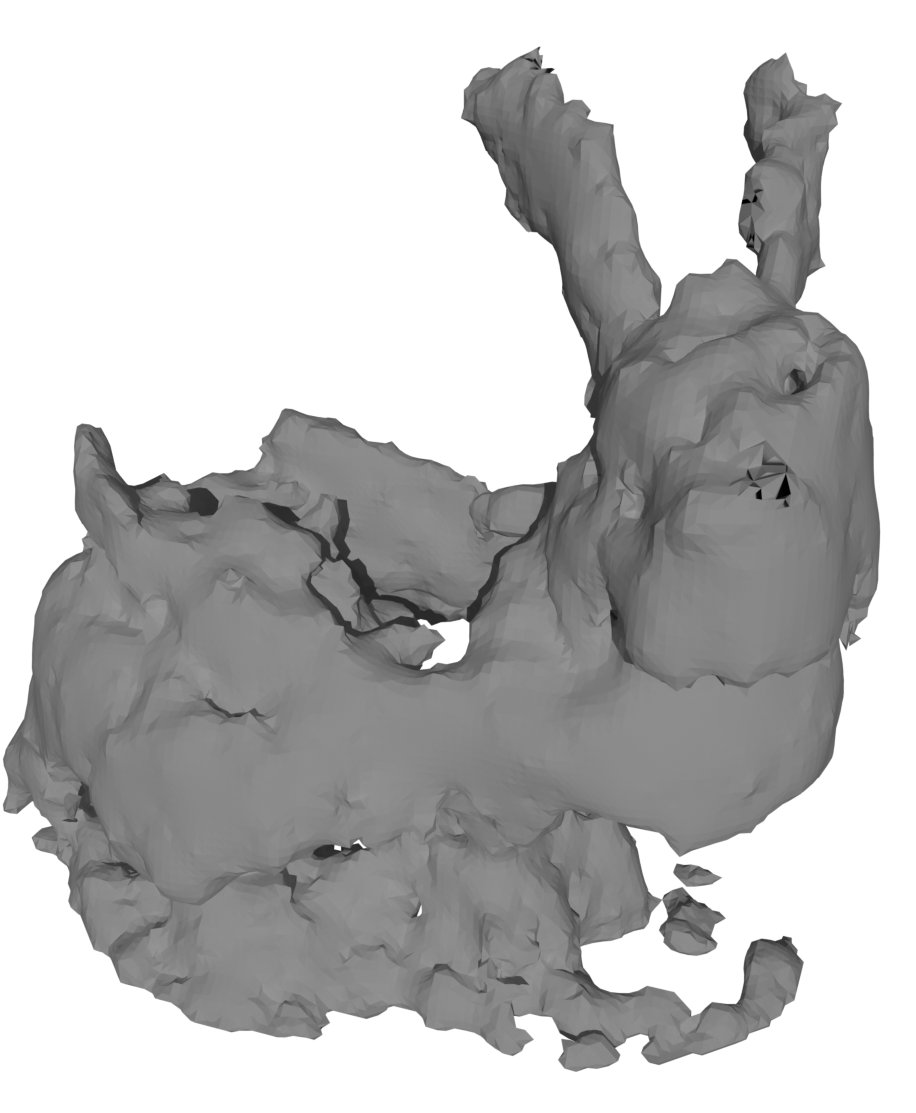}
    \end{subfigure}
    
    \makebox[0.30\columnwidth]{\footnotesize Backprojection}
    \makebox[0.30\columnwidth]{\footnotesize Reed et al.~\cite{reed2023neural}}
    \makebox[0.30\columnwidth]{\footnotesize Ours}
    
    \caption{Real AirSAS results for Armadilo and Bunny at $\Delta f=20\,\mathrm{kHz}$}
    \label{fig:real_results_arma_bunny}
\end{figure}

We evaluate on real AirSAS measurements of the 3D-printed \emph{Armadillo} and \emph{Bunny}. All acquisitions use an LFM chirp with $f_c=20\,\mathrm{kHz}$ and $\Delta f=20\,\mathrm{kHz}$. ~\cref{fig:real_results_arma_bunny} compare backprojection, Reed et al.~\cite{reed2023neural}, and our method.

Qualitatively, our reconstructions align best with the ground truth: the legs and tail are better preserved and less fragmented, and radial streaking artifacts are reduced relative to backprojection. Compared to Reed et al., our method also yields sharper surfaces with fewer fill-in artifacts, especially in side and front views.

% \begin{figure*}[t]
%     \centering

%     \includegraphics[width=0.30\textwidth]{images/real_arma_sh_1.jpg}
%     \includegraphics[width=0.30\textwidth]{images/real_arma_sh_2.jpg}
%     \includegraphics[width=0.30\textwidth]{images/real_arma_sh_3.jpg}

%     \vspace{5mm}
    
%     \includegraphics[width=0.30\textwidth]{images/real_arma_sh_1_side.jpg}
%     \includegraphics[width=0.30\textwidth]{images/real_arma_sh_2_side.jpg}
%     \includegraphics[width=0.30\textwidth]{images/real_arma_sh_3_side.jpg}

%     \makebox[0.30\textwidth]{\footnotesize Level 1}
%     \makebox[0.30\textwidth]{\footnotesize Level 2}
%     \makebox[0.30\textwidth]{\footnotesize Level 3}

%     \caption{Albation on increase in SH levels}
%     \label{fig:sh_level_increase}
% \end{figure*}

\subsection{Subsampling the Synthetic Aperture Views}

\begin{figure}[t]
  \centering
  \setlength{\tabcolsep}{2pt}
  \begin{tabular}{@{}c c c@{}}
    % ---- Row 1: Front view ----
    \includegraphics[width=0.32\columnwidth]{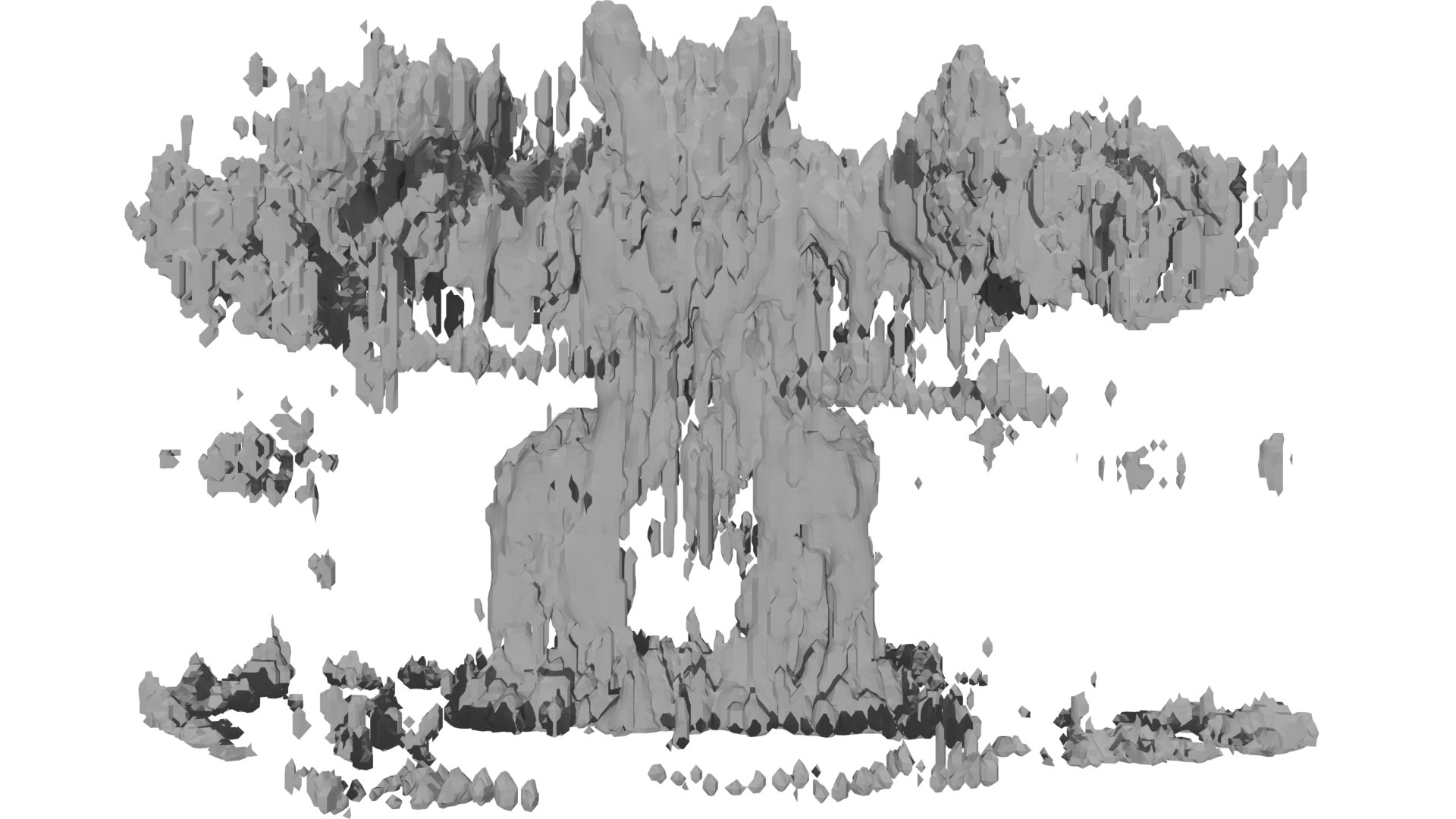} &
    \includegraphics[width=0.32\columnwidth]{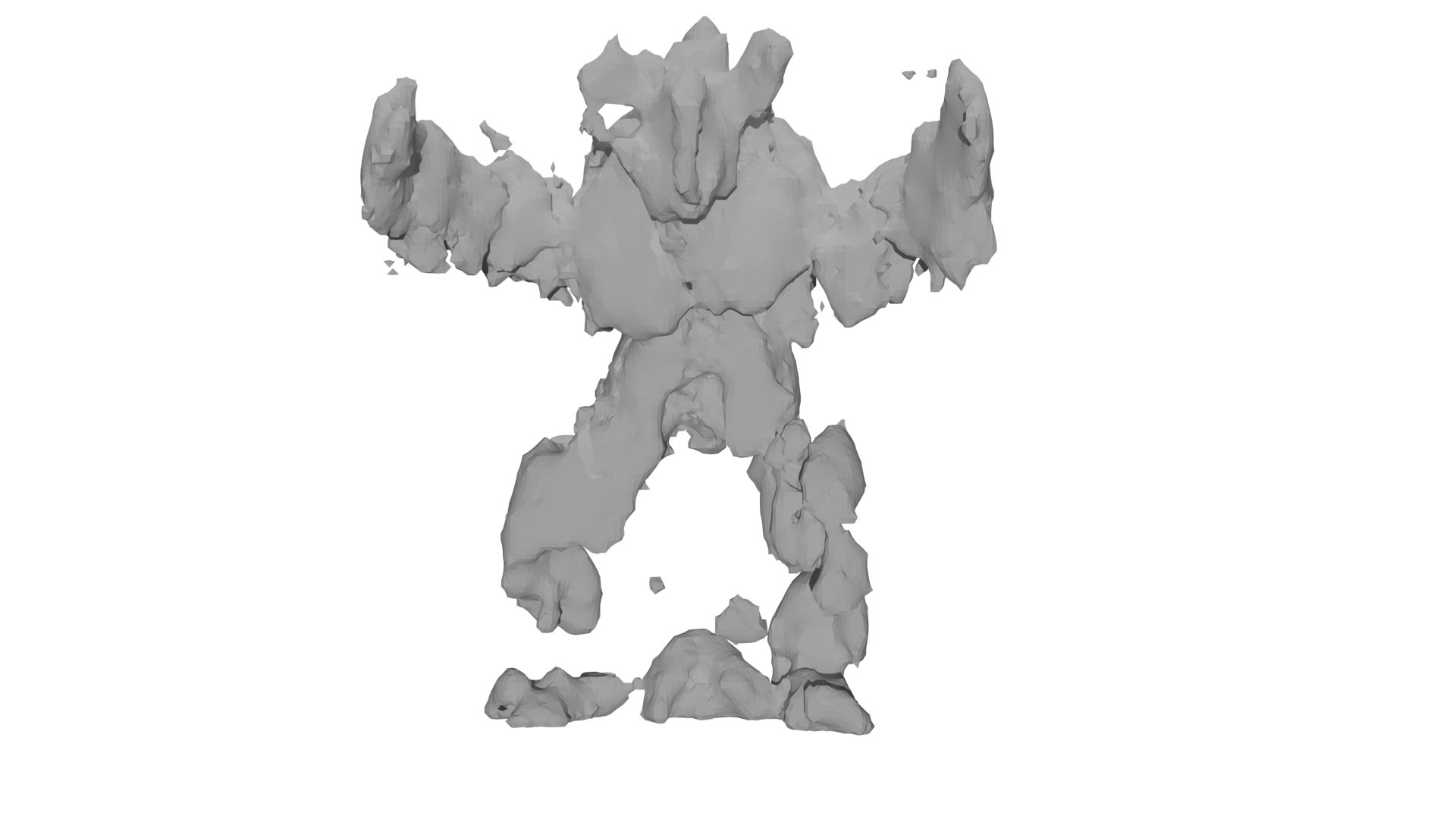} &
    \includegraphics[width=0.32\columnwidth]{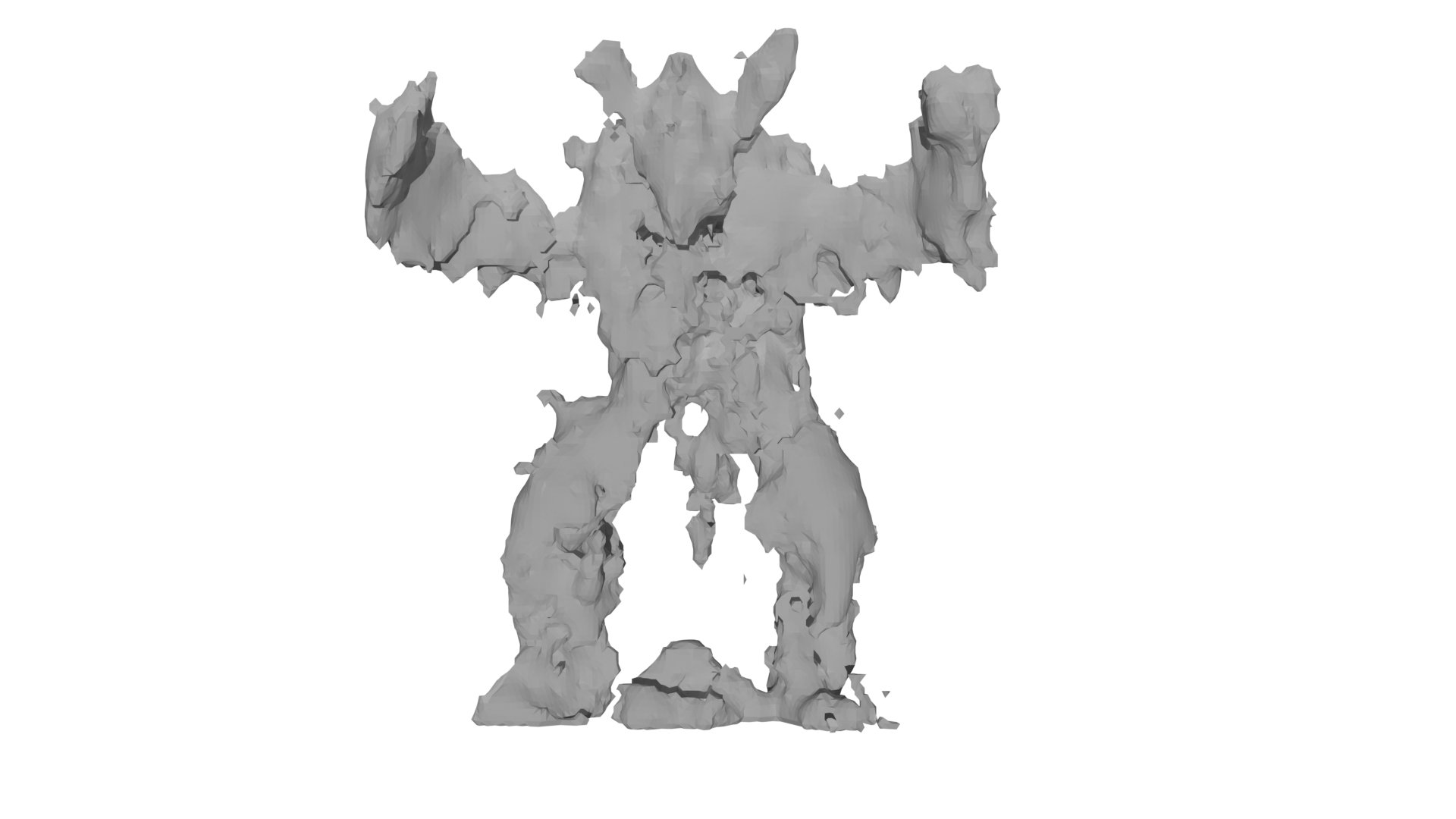} \\[-1pt]
    % ---- Row 2: Side view ----
    \includegraphics[width=0.32\columnwidth]{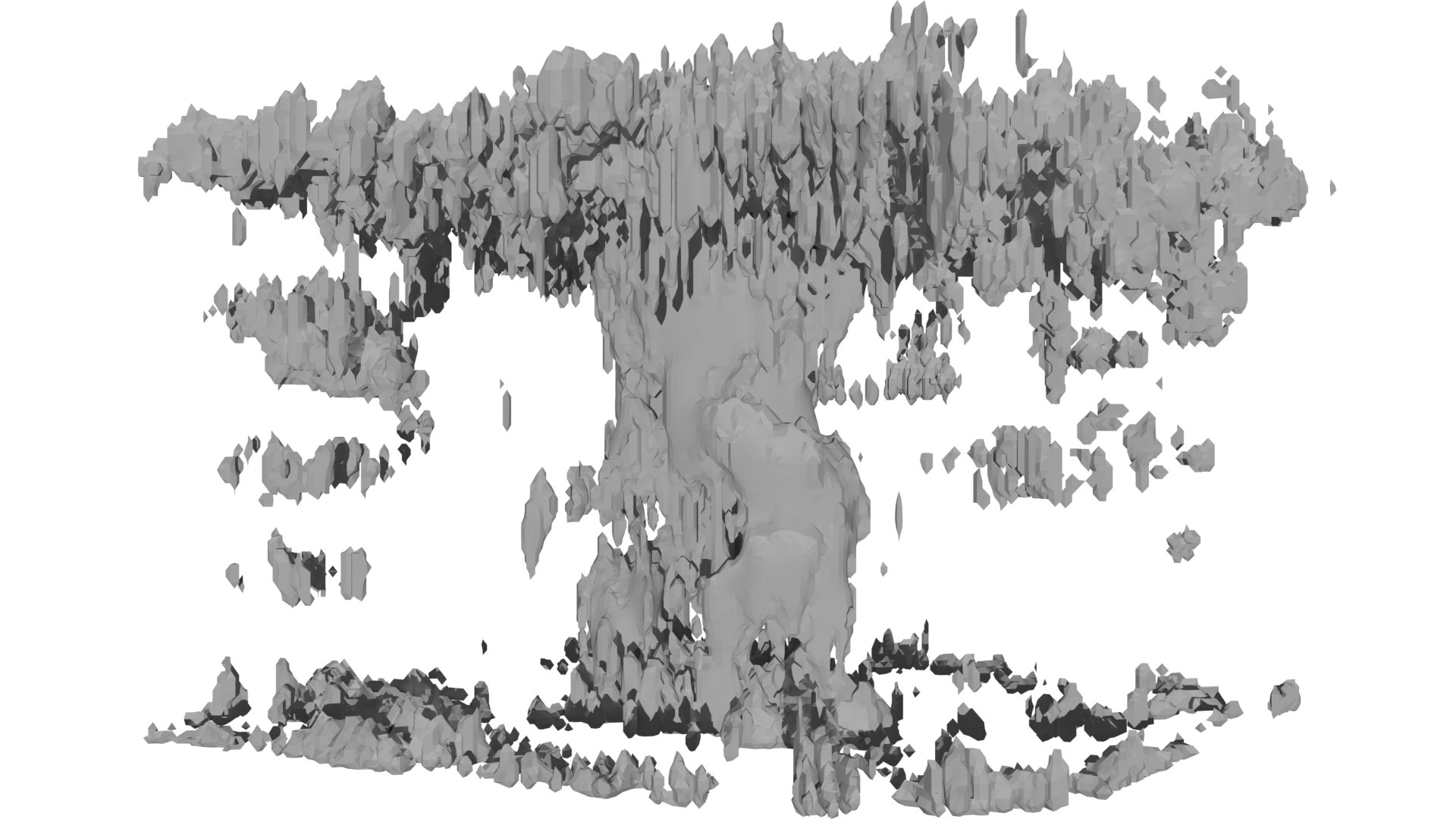} &
    \includegraphics[width=0.32\columnwidth]{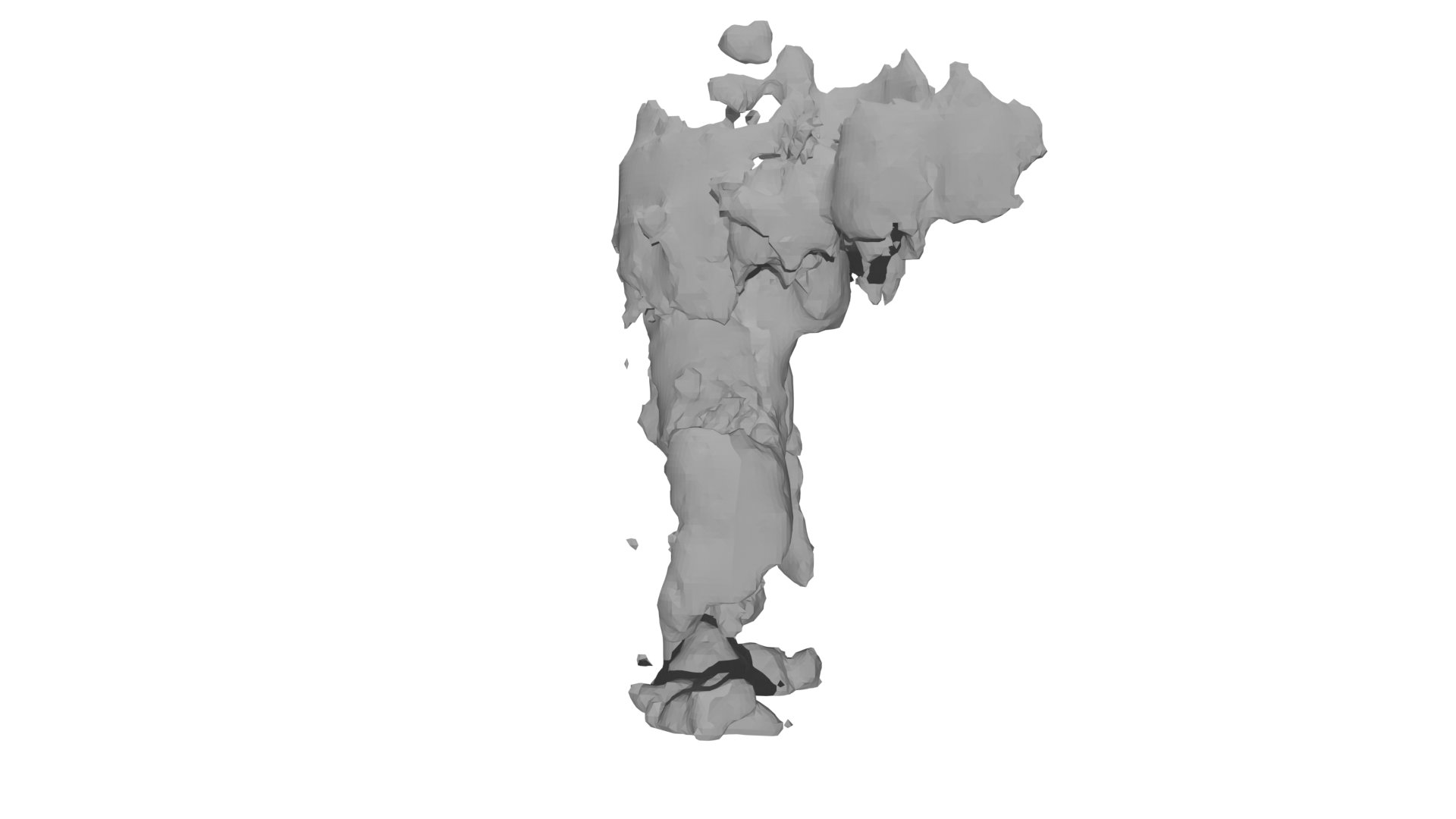} &
    \includegraphics[width=0.32\columnwidth]{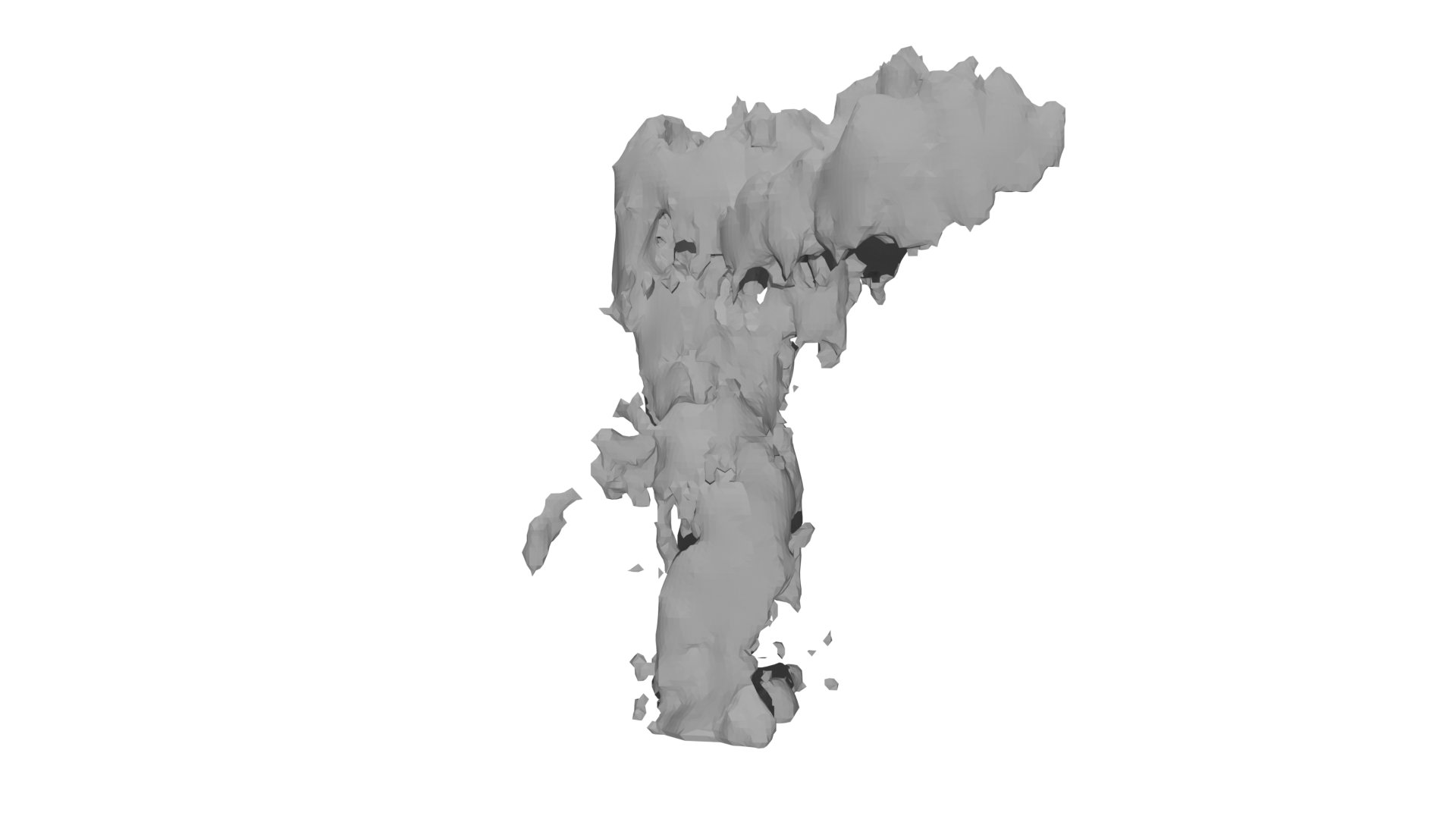} \\
  \end{tabular}

  % method labels under the three columns
  \vspace{2pt}
  \makebox[0.32\columnwidth]{\footnotesize Backprojection}%
  \makebox[0.32\columnwidth]{\footnotesize Reed et al.~\cite{reed2023neural}}%
  \makebox[0.32\columnwidth]{\footnotesize Ours}

  \caption{Sparse subsampling with 20\% of views. Top: front view. Bottom: side view.}
  \label{fig:subsampling_views}
\end{figure}

In addition to fully sampled reconstructions, we evaluate our method under sparse view sampling, where only 20\% of the available views are used. As expected, backprojection under sparse sampling exhibits strong radial streaking artifacts due to missing angular information. In contrast, our method produces reconstructions that are both sharper and more faithful to the ground truth. Notably, ~\cref{fig:subsampling_views} shows that our method outperform both backprojection and the approach of Reed et al.~\cite{reed2023neural} on the tested dataset, demonstrating that our representation can robustly handle limited angular coverage.

\subsection{Latency and Convergence}
All timings were measured on an RTX 4090 GPU; see ~\cref{tab:train-time} for dataset-level iteration counts and wall-clock times. Our \emph{per-iteration} latency—measured end-to-end, including surface-normal computation, transmission-probability accumulation, and ray integration—averages \(140\,\mathrm{ms}\) for Reed et al.~\cite{reed2023neural} versus \(125\,\mathrm{ms}\) for ours. On simulated data we also reach target quality in fewer iterations, yielding shorter overall training; on AirSAS and SVSS, where we match the iteration budget for fairness, our wall-clock time remains lower. ~\cref{fig:iter_comparison} illustrates that our reconstructions achieve higher quality earlier in training.

\begin{table}[t]
    \centering
    \setlength{\tabcolsep}{6pt}
    \caption{Training iterations and wall-clock time on an RTX 4090.}
    \label{tab:train-time}
    \begin{tabular}{@{}l l r l@{}}
    \toprule
    Dataset & Method & Iterations & Time \\
    \midrule
    \multirow{2}{*}{Simulated}
      & Reed et al.~\cite{reed2023neural} & 100{,}000 & 2 h 10 min \\
      & Ours                               & 50{,}000  & 44 min      \\
    \midrule
    \multirow{2}{*}{AirSAS}
      & Reed et al.~\cite{reed2023neural} & 26{,}000 & 1 h 38 min \\
      & Ours                               & 26{,}000 & 1 h 15 min \\
    \midrule
    \multirow{2}{*}{SVSS}
      & Reed et al.~\cite{reed2023neural} & 6{,}000 & 38 min \\
      & Ours                               & 6{,}000 & 30 min \\
    \bottomrule
    \end{tabular}
\end{table}

% \vspace{-10pt}

\begin{figure}[t]
  \centering
  \setlength{\tabcolsep}{2pt}
  \begin{tabular}{@{}c c@{}}
    % Row 1
    \includegraphics[height=0.16\textheight]{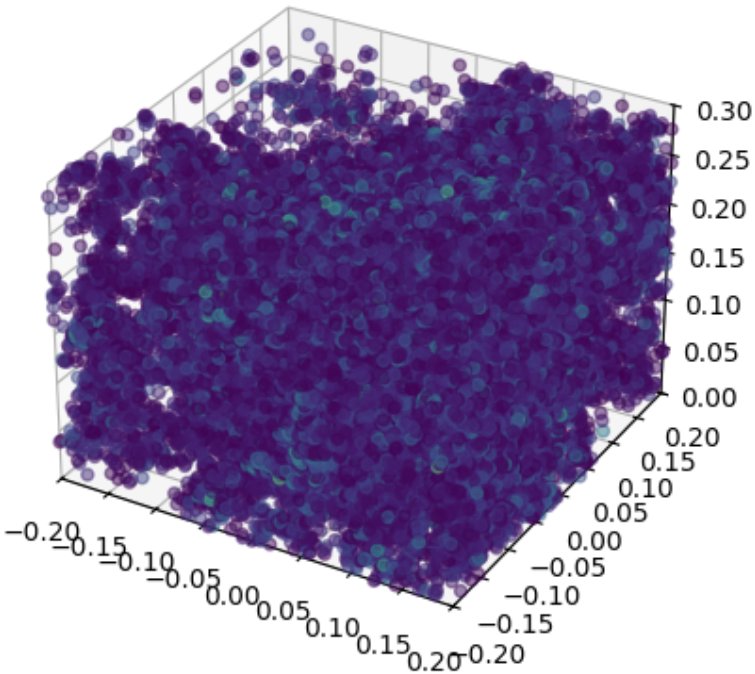} &
    \includegraphics[height=0.16\textheight]{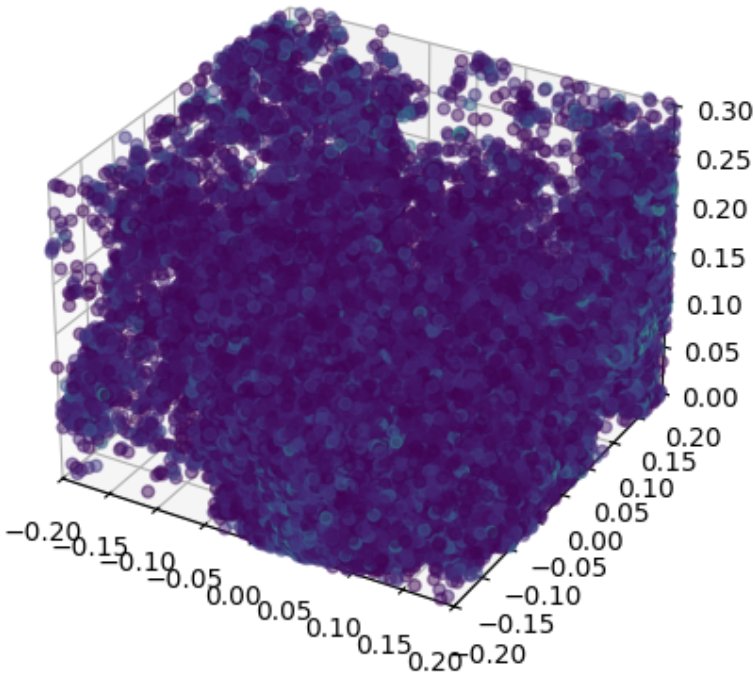} \\
    \multicolumn{2}{c}{\scriptsize \textbf{0 iterations}} \\
    % Row 2
    \includegraphics[height=0.16\textheight]{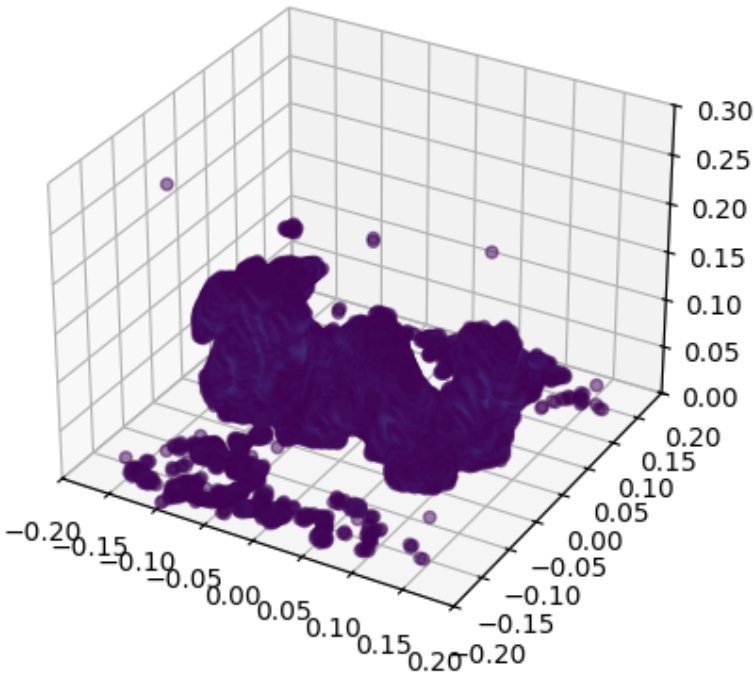} &
    \includegraphics[height=0.16\textheight]{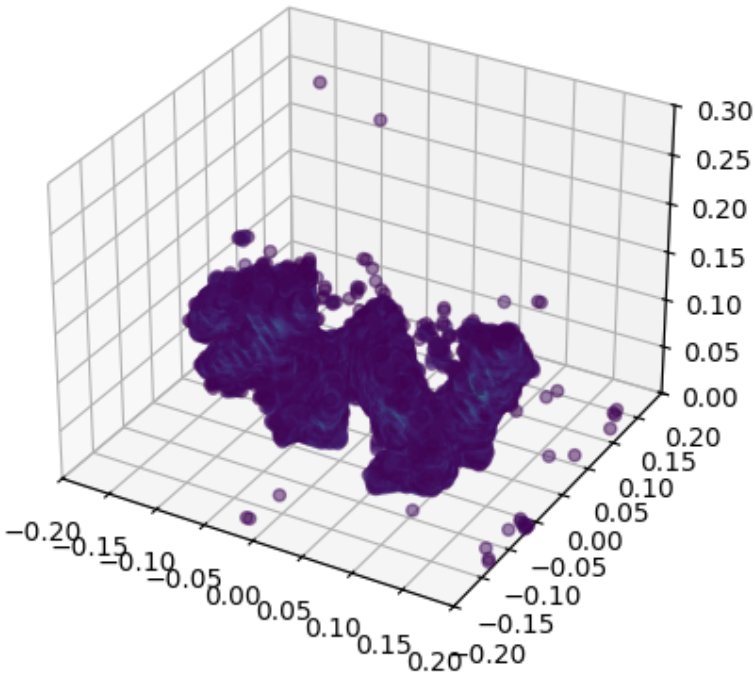} \\
    \multicolumn{2}{c}{\scriptsize \textbf{10k iterations}} \\
    % Row 3
    \includegraphics[height=0.16\textheight]{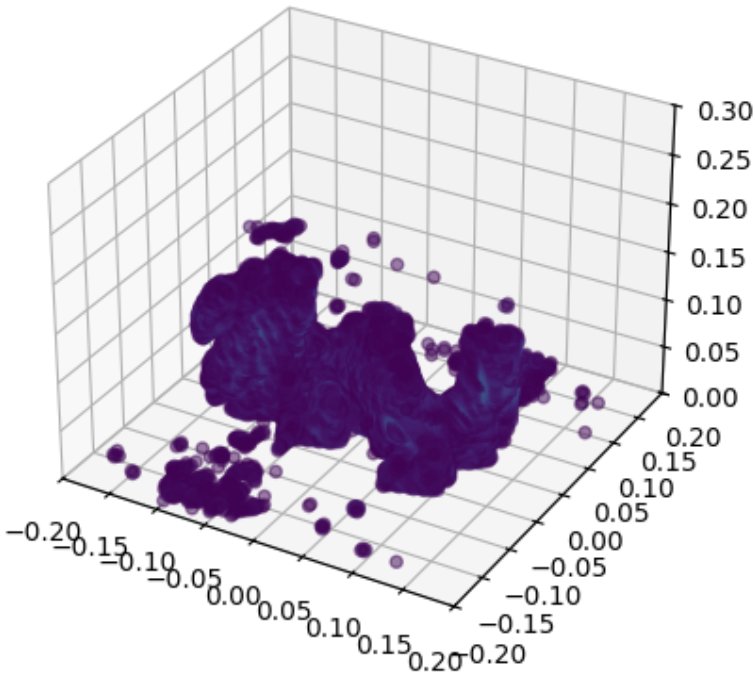} &
    \includegraphics[height=0.16\textheight]{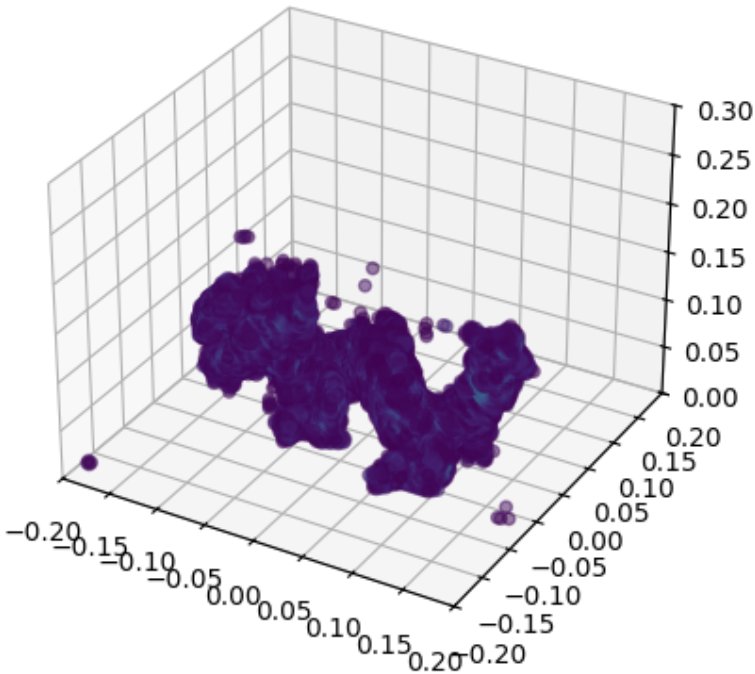} \\
    \multicolumn{2}{c}{\scriptsize \textbf{20k iterations}} \\
    % Row 4
    \includegraphics[height=0.16\textheight]{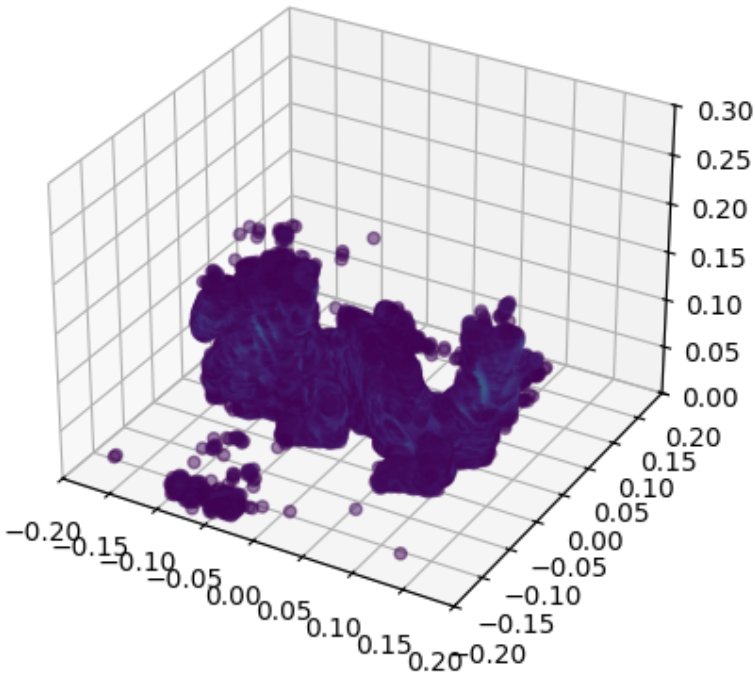} &
    \includegraphics[height=0.16\textheight]{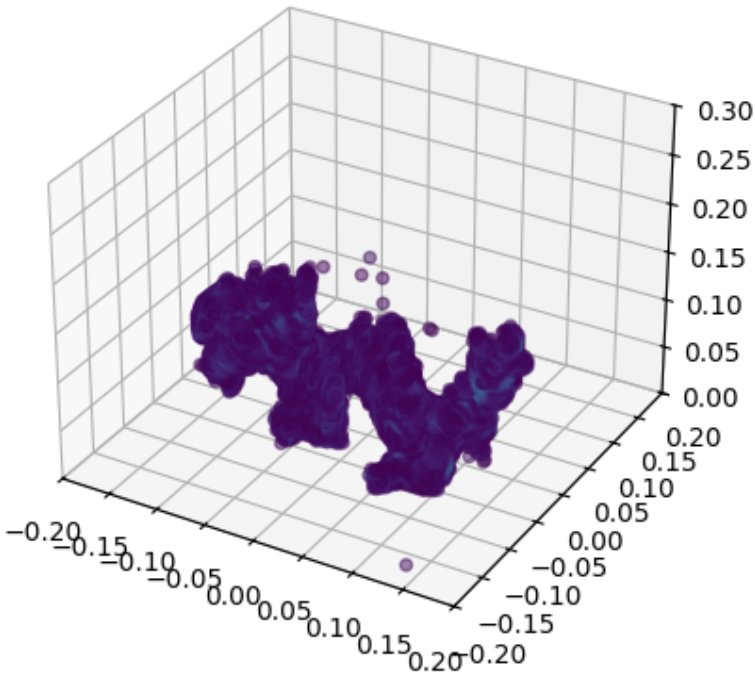} \\
    \multicolumn{2}{c}{\scriptsize \textbf{30k iterations}} \\
  \end{tabular}

  % \vspace{2pt}
  \makebox[0.48\columnwidth]{\footnotesize Reed et al.~\cite{reed2023neural}}%
  \hfill
  \makebox[0.48\columnwidth]{\footnotesize Ours}

  \caption{Convergence comparison on XYZ Dragon. Our method reconstructs fine-scale structure more quickly and achieves higher quality at earlier iterations compared to Reed et al.~\cite{reed2023neural}.}
  \label{fig:iter_comparison}
\end{figure}

% \begin{figure*}[t!]
%     \centering

%     \includegraphics[width=0.23\textwidth]{images/albert_xyz_dragon_comp_albedo_0.jpg}
%     \includegraphics[width=0.23\textwidth]{images/albert_xyz_dragon_comp_albedo_10000.jpg}
%     \includegraphics[width=0.23\textwidth]{images/albert_xyz_dragon_comp_albedo_20000.jpg}
%     \includegraphics[width=0.23\textwidth]{images/albert_xyz_dragon_comp_albedo_30000.jpg}

%     \vspace{5mm}
    
%     \includegraphics[width=0.23\textwidth]{images/ours_xyz_dragon_comp_albedo_0.jpg}
%     \includegraphics[width=0.23\textwidth]{images/ours_xyz_dragon_comp_albedo_10000.jpg}
%     \includegraphics[width=0.23\textwidth]{images/ours_xyz_dragon_comp_albedo_20000.jpg}
%     \includegraphics[width=0.23\textwidth]{images/ours_xyz_dragon_comp_albedo_30000.jpg}

%     \caption{Convergence speed at different iterations of training. Object: XYZ dragon. (left to right): 0, 10k, 20k, 30k iterations. We show that our method converges faster than Reed et. al \cite{reed2023neural}.}
%     \label{fig:iter_comparison}
% \end{figure*}

\subsection{Ablation Studies}
\label{sec:ablation}
We provide ablations and additional results in the supplemental material. First, we include additional simulated results with qualitative comparisons and per-object views. Second, an SH-level ablation varies the maximum degree $L\in\{1,2,3\}$ and reports qualitative trends; we omit $L{=}0$ since it matches the isotropic baseline~\cite{reed2023neural} in the main text. Third, a threshold sensitivity study compares iso-surface thresholds for marching cubes to illustrate reconstruction stability. Finally, signal fitting and novel-view synthesis present ToF fitting curves on training baselines and held-out TX–RX pairs to demonstrate generalization.

\subsection{Underwater Reconstruction from SVSS}
\begin{figure*}[!t]
  \centering
  \includegraphics[width=\textwidth]{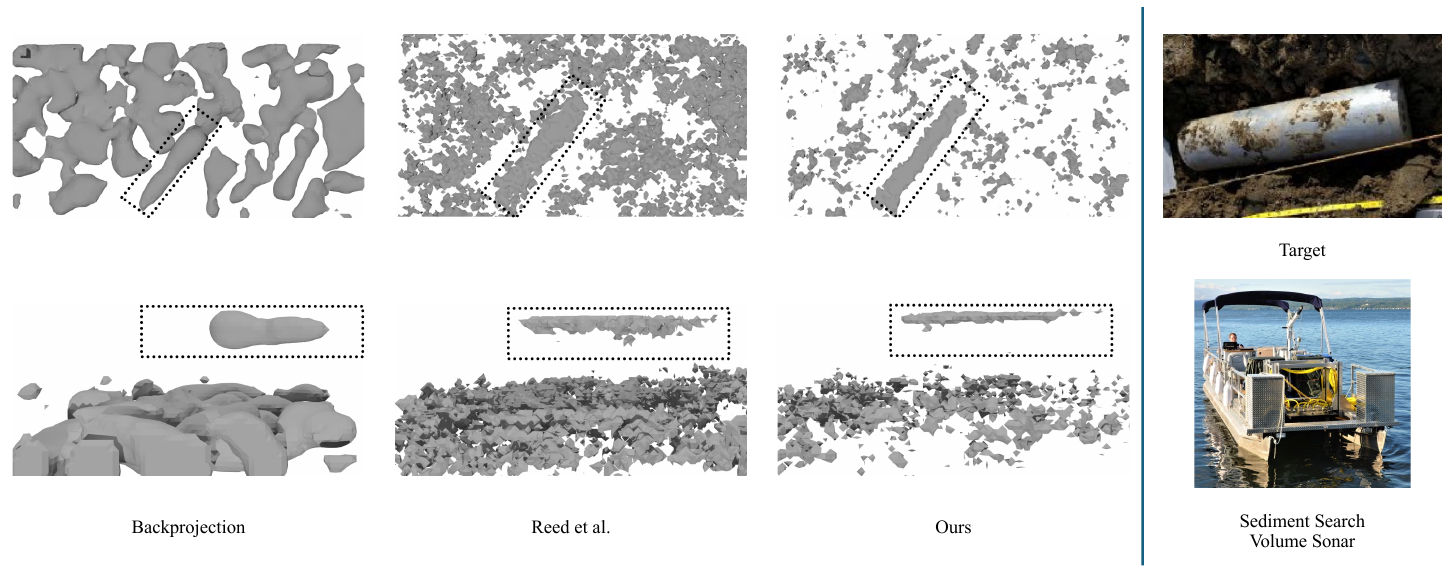}
  \caption{Underwater reconstruction from SVSS. Compared to backprojection and Reed et al.~\cite{reed2023neural}, our method recovers the expected flat side profile and rectangular footprint of the cylindrical pipe, while suppressing spurious returns below the target.}
  \label{fig:svss_results}
\end{figure*}

The SVSS target is a cylindrical pipe placed on the lakebed. Because the acquisition provides a single-look, grazing-angle view, only the near (exposed) side of the cylinder is insonified; the underside is fully shadowed. Consequently, the physically consistent reconstruction should exhibit a \emph{flattened} side profile and a footprint whose apparent cross-section is a rectangular slab corresponding to the visible face, with no structure below it. In~\cref{fig:svss_results}, backprojection produces rounded sidewalls with streaking and spurious energy beneath the target, and Reed et al.~\cite{reed2023neural} reduces but does not remove this curvature and bleed-through. Our method yields the desired flat side view and a clean rectangular cross-section while suppressing returns below the object, aligning with the one-sided visibility of the SVSS geometry (dashed boxes).

\section{Discussion}
\label{sec:discussion}

\paragraph{Discussion.}
To our knowledge, this is the first SAS reconstruction method that parameterizes the \emph{complex} acoustic scattering field with spherical harmonics (SH) inside an implicit neural representation trained directly from 1D ToF signals. Although standard in optical neural rendering and Gaussian splatting, this design choice empirically improves 3D reconstruction in the acoustic modality. Across simulated, in-air (AirSAS), and underwater (SVSS) experiments, modeling view dependence via higher-order SH consistently outperforms isotropic INR baselines (e.g., \cite{reed2023neural}). The DC term provides a stable isotropic density proxy, while higher orders compactly capture directional scattering—yielding sharper geometry under sparse views and fewer artifacts (e.g., better leg/tail preservation on real objects and a flatter side profile with rectangular footprint on the SVSS cylinder, consistent with one-sided visibility).

Our approach also converges faster in practice than isotropic INR while maintaining similar per-iteration cost (~\cref{sec:results}). These results suggest that directional parameterizations are a better inductive bias for SAS than purely isotropic scattering.

\paragraph{Limitations:} Our approach assumes moderate SNR and sufficient bandwidth. In very noisy acquisitions, the higher-order SH terms ($\ell>0$) act like high-frequency angular basis functions that readily fit noise, introducing spurious anisotropy and ripples that corrupt the geometry. With narrowband signals the problem worsens: range resolution degrades as $\Delta R \approx c/(2\Delta f)$ and pulse-compression gain scales with the time–bandwidth product, so returns from neighboring depths blur together and the directional components become weakly identifiable, making the SH fit ill-conditioned. Finally, our method does not model complex acoustic wave phenomena (e.g., diffraction, interference, and multiple reflections/scattering); incorporating these effects would require differentiable forward models integrated into our framework and is left for future work.

\paragraph{Future work:}
% A promising direction is a \emph{Gaussian splatting}~\cite{kerbl2023gaussian} formulation for SAS with directional scattering:
% \begin{itemize}[leftmargin=10pt,itemsep=2pt,topsep=2pt]
%   \item \textbf{SH-Gaussian primitives:} Represent the scene as a set of 3D anisotropic Gaussians, each storing complex SH coefficients (DC as density/opacity, higher orders as angular scattering). 
%   \item \textbf{Bistatic splat rendering:} Replace volumetric integration with ordered alpha-compositing of Gaussians along the bistatic TX–RX path with time-gated contributions to match ToF bins; include an explicit transmittance/occlusion term.
%   \item \textbf{Efficiency:} Expect far fewer primitives than voxel fields, enabling faster training/inference and easier incremental updates.
% \end{itemize}
Beyond SAS, we plan to test this technique on other time-of-flight modalities—particularly radar and LiDAR—by swapping in modality-specific forward models (e.g., wavelength and antenna/polarization effects for radar; visibility/BRDF terms for LiDAR). We hypothesize that the SH-based view-dependent modeling that improves SAS will likewise benefit these sensors. In addition, we plan to explore 3D diffusion priors as learned regularizers: beyond sparsity, phase smoothness, and bounded density, a diffusion prior could suppress artifacts under sparse views. Because public SVSS/AirSAS data are limited, one might be able to pre-train such priors on large-scale synthetic data from our simulator and then adapt to real data with lightweight fine-tuning.

% In these regimes we set $L{=}0$ and recover performance on par with Reed et al.~\cite{reed2023neural}; this is expected because the zeroth-order term $\hat{\sigma}_{s,\mathrm{DC}}=\frac{1}{\sqrt{4\pi}}c_{0,0}$ is an isotropic density, i.e., effectively a scaled version of the isotropic scatterer model in \cite{reed2023neural}.

\section{Acknowledgement}
This research was supported in part by ONR grant N00014-23-1-2406, NSF CCF-2326904, NSF CCF-2326905, and a gift by RTX, Inc. GPU support for this research was provided by ASU Research Computing through the SOL supercomputer~\cite{jennewein2023sol}.

{
    \small
    \bibliographystyle{ieeenat_fullname}
    \bibliography{main}
}
\clearpage
\setcounter{page}{1}
\maketitlesupplementary

The supplemental material contains several ablations and additional results that extend the findings of the main manuscript. We provide further simulated reconstructions with qualitative comparisons and per-object visualizations. An SH-level ablation study explores the effect of varying the maximum degree $L \in \{1,2,3\}$. We also present a threshold sensitivity analysis of iso-surface extraction with marching cubes, highlighting reconstruction stability under different thresholds. In addition, we include signal fitting and novel-view synthesis results, showing time-of-flight fitting curves. The supplemental zip file also contains our implementation code, a sample dataset to facilitate reproducibility, and a short video teaser that visually summarizes the main contributions.

\section{Ellipsoidal Sampling: Derivation and Ray–Ellipsoid Intersection}
\label{supp:ellipsoid}
Our measurements are bistatic time-of-flight (ToF), so each sample constrains contributing scene points to a two-focus ellipsoid with foci at the TX/RX. The correct way to associate a ToF bin with spatial hypotheses is therefore \emph{ellipsoidal sampling}: for a given bin, we sample points where a TX ray intersects its constant-ToF ellipsoid. The derivation below (i) parameterizes that ellipsoid from the bin center, (ii) introduces a canonical “ellipsoid frame’’ that accommodates arbitrary TX/RX poses, and (iii) provides a closed-form ray–ellipsoid intersection used throughout our pipeline.
\paragraph{Constant-ToF ellipsoid.}
Let $\mathbf{o}_T,\mathbf{o}_R\in\mathbb{R}^3$ be TX/RX positions, $d=\|\mathbf{o}_T-\mathbf{o}_R\|$ their separation, $c$ the sound speed, and $t$ the ToF of a sample. The locus of points with
\[
t=\frac{R_T+R_R}{c},\quad R_T=\|\mathbf{x}-\mathbf{o}_T\|,\ R_R=\|\mathbf{x}-\mathbf{o}_R\|
\]
is an ellipsoid with foci at $\mathbf{o}_T,\mathbf{o}_R$ and semi-major
\[
a = \frac{ct}{2},\qquad b=c=\sqrt{a^2-(d/2)^2}.
\]

\paragraph{Ellipsoid frame.}
Let $\mathbf{m}=\tfrac{1}{2}(\mathbf{o}_T+\mathbf{o}_R)$ and choose $R\in SO(3)$ so that $R^\top(\mathbf{o}_R-\mathbf{o}_T)=(d,0,0)^\top$. Define ellipsoid-frame coordinates
\[
\tilde{\mathbf{x}} = R^\top(\mathbf{x}-\mathbf{m}),
\quad
\tilde{\mathbf{o}}_T = R^\top(\mathbf{o}_T-\mathbf{m}),
\quad
\tilde{\mathbf{d}}_T = R^\top\mathbf{d}_T,
\]
where $\mathbf{d}_T$ is a unit TX ray direction. In this frame, the ellipsoid is axis-aligned:
\[
\frac{\tilde{x}^2}{a^2} + \frac{\tilde{y}^2}{b^2} + \frac{\tilde{z}^2}{b^2} = 1.
\]

\paragraph{Ray–ellipsoid intersection.}
A TX-emitted ray is
\[
\tilde{\mathbf{x}}(l)=\tilde{\mathbf{o}}_T + l\,\tilde{\mathbf{d}}_T,\quad l\ge 0.
\]
Plugging into the quadric gives $a_0 l^2 + b_0 l + c_0 = 0$ with
\[
\begin{aligned}
a_0 &= \frac{\tilde{d}_{T,x}^2}{a^2} + \frac{\tilde{d}_{T,y}^2}{b^2} + \frac{\tilde{d}_{T,z}^2}{b^2},\\
b_0 &= 2\!\left(\frac{\tilde{o}_{T,x}\tilde{d}_{T,x}}{a^2} + \frac{\tilde{o}_{T,y}\tilde{d}_{T,y}}{b^2} + \frac{\tilde{o}_{T,z}\tilde{d}_{T,z}}{b^2}\right),\\
c_0 &= \frac{\tilde{o}_{T,x}^2}{a^2} + \frac{\tilde{o}_{T,y}^2}{b^2} + \frac{\tilde{o}_{T,z}^2}{b^2} - 1.
\end{aligned}
\]
If $\Delta=b_0^2-4a_0c_0<0$, there is no intersection for that ToF. Otherwise,
\[
l_{\pm}=\frac{-b_0\pm\sqrt{\Delta}}{2a_0},\qquad
l^\star=\min\{l_{\pm}\mid l_{\pm}>0\},
\]
and the sample location is $\mathbf{x}^\star=\mathbf{m}+R\,\tilde{\mathbf{x}}(l^\star)$.
(Use the smallest positive root since the ray originates at the TX.)

% \paragraph{Notes.}
% (i) The formulation naturally supports bistatic geometries and arbitrary TX–RX poses via the ellipsoid frame. (ii) For discretized ToF bins, set $a=\tfrac{c}{2}\,t_{\text{bin}}$. (iii) Degenerate cases with $a\le d/2$ (pre-focus bins) yield no real intersection and are skipped.

\section{Simulations—Additional Results}
\begin{figure*}[p]
    \centering
    \includegraphics[width=0.23\textwidth]{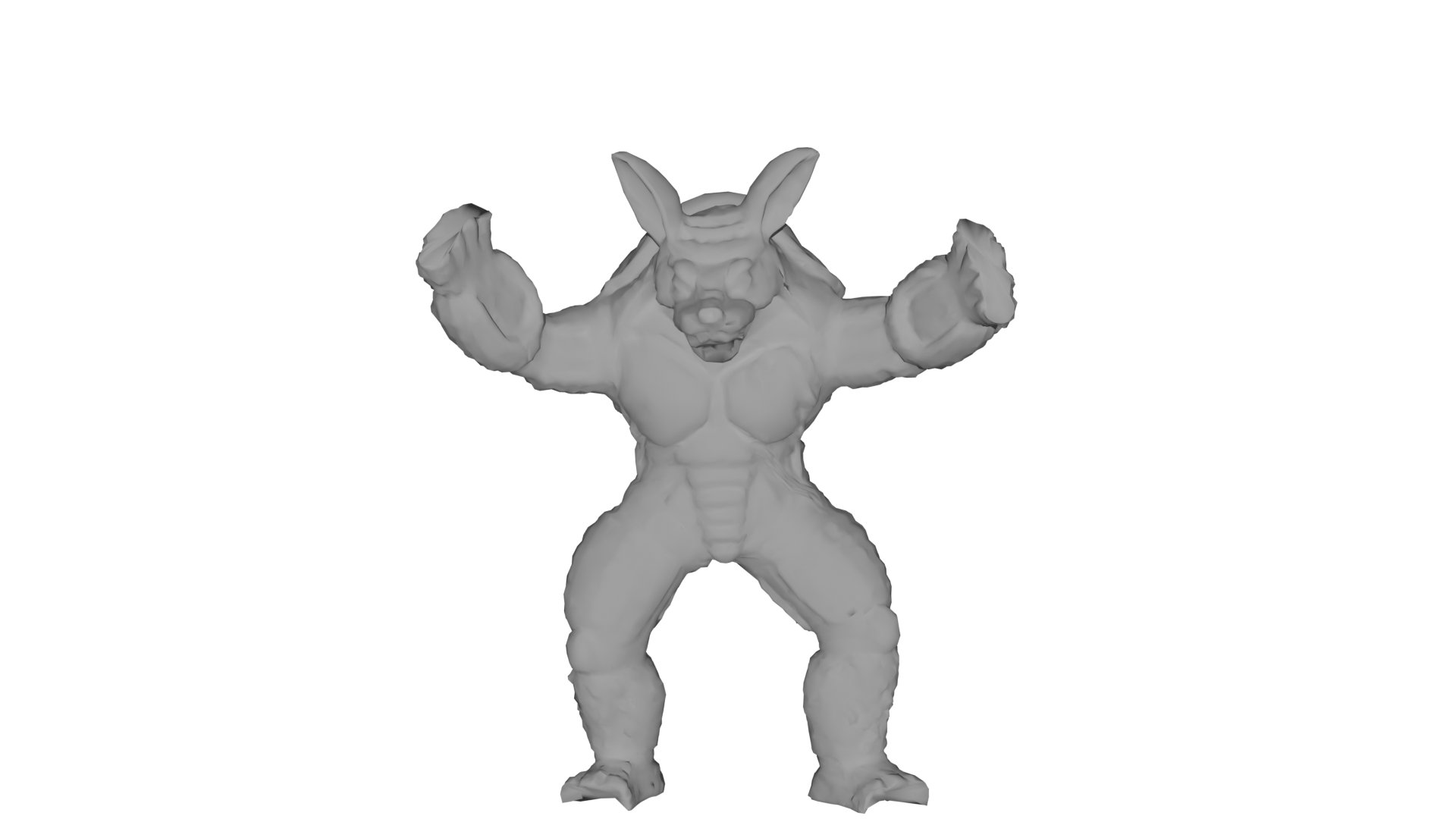}
    \includegraphics[width=0.23\textwidth]{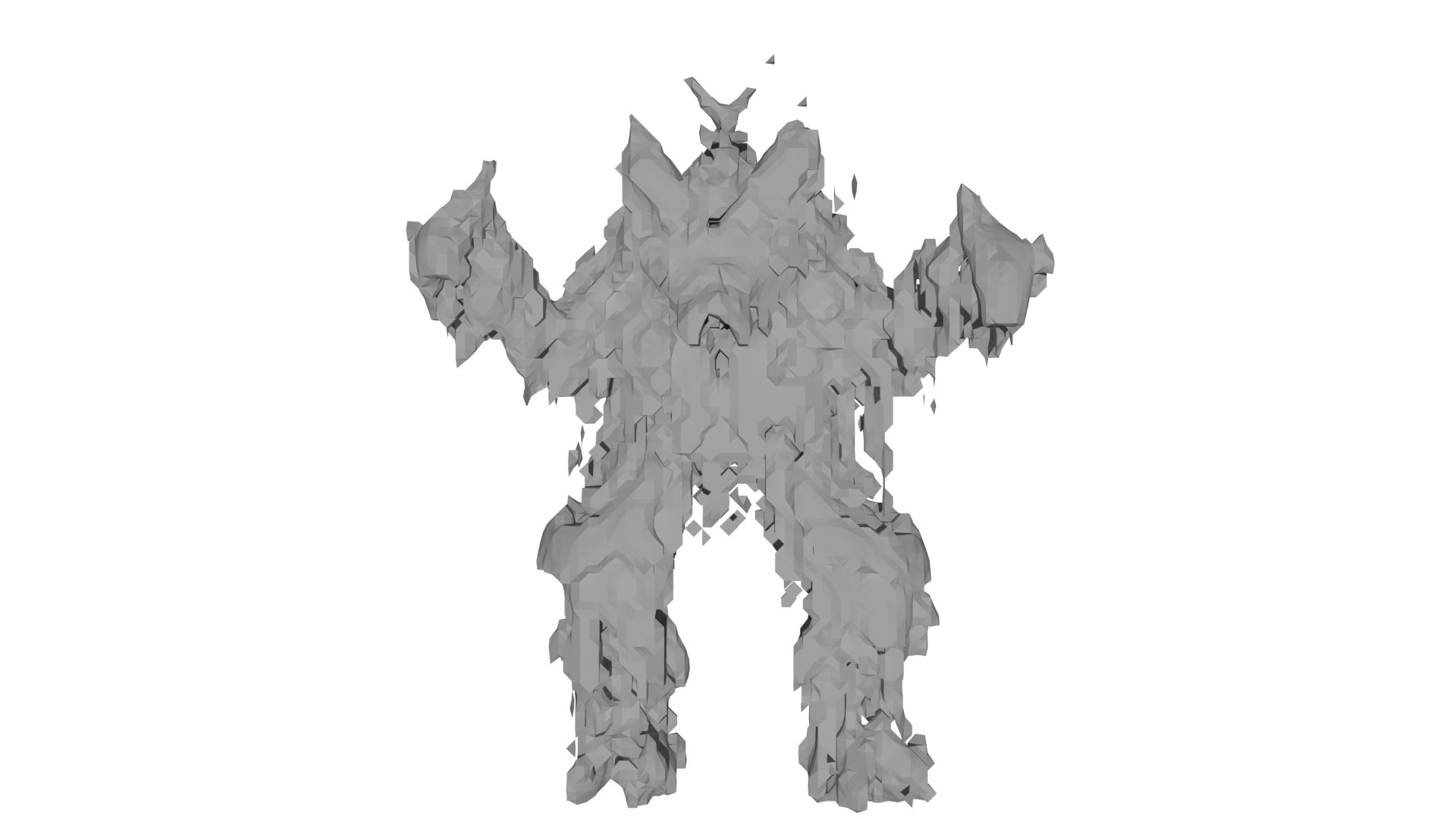}
    \includegraphics[width=0.23\textwidth]{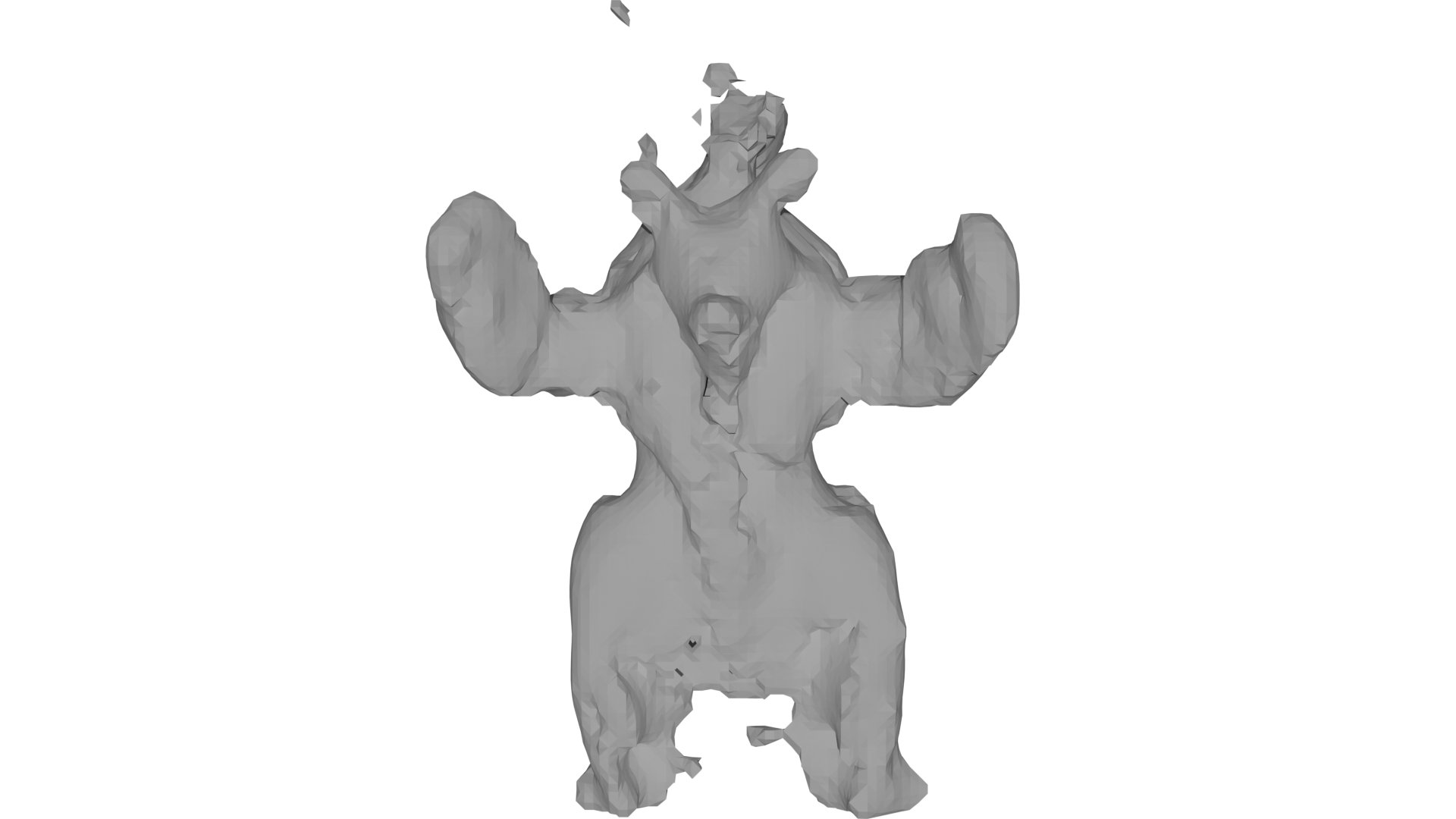}
    \includegraphics[width=0.23\textwidth]{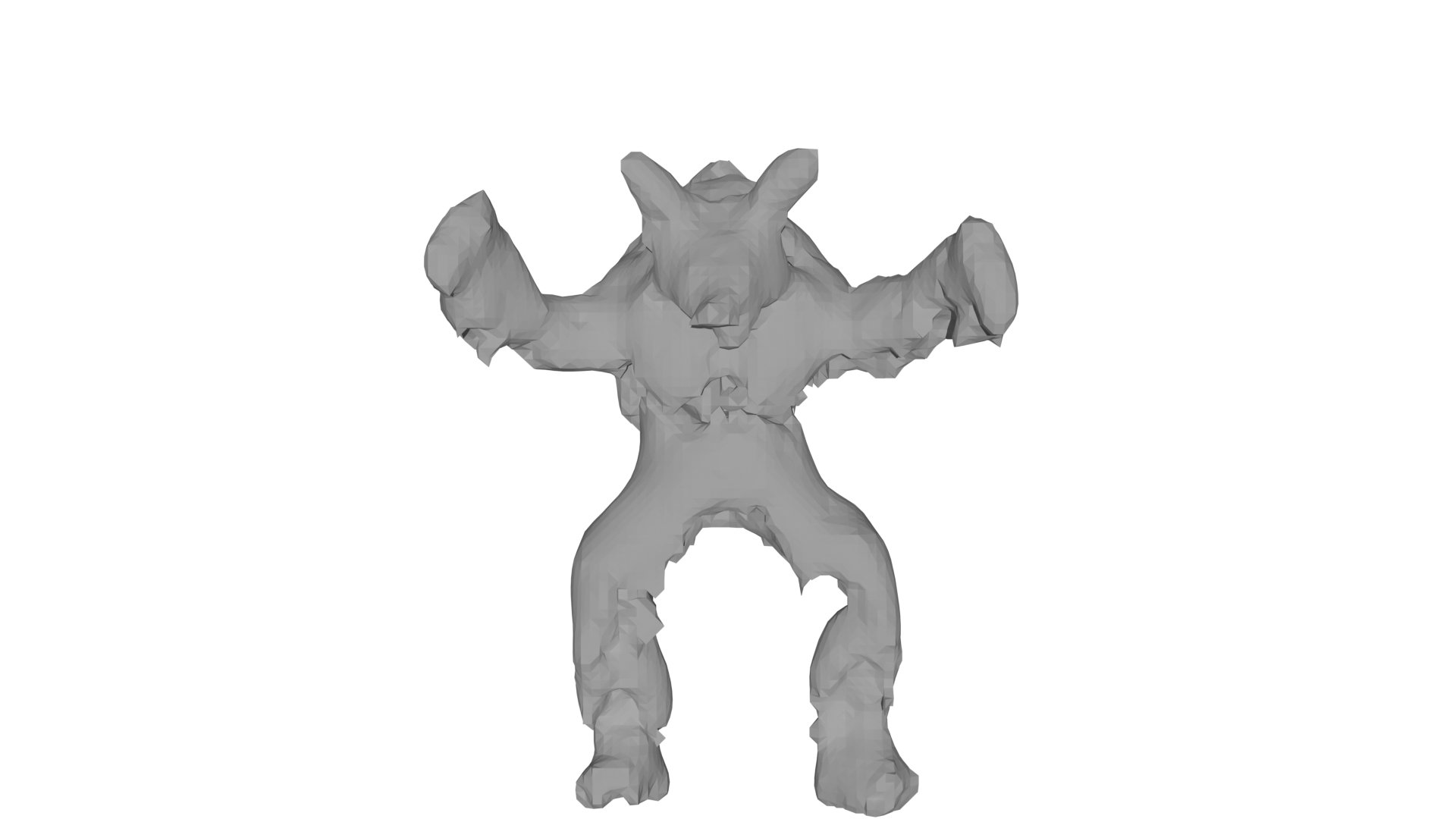}

    \vspace{2mm}

    \includegraphics[width=0.23\textwidth]{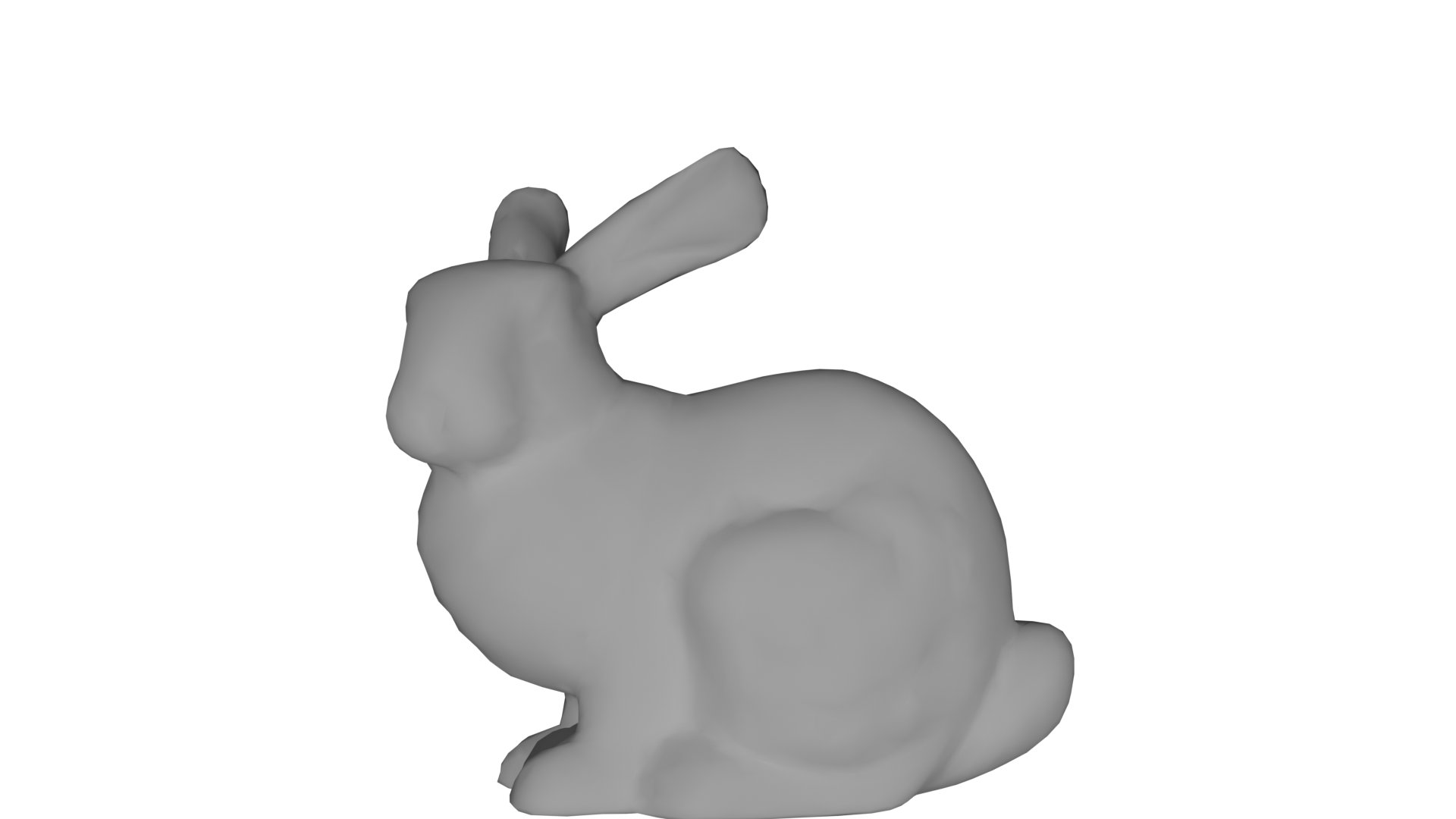}
    \includegraphics[width=0.23\textwidth]{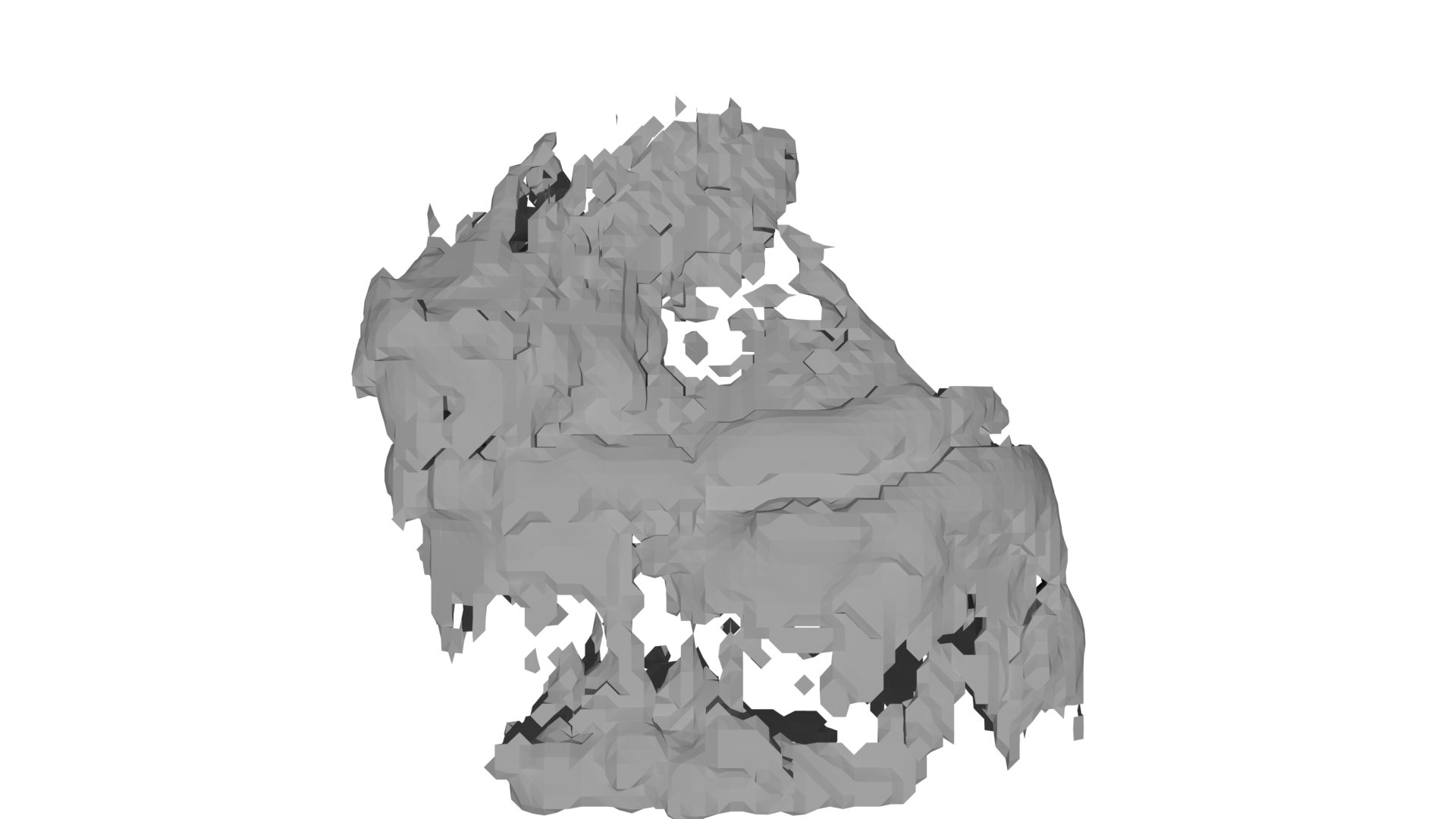}
    \includegraphics[width=0.23\textwidth]{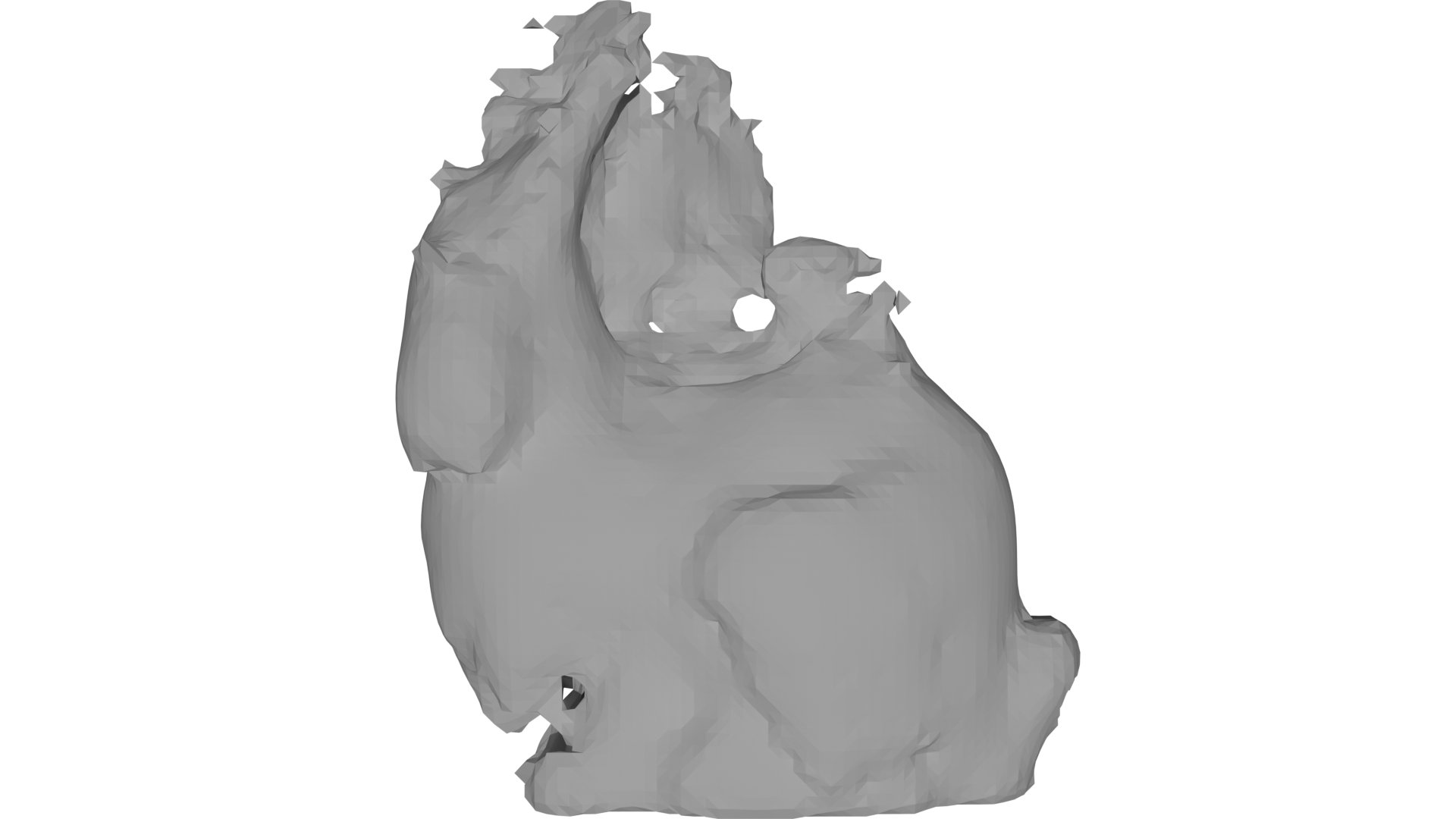}
    \includegraphics[width=0.23\textwidth]{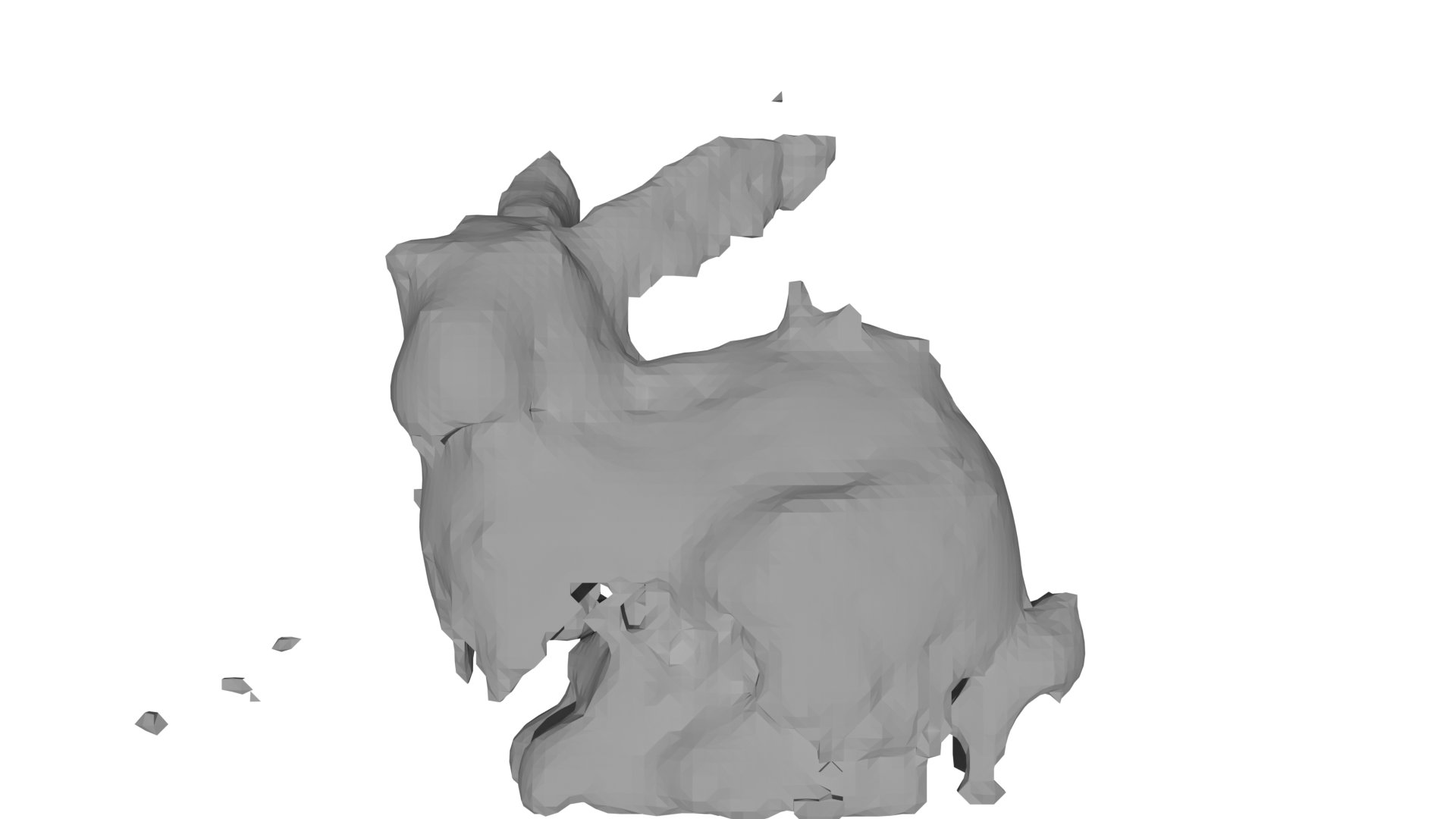}

    \vspace{2mm}

    \includegraphics[width=0.23\textwidth]{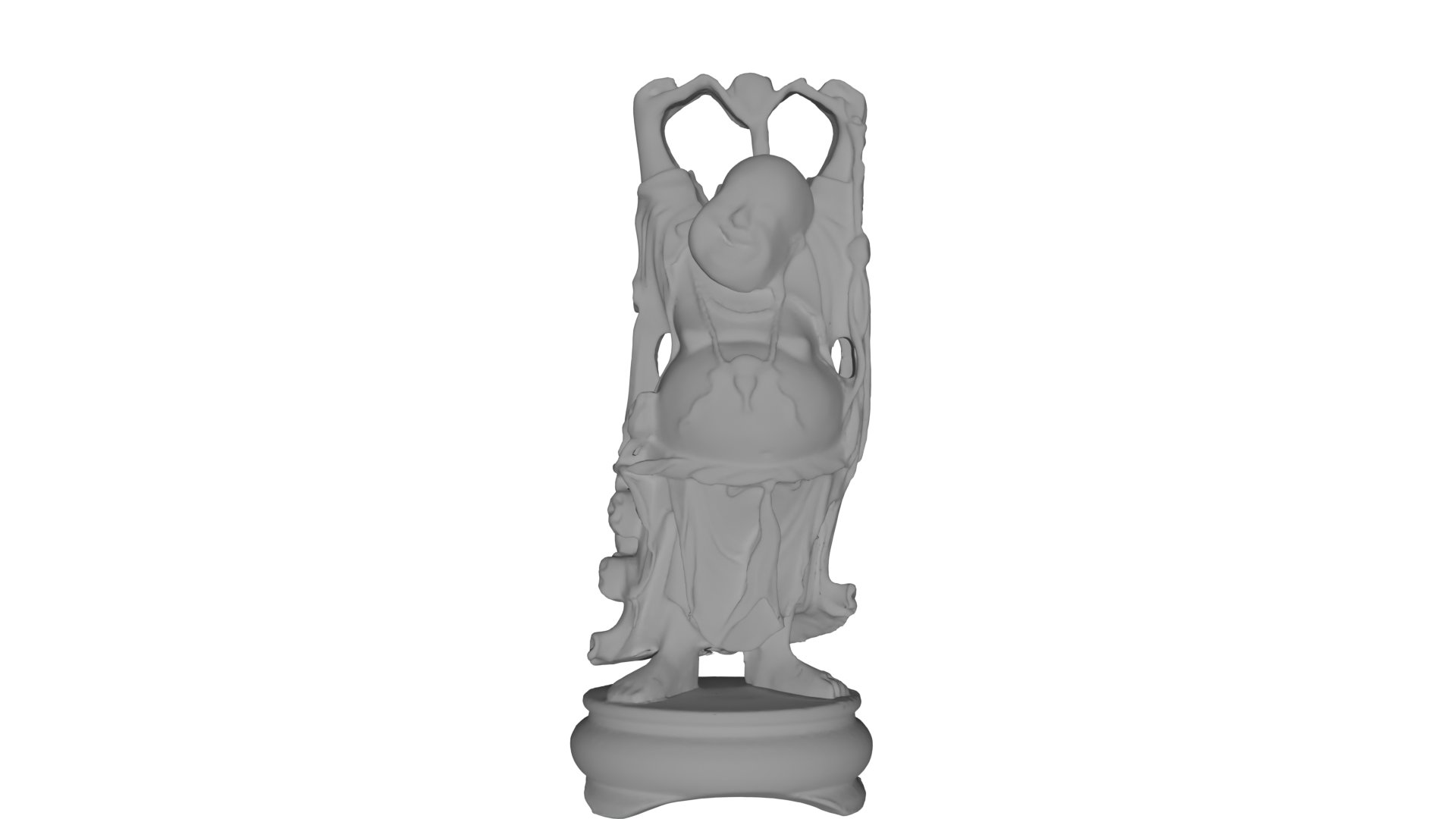}
    \includegraphics[width=0.23\textwidth]{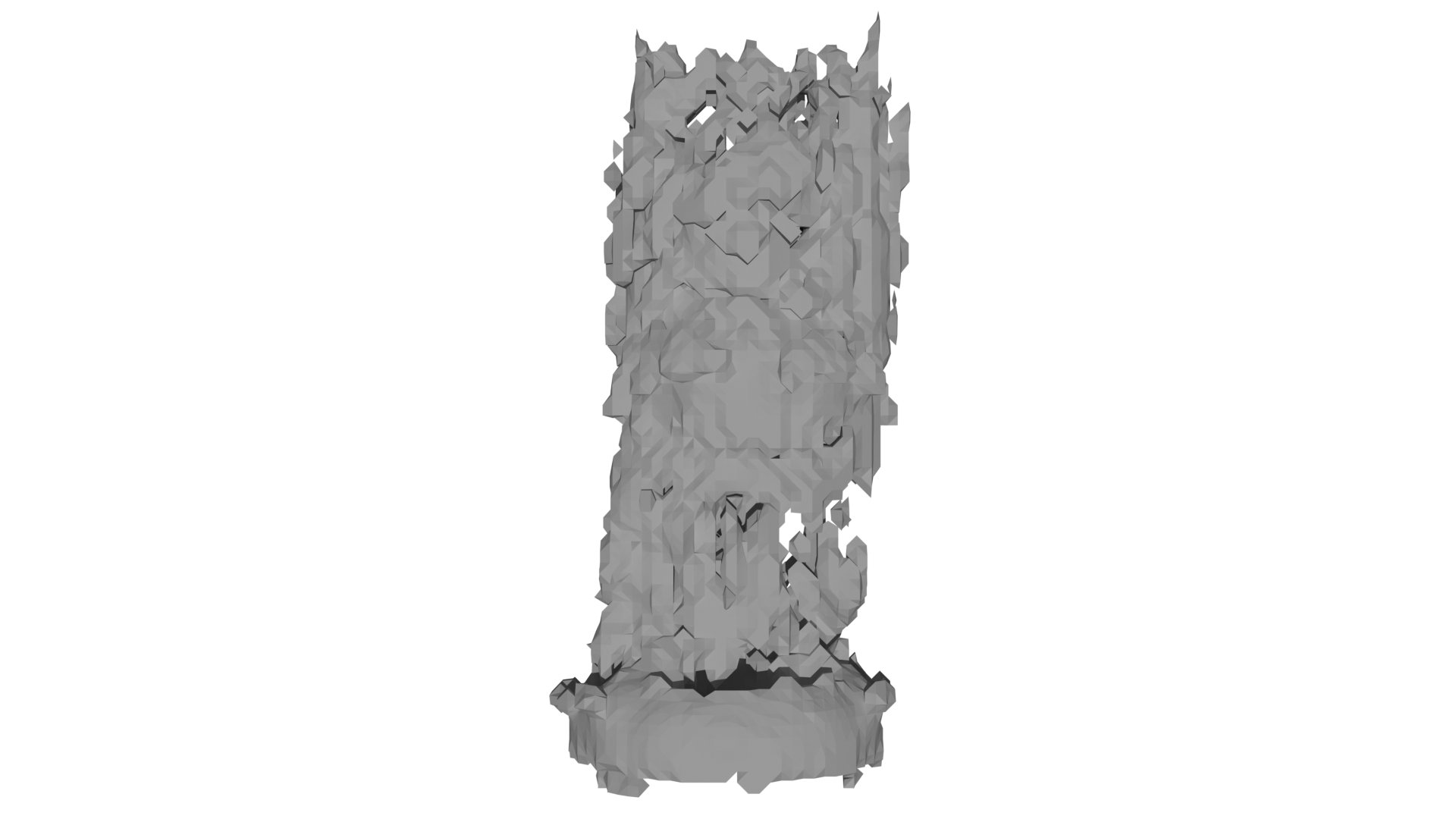}
    \includegraphics[width=0.23\textwidth]{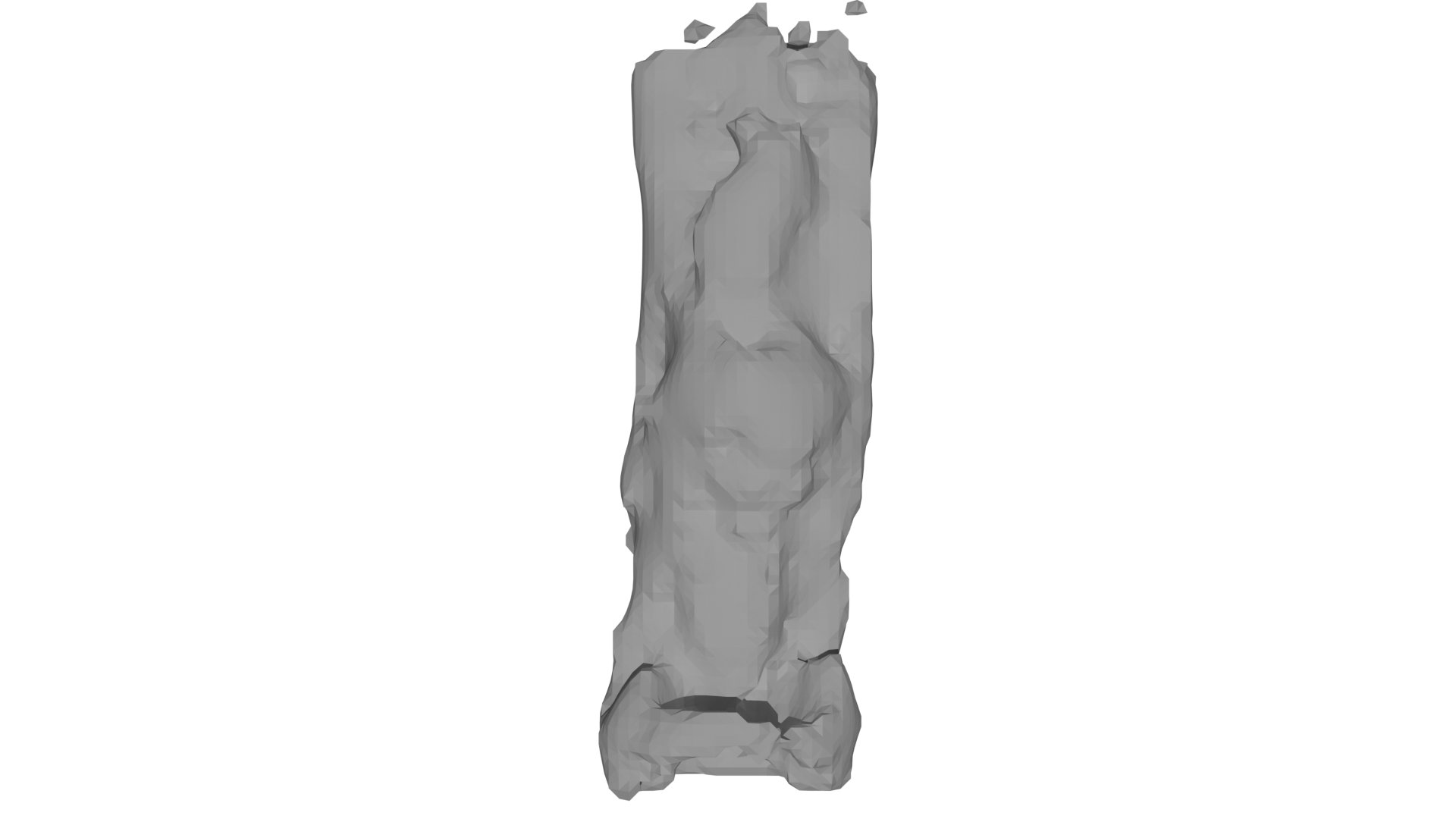}
    \includegraphics[width=0.23\textwidth]{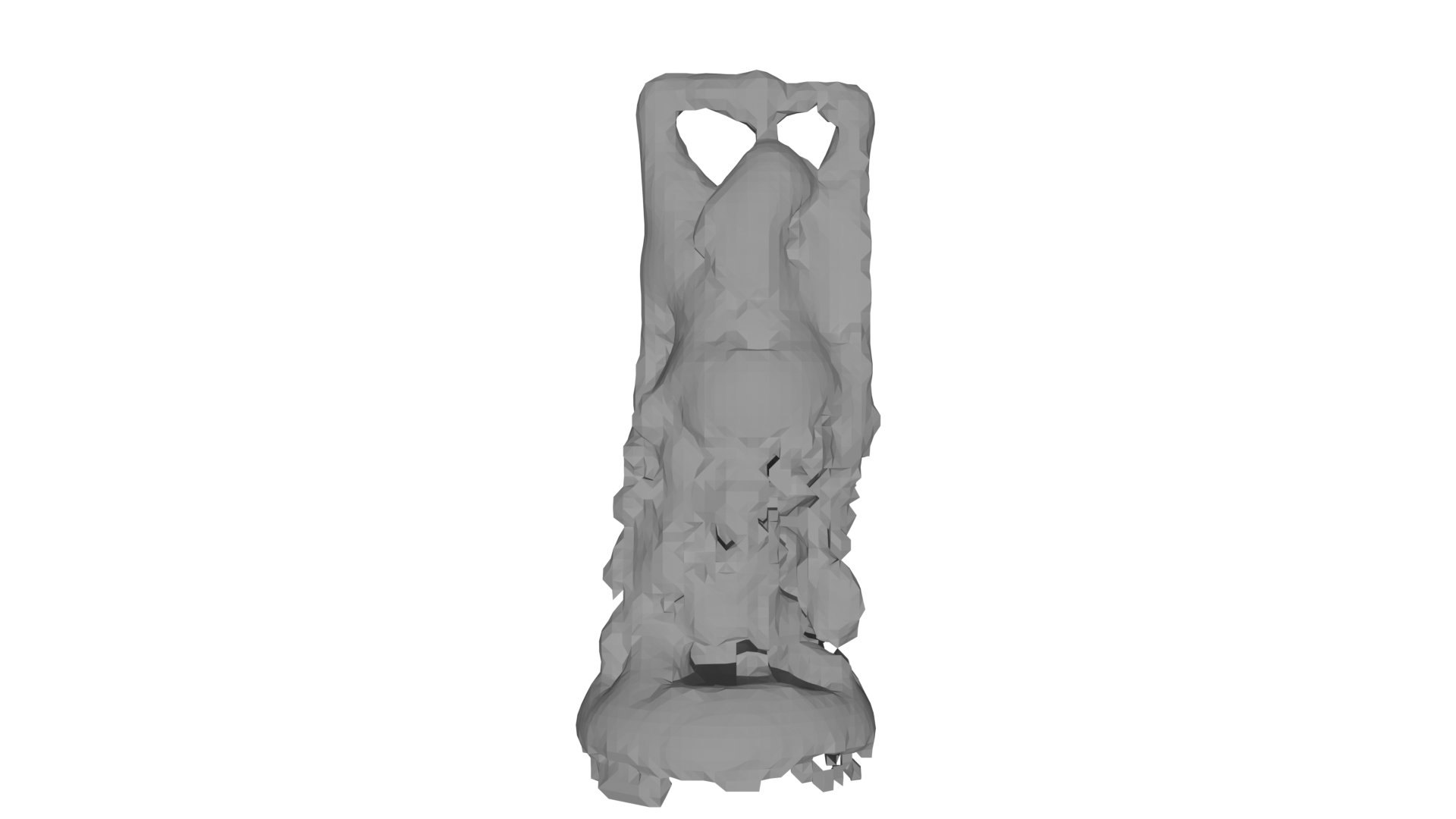}

    \vspace{2mm}

    \includegraphics[width=0.23\textwidth]{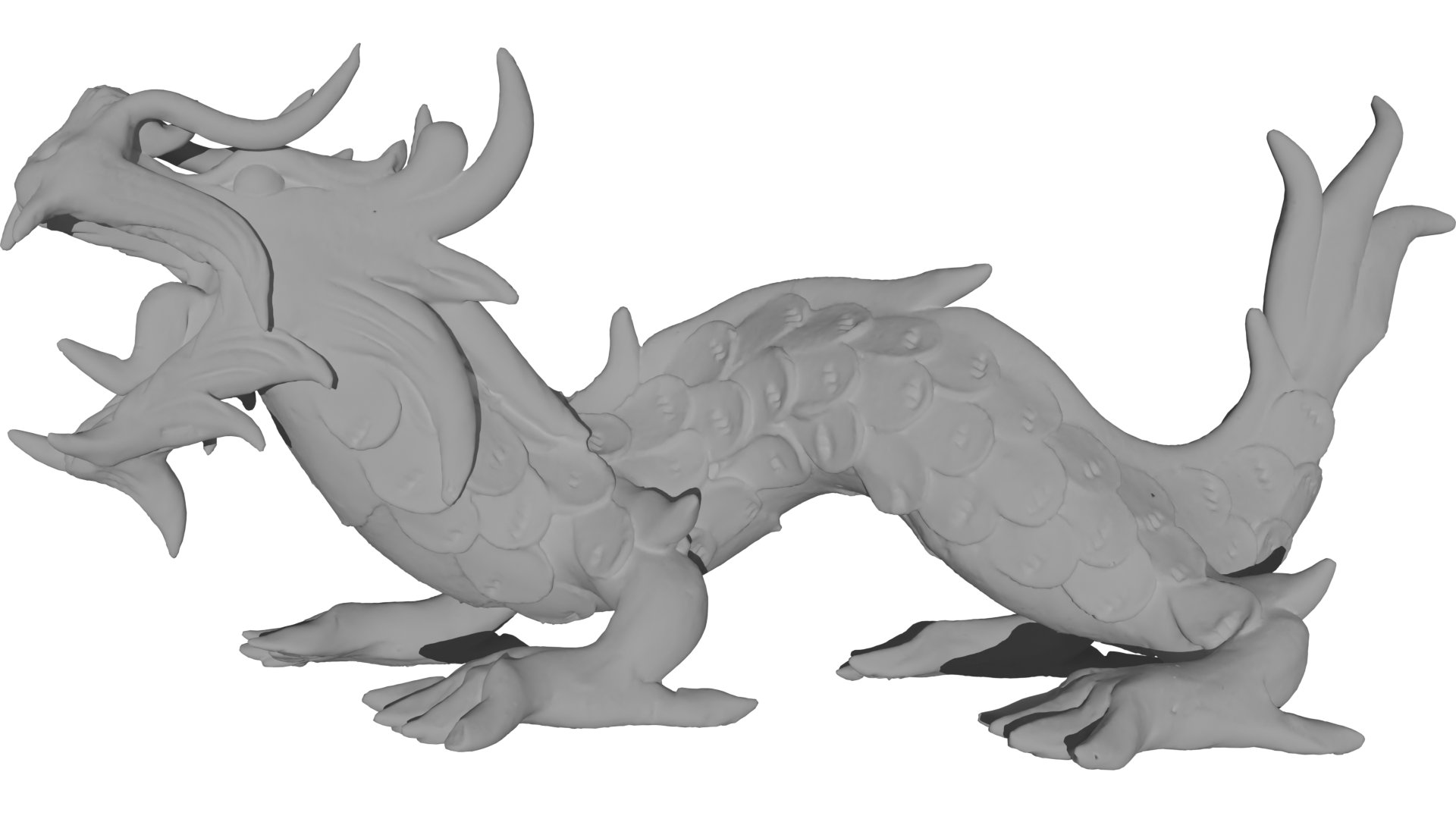}
    \includegraphics[width=0.23\textwidth]{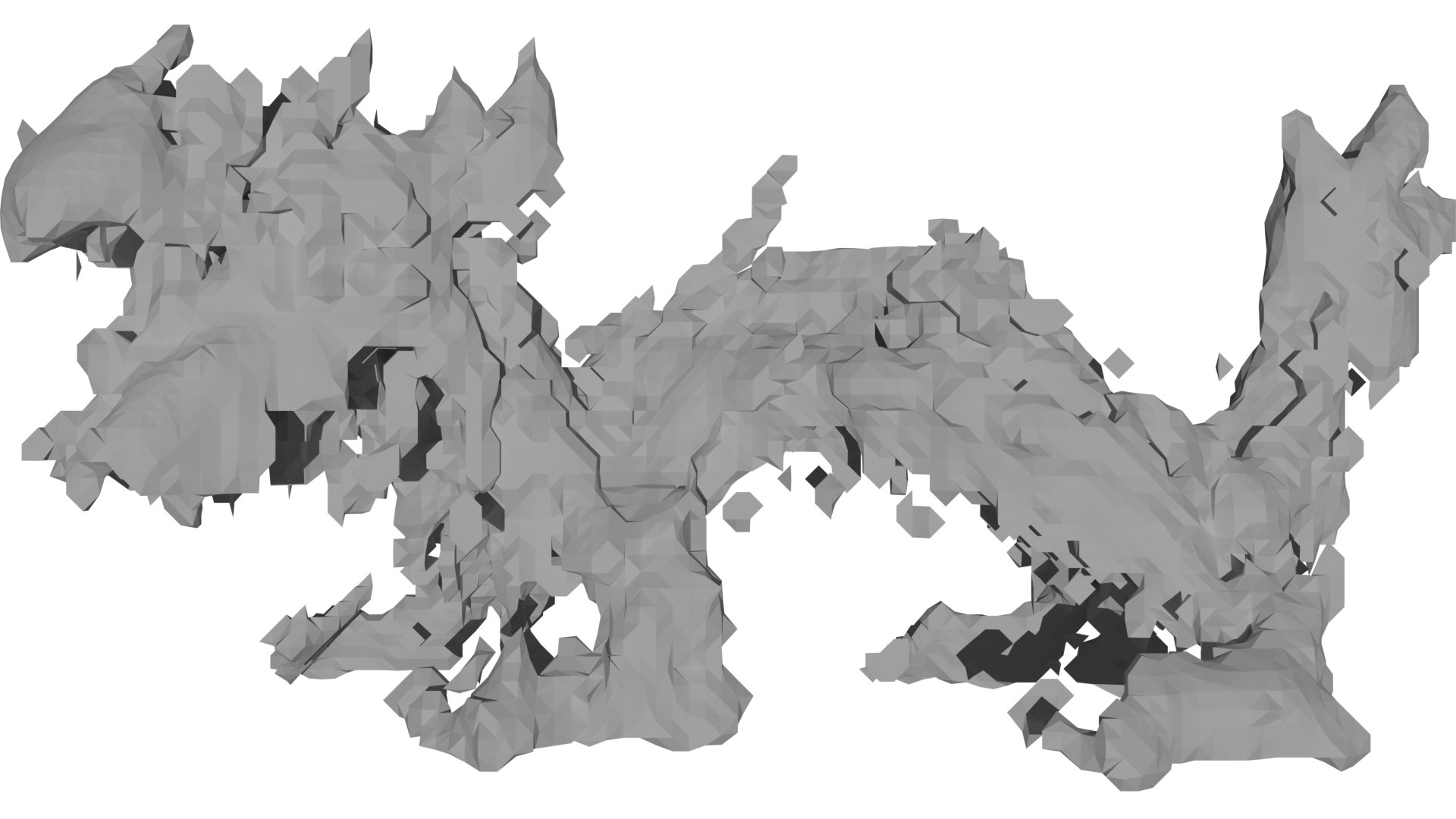}
    \includegraphics[width=0.23\textwidth]{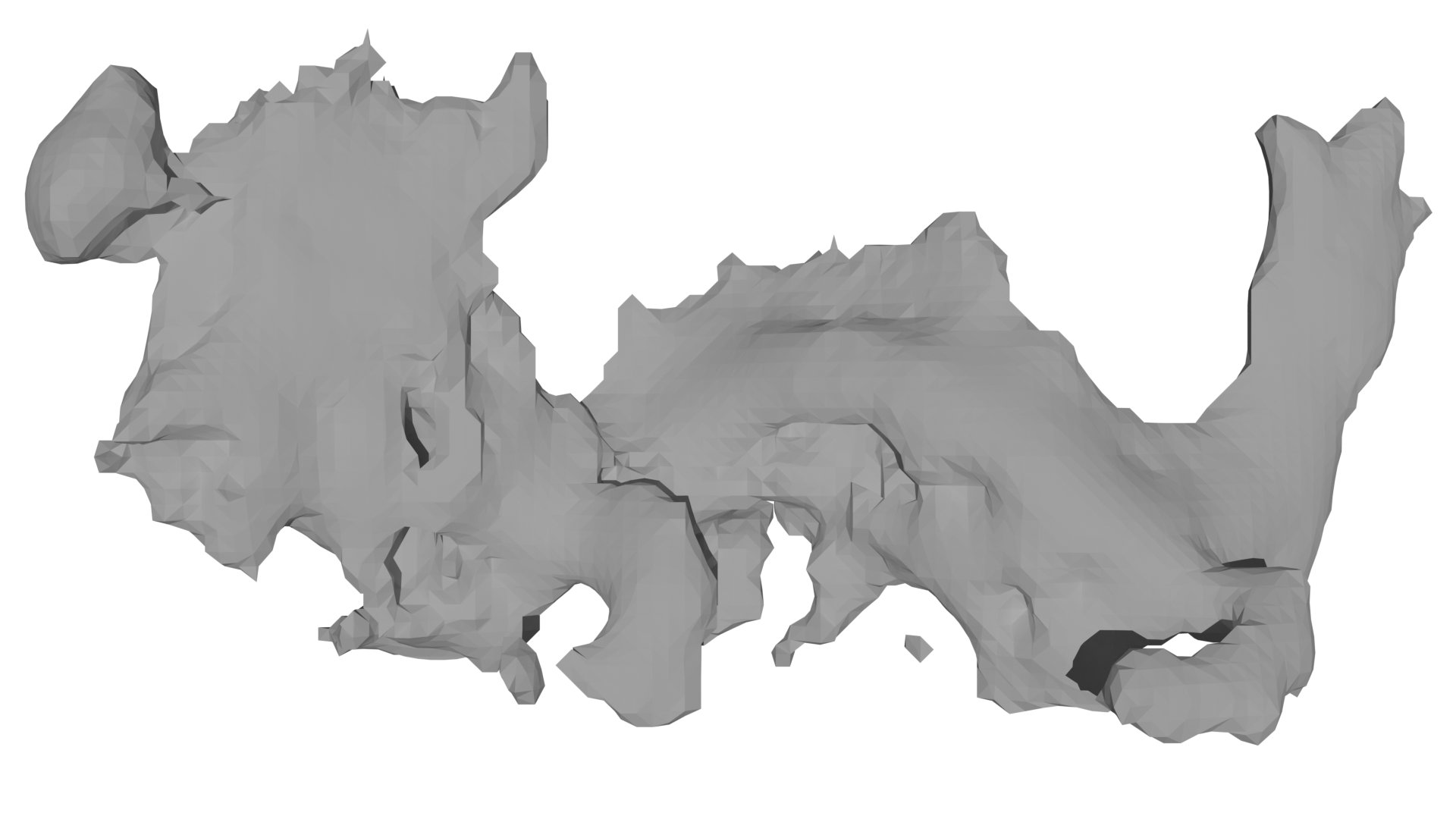}
    \includegraphics[width=0.23\textwidth]{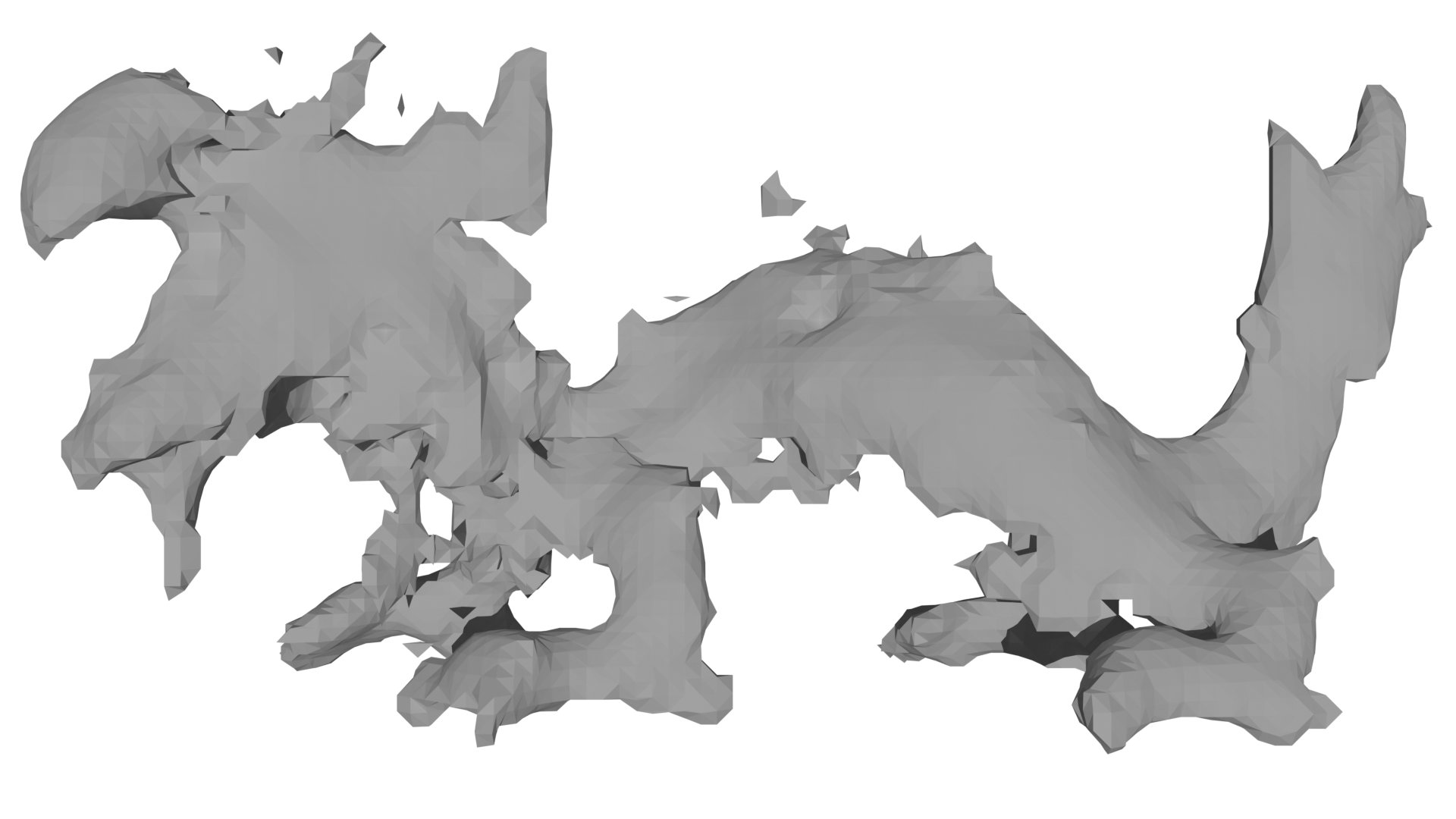}

    \vspace{2mm}

    \includegraphics[width=0.23\textwidth]{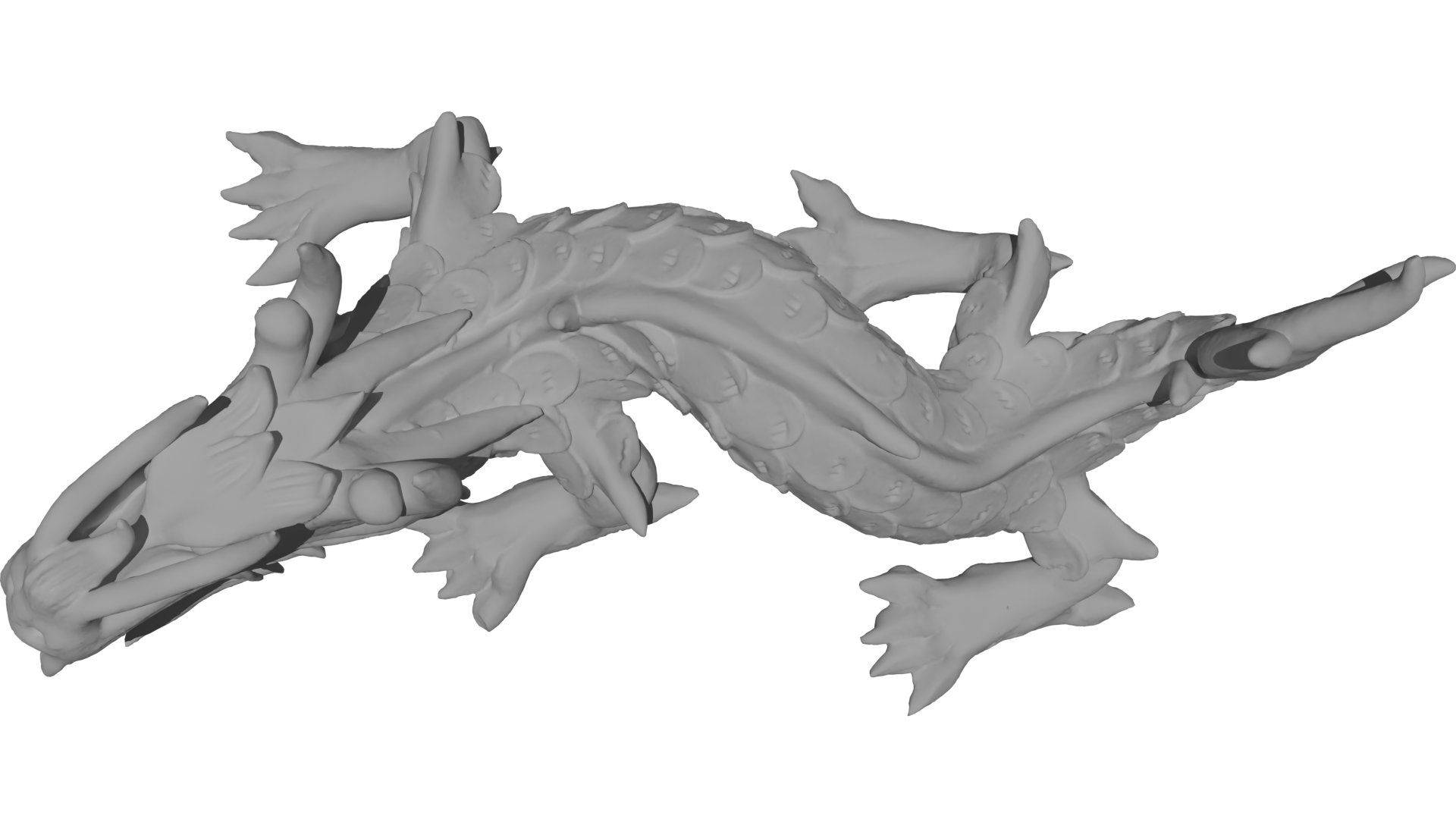}
    \includegraphics[width=0.23\textwidth]{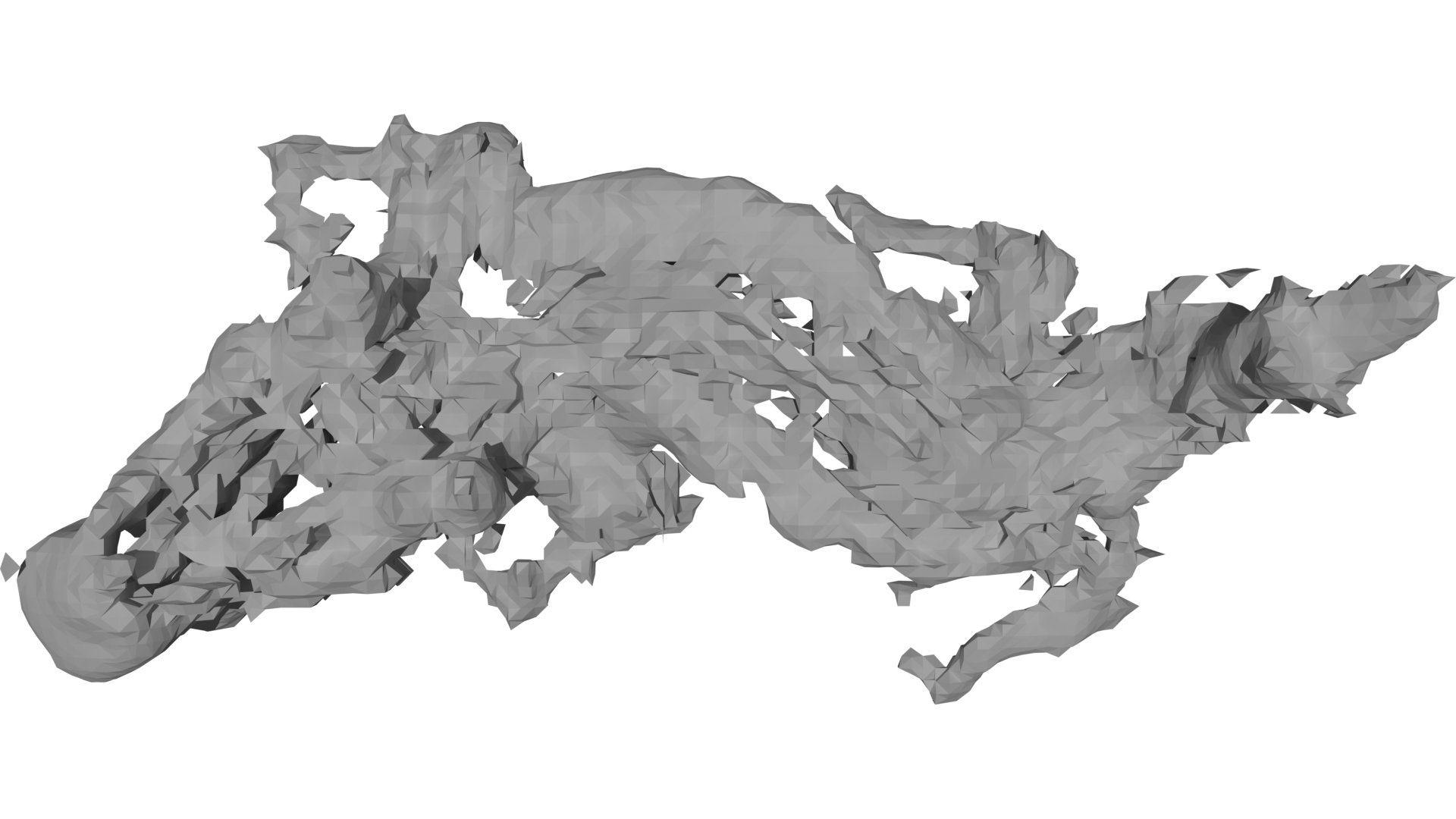}
    \includegraphics[width=0.23\textwidth]{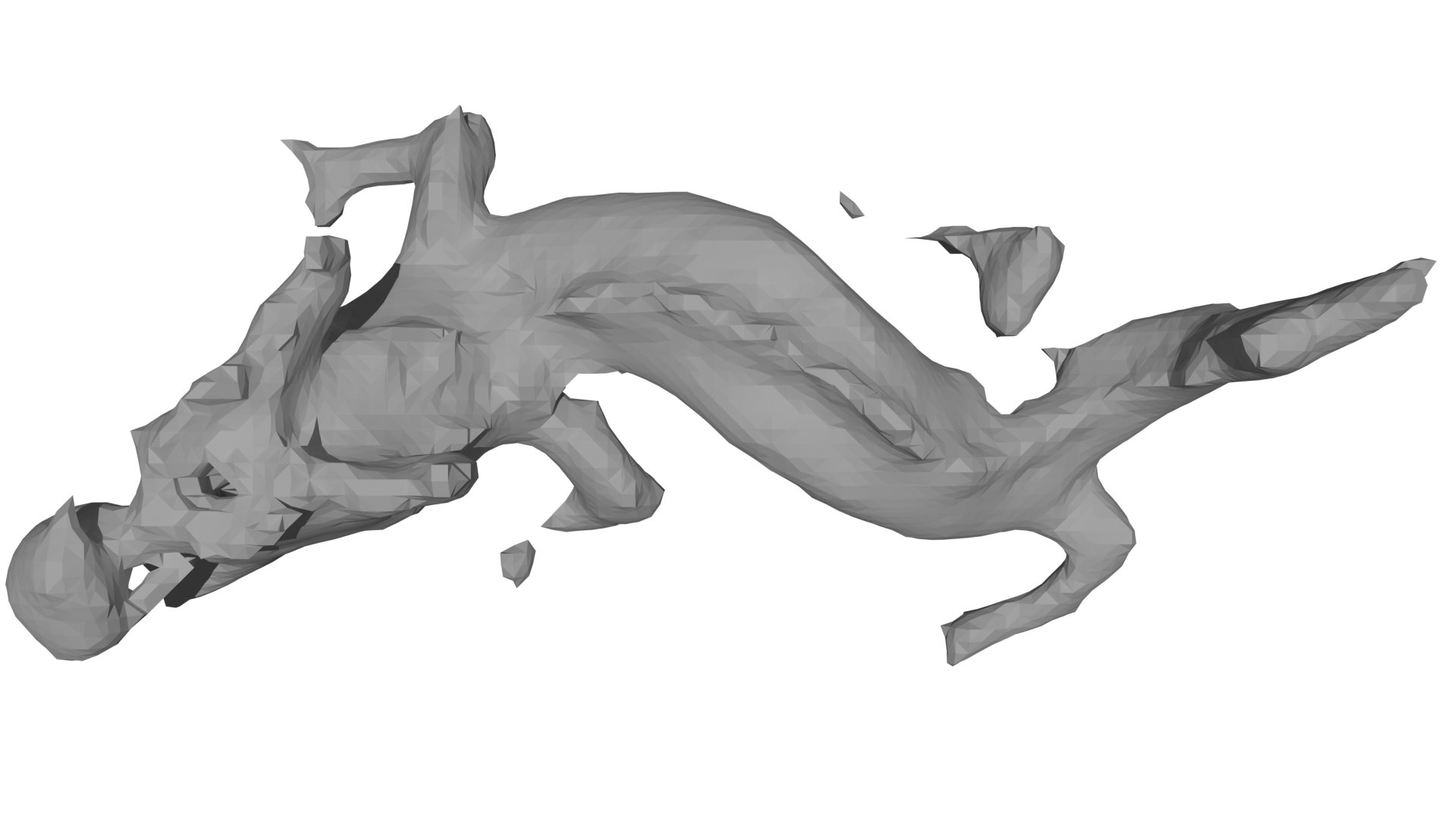}
    \includegraphics[width=0.23\textwidth]{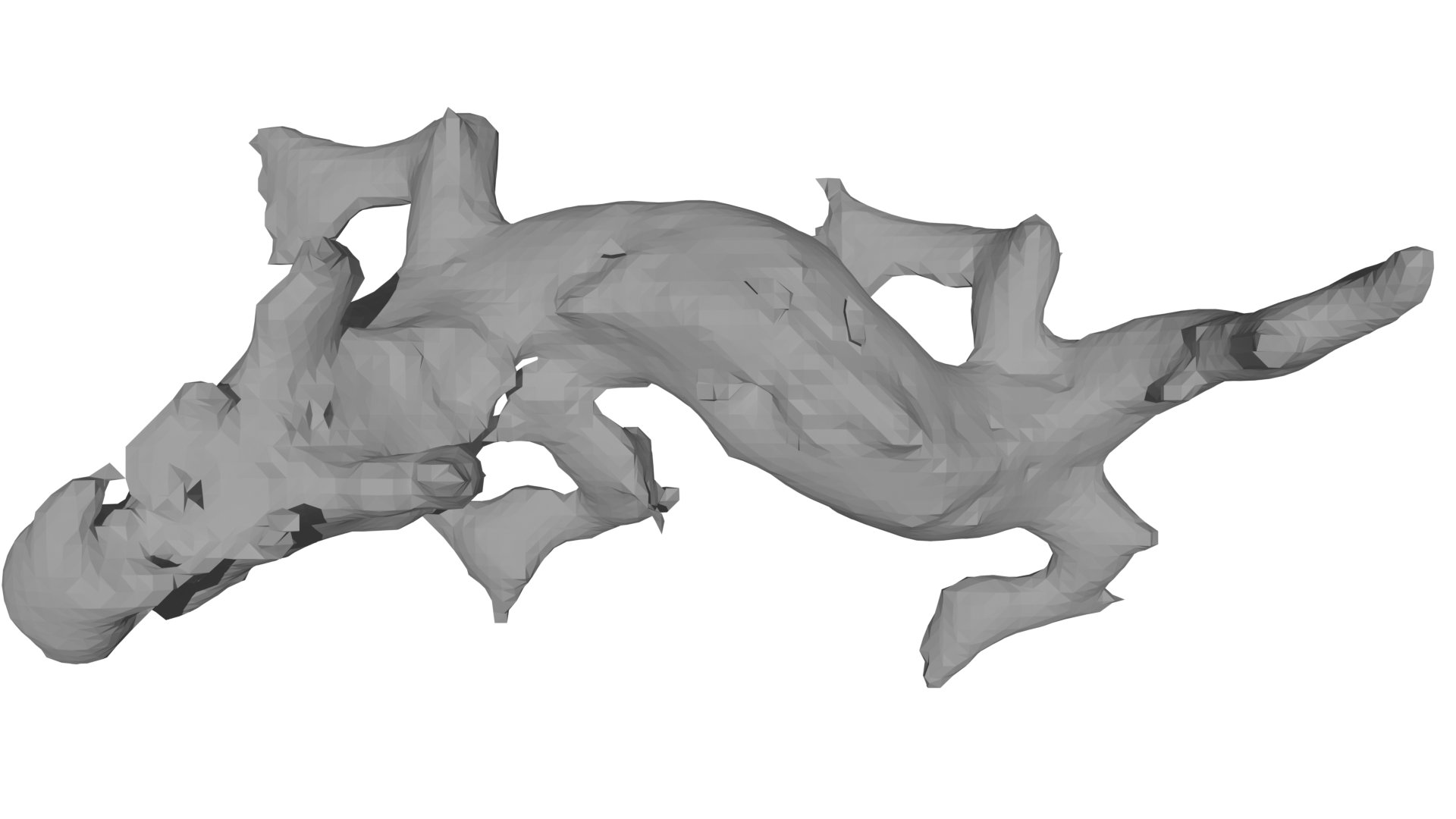}

    \caption{Mesh reconstructions of synthetic data: Armadillo, Bunny, Happy Buddha, XYZ Dragon (side, top) scenes using (from left to right): Ground Truth, Backprojection, Reed et. al \cite{reed2023neural}, and Ours.}
    \label{fig:synthetic_data_comparison}
\end{figure*}

Qualitatively, our reconstructions are cleaner and more complete across all four synthetic scenes (Armadillo, Bunny, Happy Buddha, XYZ Dragon)~\cref{fig:synthetic_data_comparison}: backprojection exhibits speckle and fragmentation, while Reed et~al.\ tends to over-smooth and lose thin structures; our method recovers sharper geometry with fewer holes and floaters. Quantitatively, our approach ranks best on \emph{most} metrics and is never worst. We evaluate the methods on the Chamfer distance, IoU, precision, and F1 score. On average, our method outperforms the baseline methods as shown in~\cref{tab:objects-pc-vs-mesh-colored}.

\section{Ablation Study: SH levels}

\paragraph{Background on spherical harmonics.}
Spherical harmonics (SH) form an orthonormal basis on the unit sphere $\mathbb{S}^2$ and let us represent directional or view-dependent terms (e.g., scattering/BRDF lobes) as a band-limited expansion. An SH \emph{level} $L$ includes all degrees $\ell=0,\dots,L$, for a total of $(L+1)^2$ coefficients per scalar field; increasing $L$ increases the angular bandwidth that can be expressed. Intuitively, $L{=}1$ captures a diffuse term plus first-order lobes, $L{=}2$ adds quadratic variation, and $L{=}3$ can express noticeably sharper, asymmetric lobes. In our pipeline, these coefficients are learned at each 3D location and evaluated along ellipsoidal samples consistent with the bistatic ToF geometry.

\begin{figure*}[p]
    \centering

    % Level 1
    \begin{subfigure}[t]{0.22\textwidth}
        \centering
        \includegraphics[width=\linewidth]{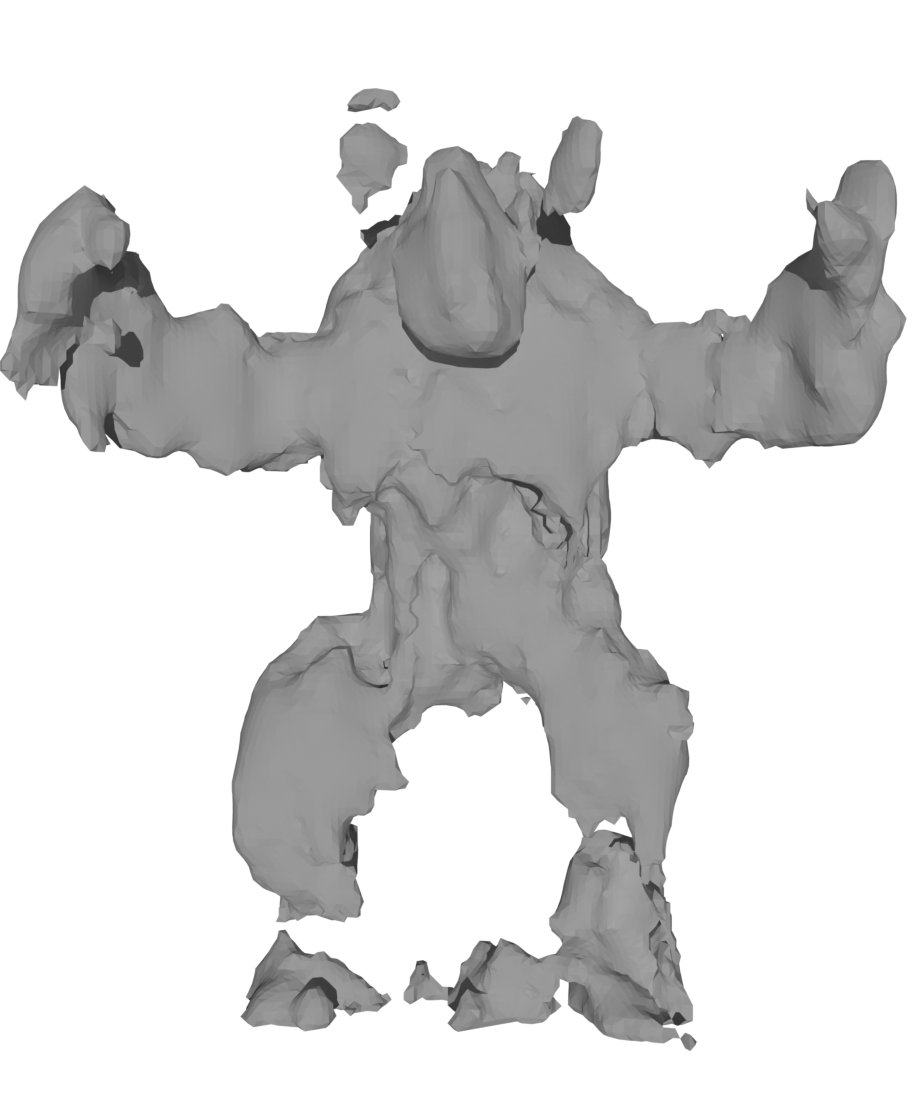}
    \end{subfigure}
    \hspace{10pt}
    \begin{subfigure}[t]{0.22\textwidth}
        \centering
        \includegraphics[width=\linewidth]{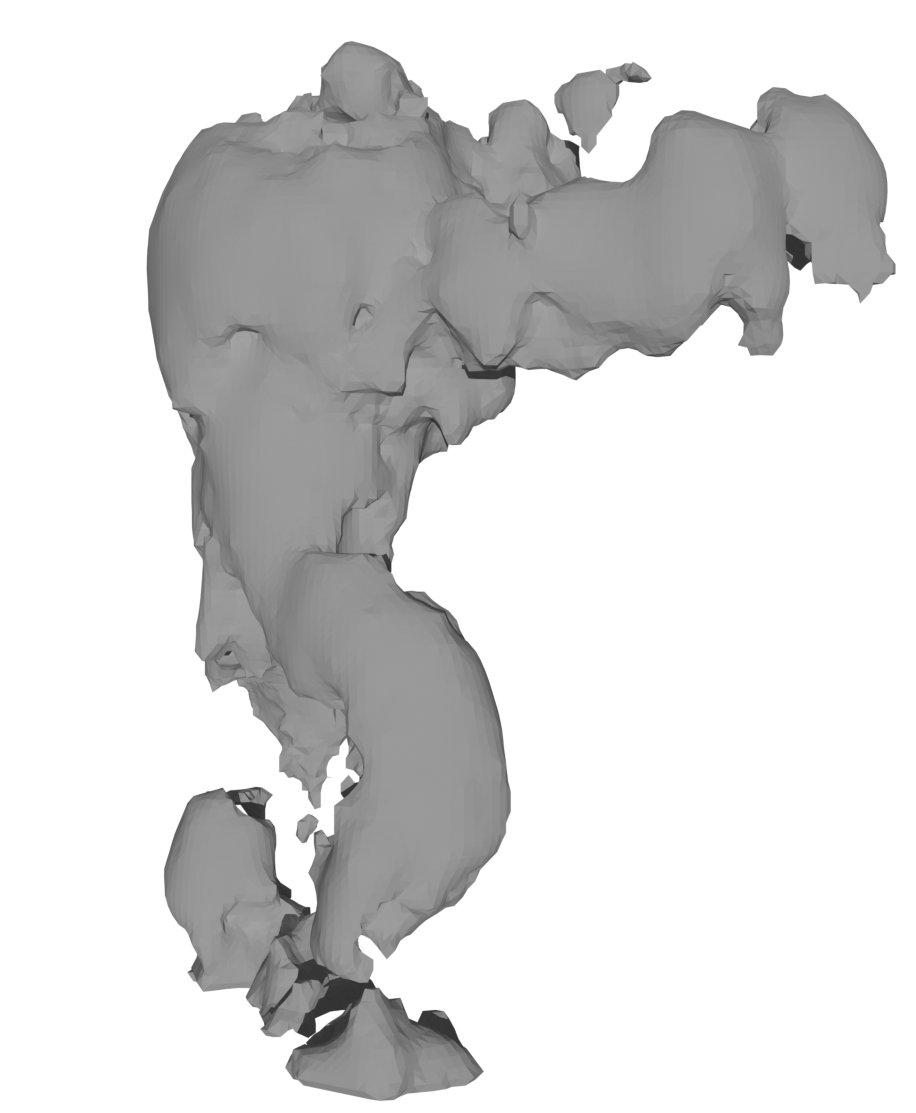}
    \end{subfigure}

    \vspace{2pt}
    {\footnotesize Level 1}
    \vspace{10pt}

    % Level 2
    \begin{subfigure}[t]{0.22\textwidth}
        \centering
        \includegraphics[width=\linewidth]{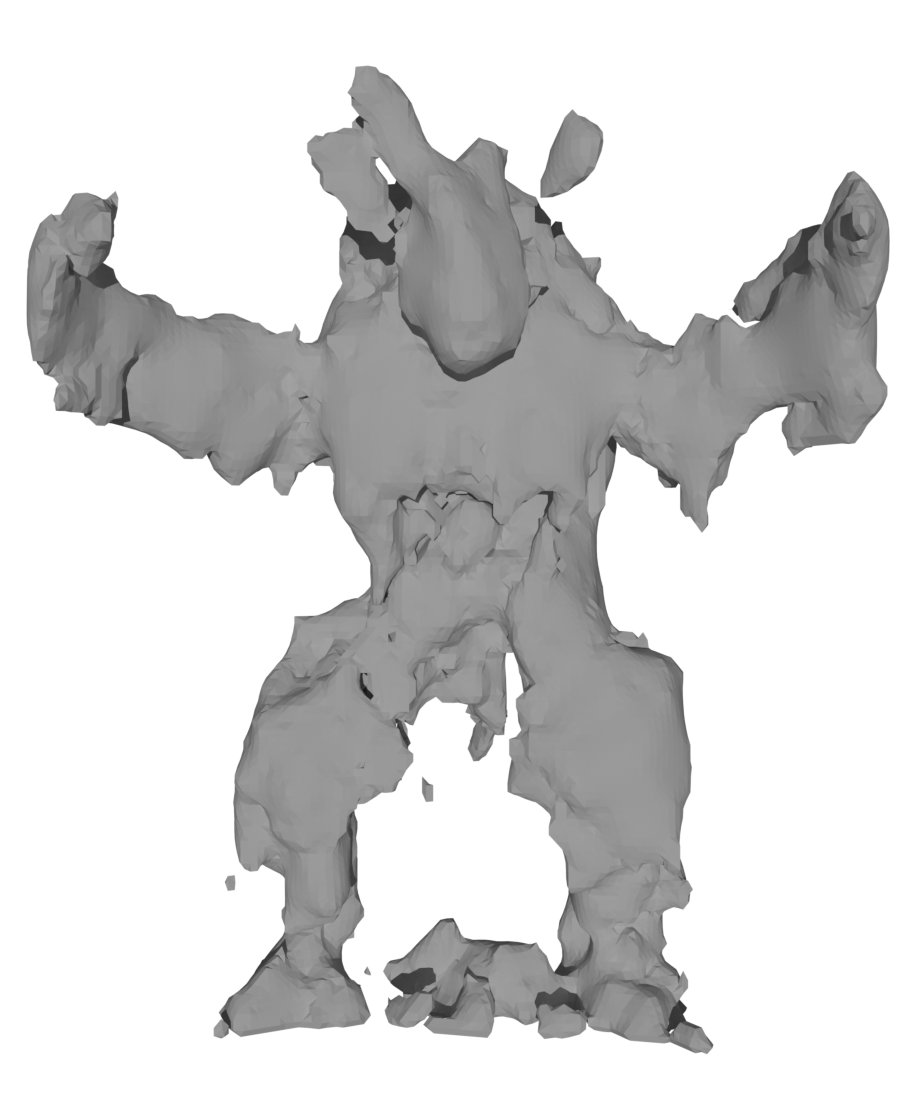}
    \end{subfigure}
    \hspace{10pt}
    \begin{subfigure}[t]{0.22\textwidth}
        \centering
        \includegraphics[width=\linewidth]{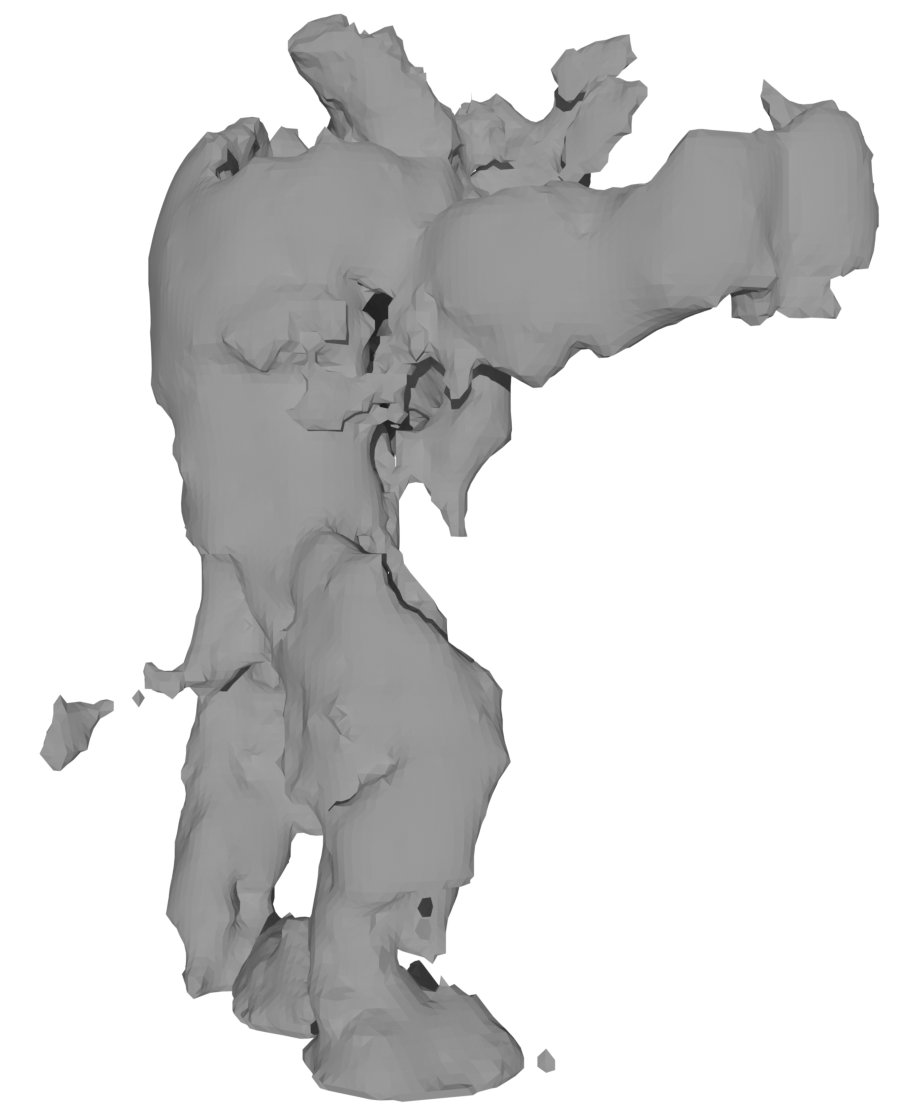}
    \end{subfigure}

    \vspace{2pt}

    {\footnotesize Level 2}
    \vspace{10pt}

    % Level 3
    \begin{subfigure}[t]{0.22\textwidth}
        \centering
        \includegraphics[width=\linewidth]{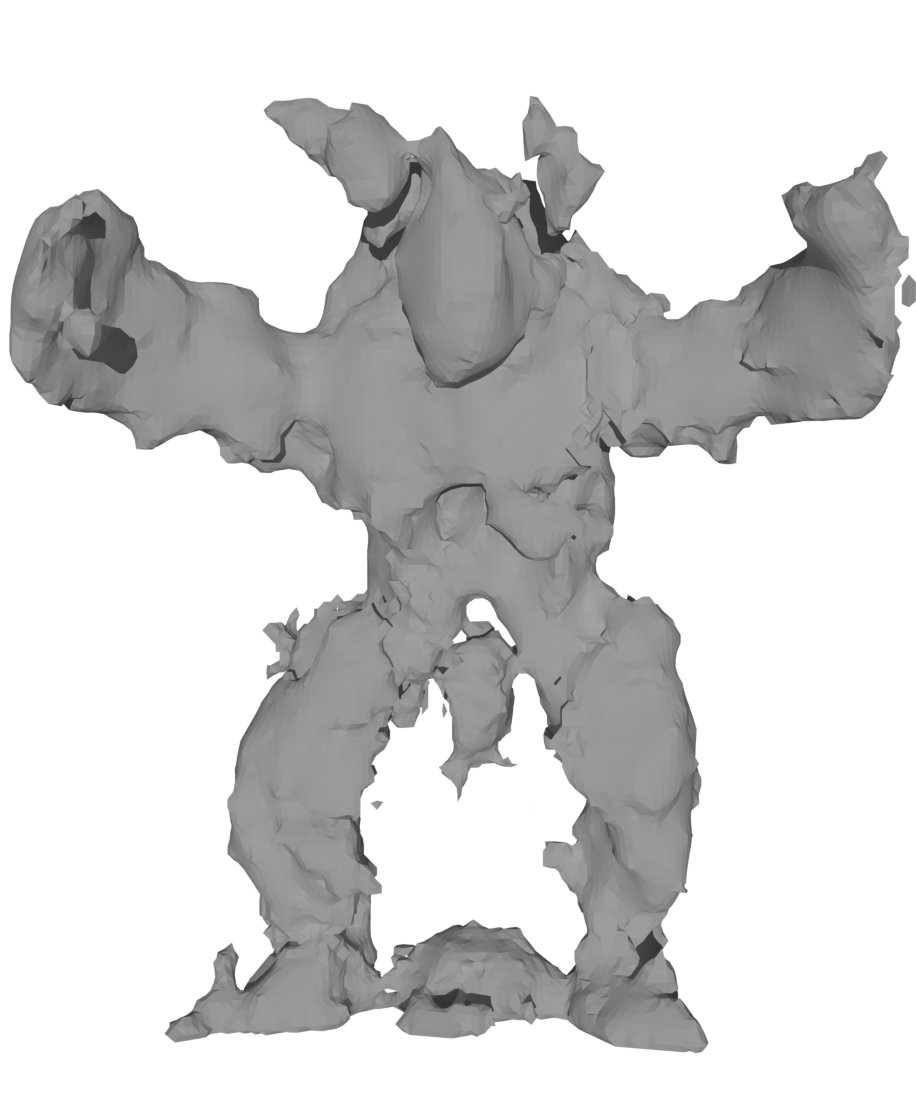}
    \end{subfigure}
    \hspace{10pt}
    \begin{subfigure}[t]{0.22\textwidth}
        \centering
        \includegraphics[width=\linewidth]{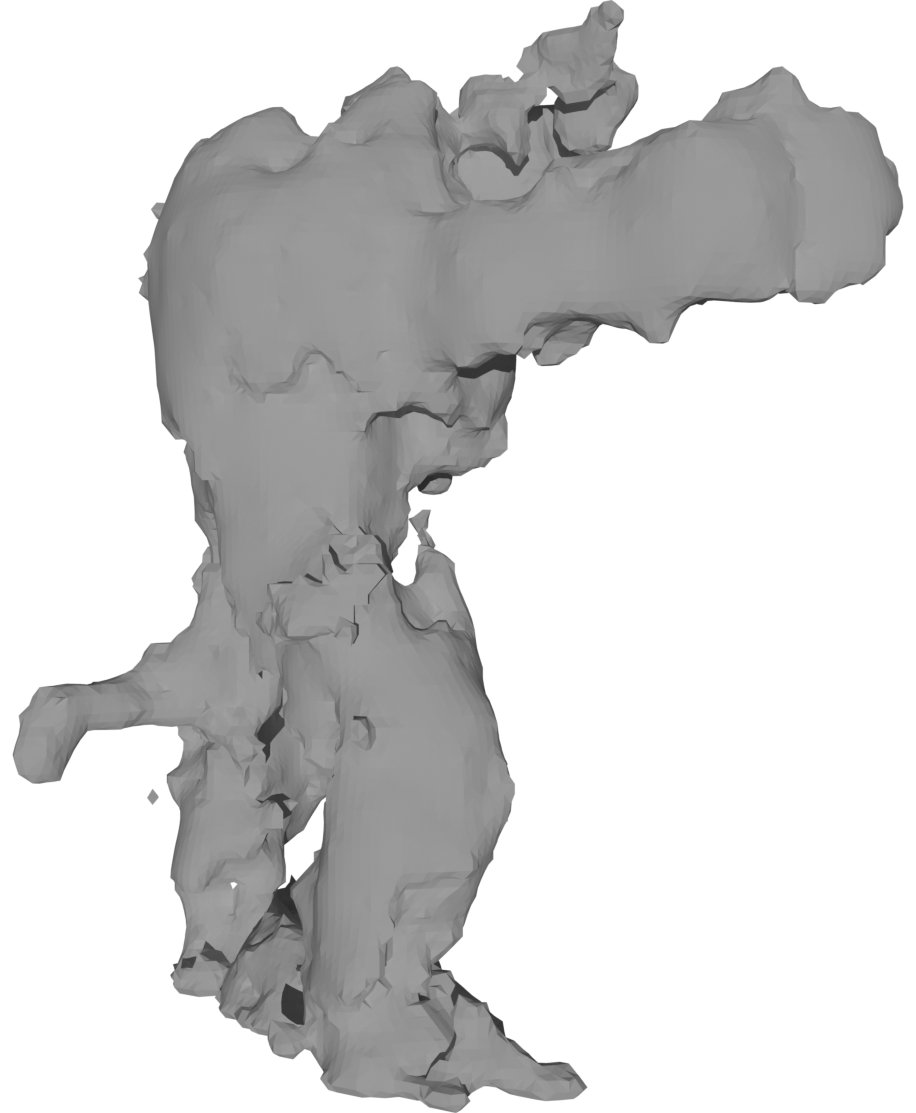}
    \end{subfigure}

    \vspace{2pt}
    {\footnotesize Level 3}

    \caption{Increase in Spherical Harmonic Levels. As the Spherical Harmonic (SH) level increases from Level 1 to Level 3, the quality of reconstruction improves significantly. Fine details, such as the legs and tail of the object, become clearer and more accurately represented at higher SH levels, demonstrating enhanced fidelity and geometric detail in the model.}
    \label{fig:sh_level_increase}
\end{figure*}

\paragraph{Ablation results.}
We ablate the SH order by increasing the level from 1 to 3 while keeping all other settings fixed. ~\cref{fig:sh_level_increase} show a clear trend: as SH level increases, reconstruction quality improves, with level 3 performing best in our setup.

% \begin{figure}[H]
%     \centering
%     \begin{subfigure}[t]{0.30\columnwidth}
%         \centering
%         \includegraphics[width=\linewidth]{images/real_arma_sh_1_vert.jpg}
%     \end{subfigure}
%     \hfill
%     \begin{subfigure}[t]{0.30\columnwidth}
%         \centering
%         \includegraphics[width=\linewidth]{images/real_arma_sh_2_vert.jpg}
%     \end{subfigure}
%     \hfill
%     \begin{subfigure}[t]{0.30\columnwidth}
%         \centering
%         \includegraphics[width=\linewidth]{images/real_arma_sh_3_vert.jpg}
%     \end{subfigure}
    
%     % \vspace{6pt} % space between rows
    
%     \begin{subfigure}[t]{0.30\columnwidth}
%         \centering
%         \includegraphics[width=\linewidth]{images/real_arma_sh_1_side_vert.jpg}
%     \end{subfigure}
%     \hfill
%     \begin{subfigure}[t]{0.30\columnwidth}
%         \centering
%         \includegraphics[width=\linewidth]{images/real_arma_sh_2_side_vert.jpg}
%     \end{subfigure}
%     \hfill
%     \begin{subfigure}[t]{0.30\columnwidth}
%         \centering
%         \includegraphics[width=\linewidth]{images/real_arma_sh_3_side_vert.jpg}
%     \end{subfigure}

%     % \vspace{6pt} % space between rows
    
%     \makebox[0.30\columnwidth]{\footnotesize Level 1}
%     \makebox[0.30\columnwidth]{\footnotesize Level 2}
%     \makebox[0.30\columnwidth]{\footnotesize Level 3}
    
%     \caption{Increase in Spherical Harmonic Levels}
%     \label{fig:sh_level_increase}
% \end{figure}

\section{Effect of Threshold on Visualization}
\begin{figure*}[p]
    \centering
    \setlength{\tabcolsep}{2pt}
    \renewcommand{\arraystretch}{0.0}
    \begin{tabular}{c c c}
        \footnotesize Backprojection & \footnotesize Reed et al.~\cite{reed2023neural} & \footnotesize Ours \\
        \vspace{10pt}
        \includegraphics[width=0.33\textwidth]{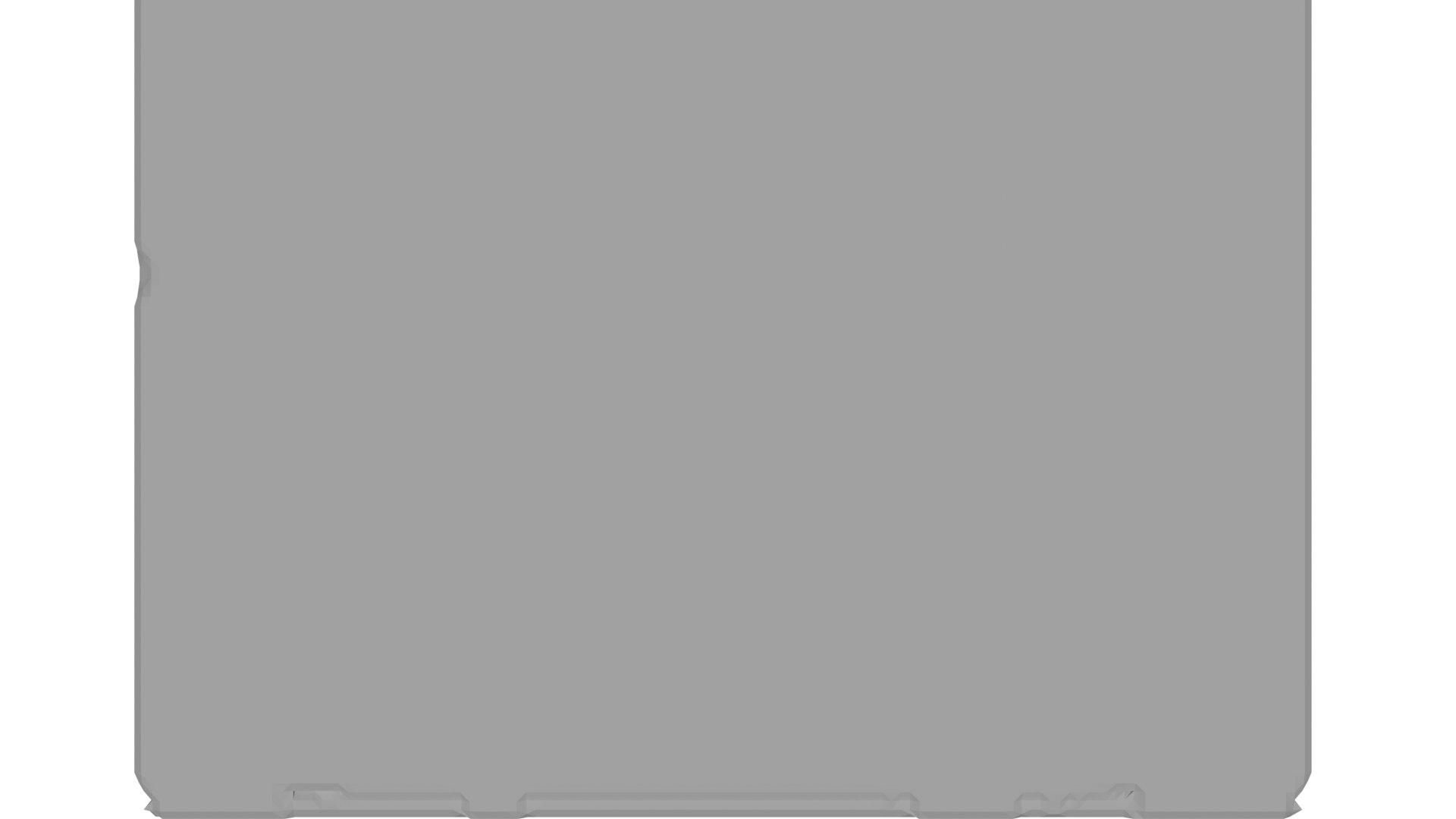} &
        \includegraphics[width=0.33\textwidth]{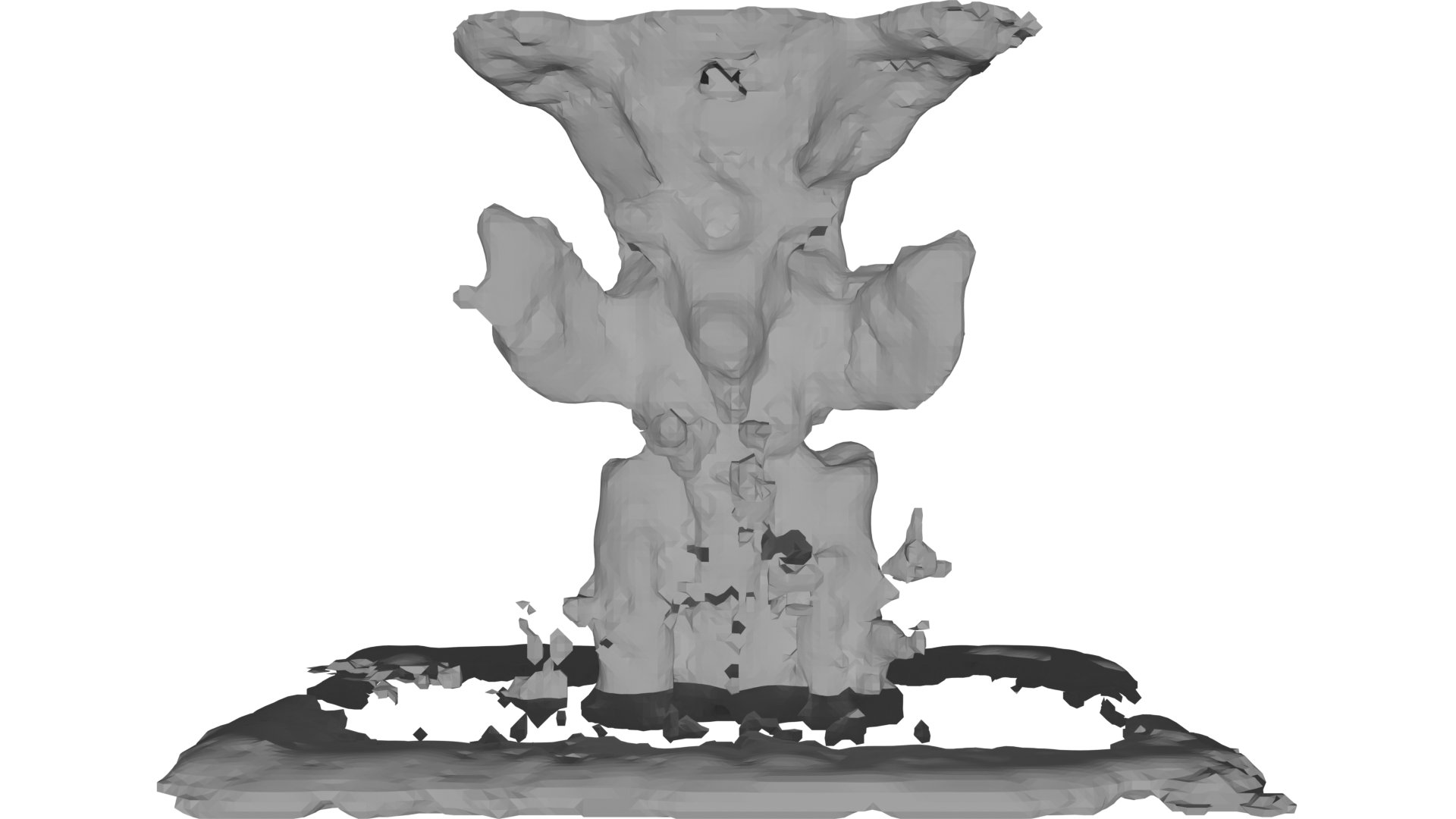} &
        \includegraphics[width=0.33\textwidth]{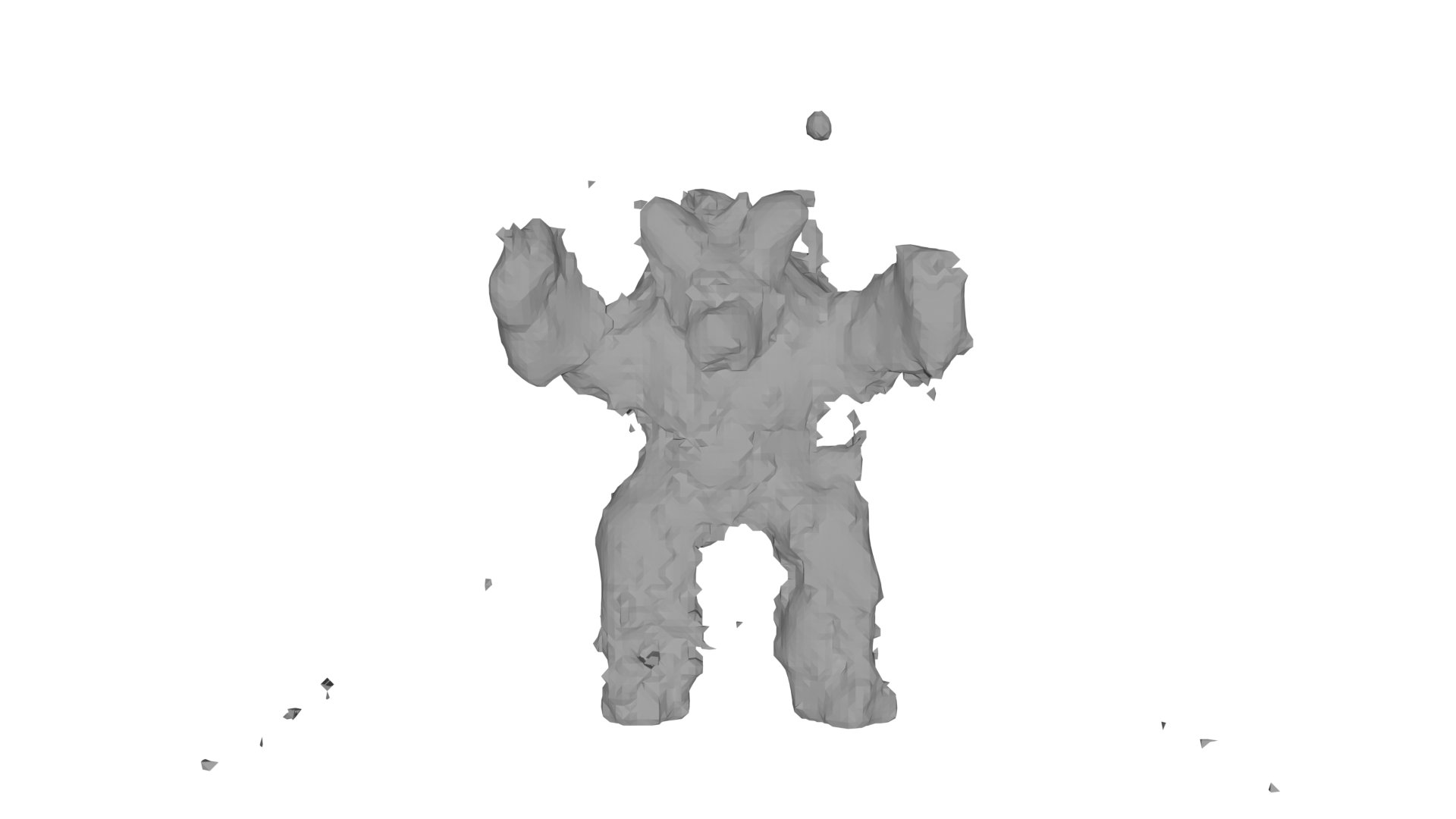} \\
        \vspace{10pt}
        \includegraphics[width=0.33\textwidth]{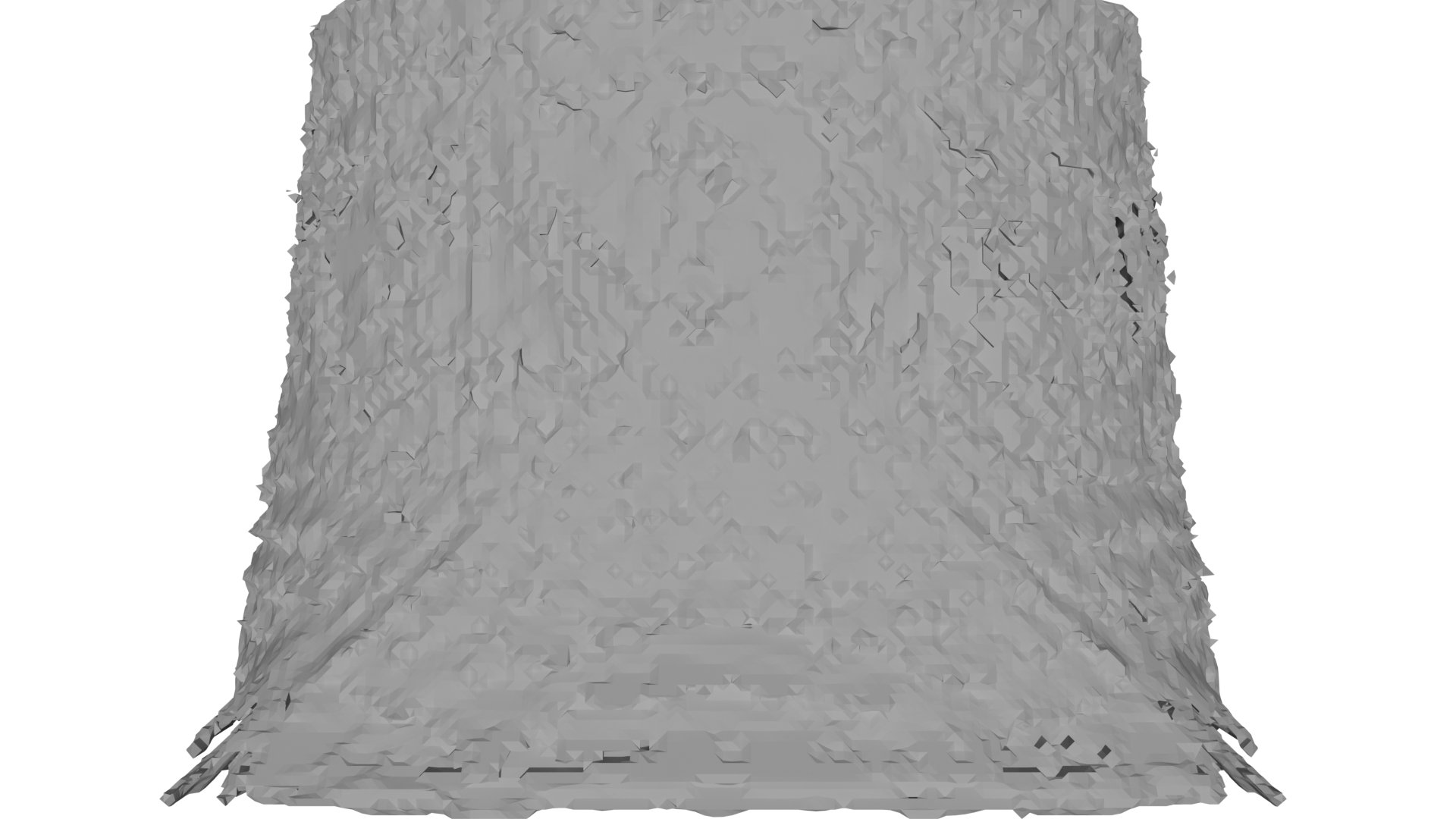} &
        \includegraphics[width=0.33\textwidth]{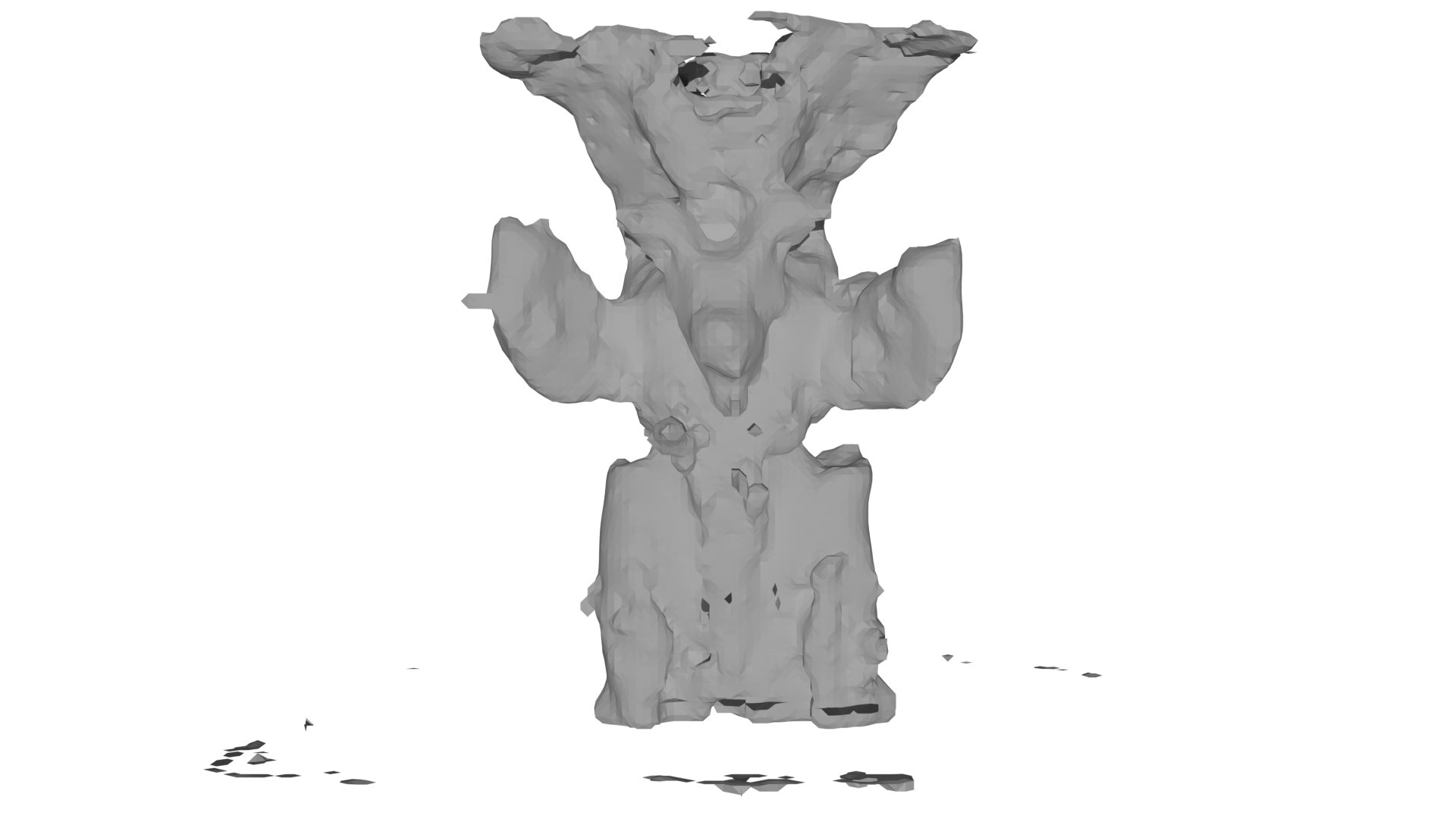} &
        \includegraphics[width=0.33\textwidth]{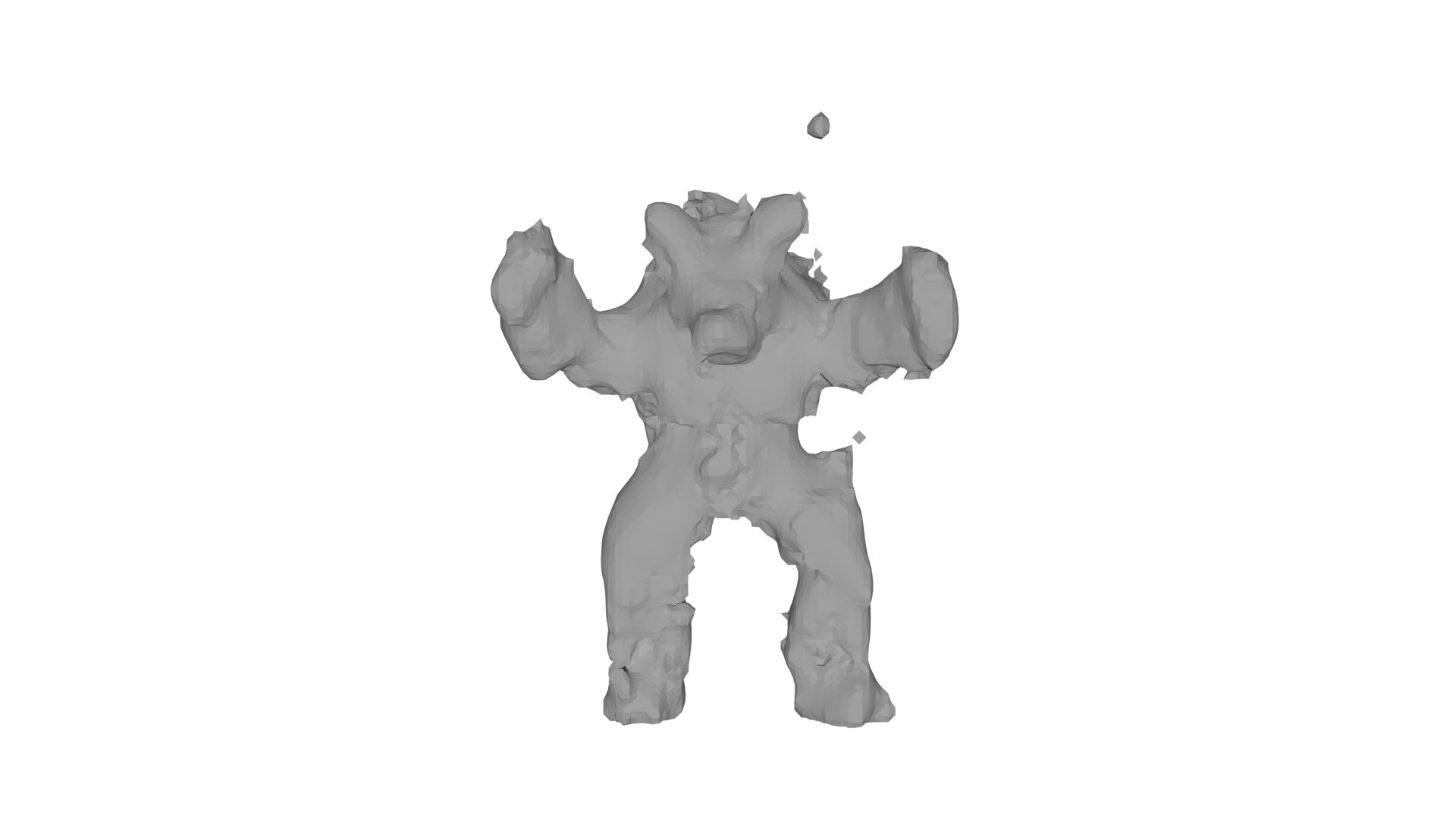} \\
        \vspace{10pt}
        \includegraphics[width=0.33\textwidth]{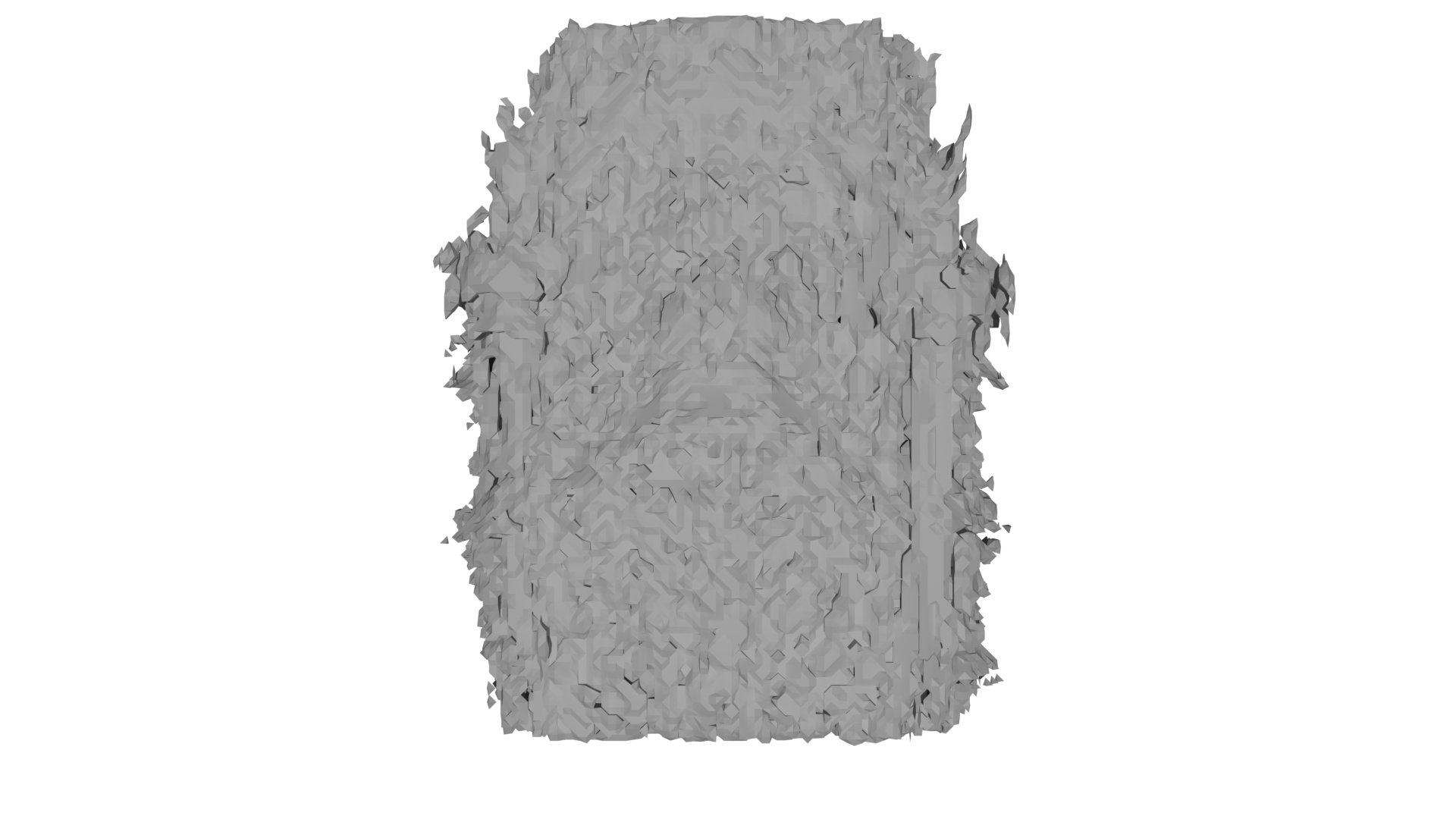} &
        \includegraphics[width=0.33\textwidth]{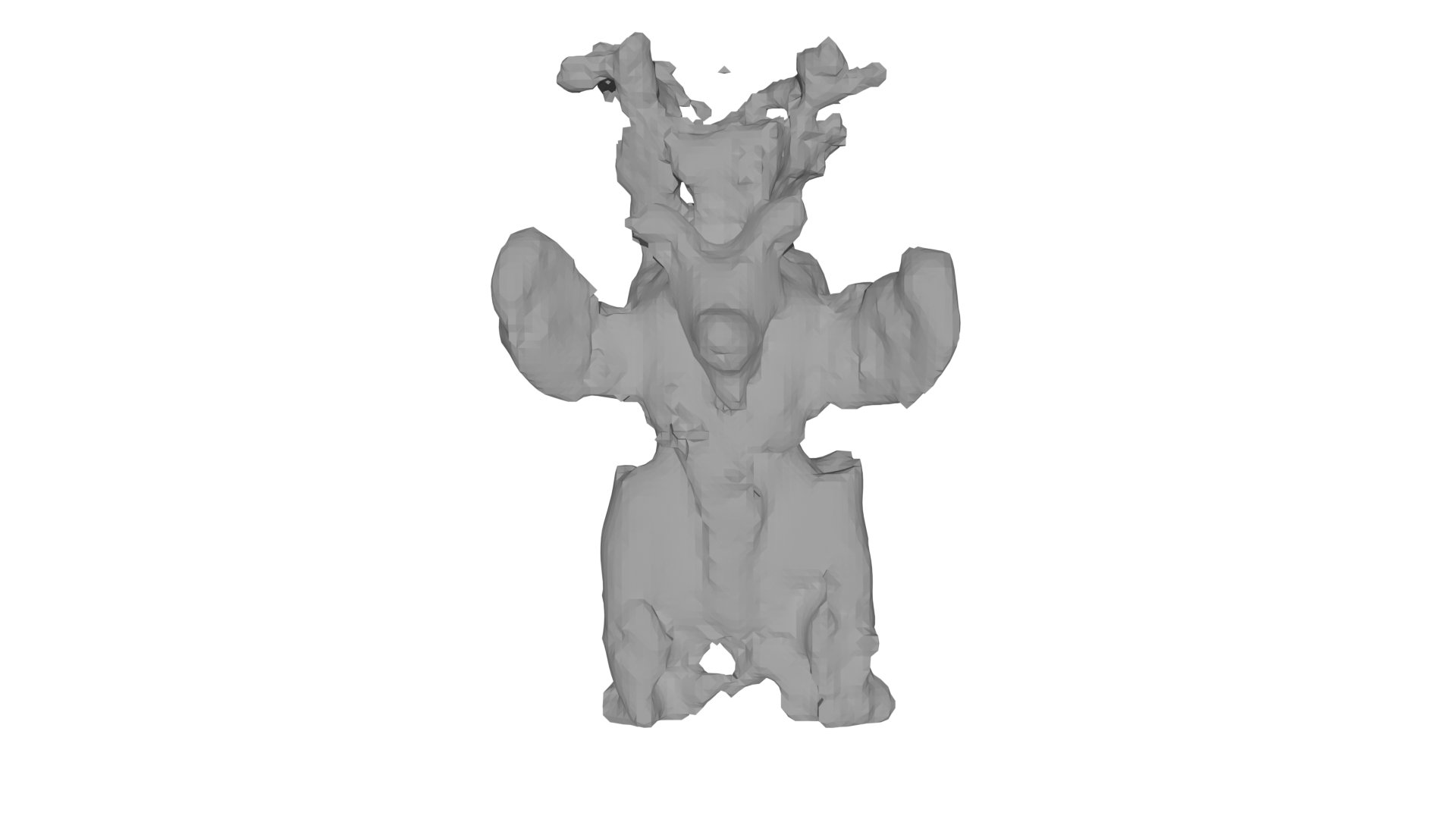} &
        \includegraphics[width=0.33\textwidth]{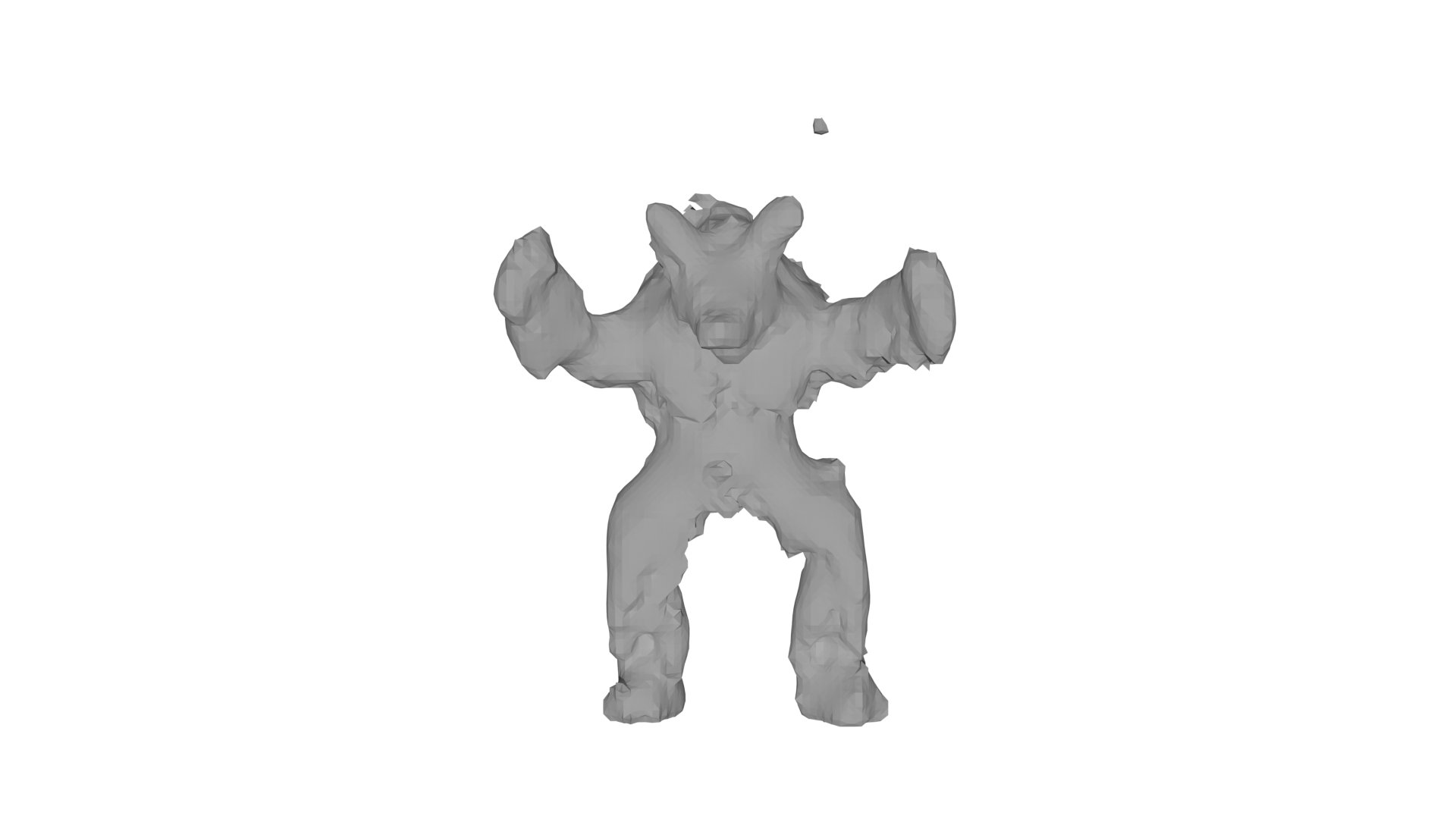} \\
        \vspace{10pt}
        \includegraphics[width=0.33\textwidth]{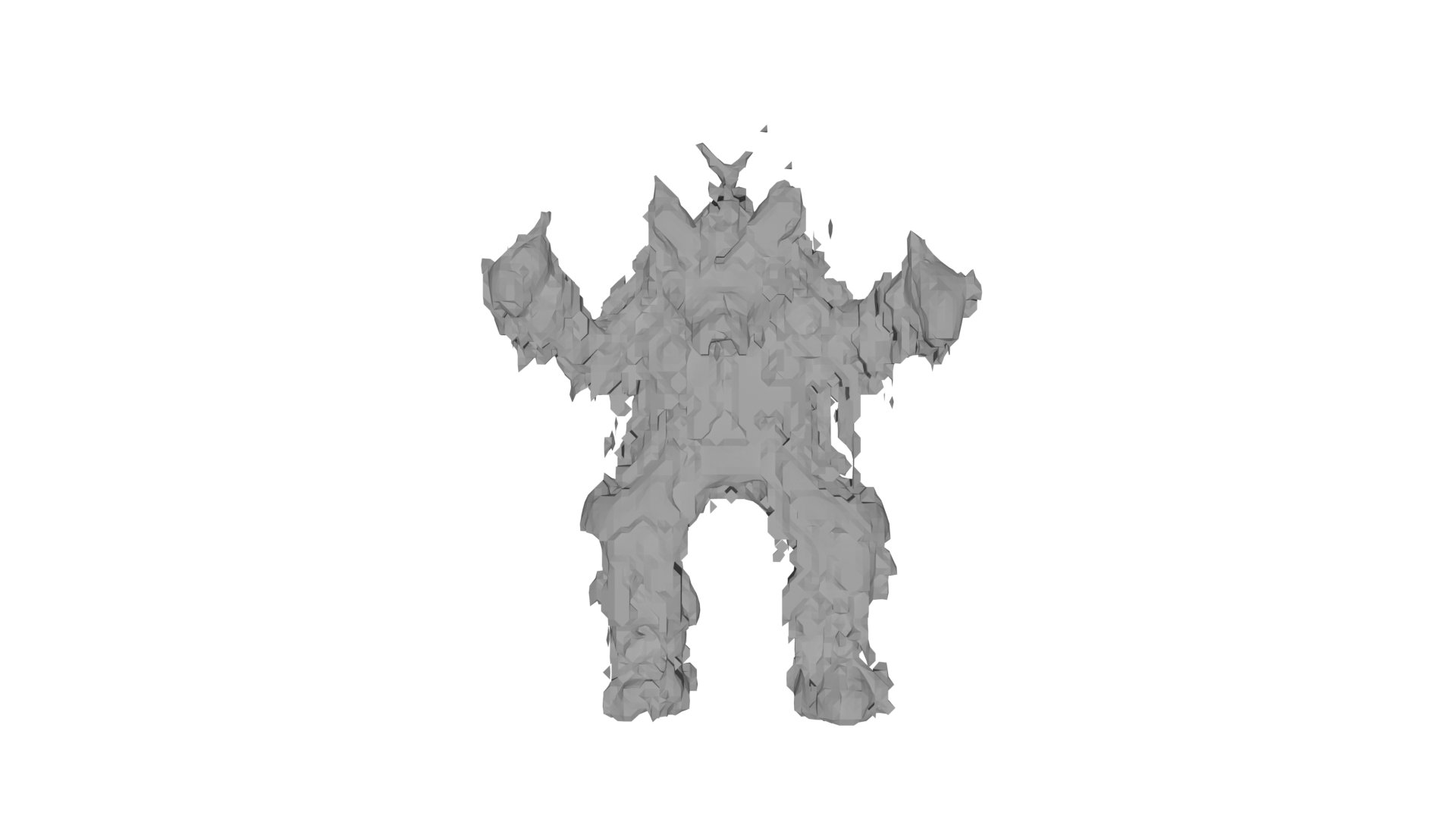} &
        \includegraphics[width=0.33\textwidth]{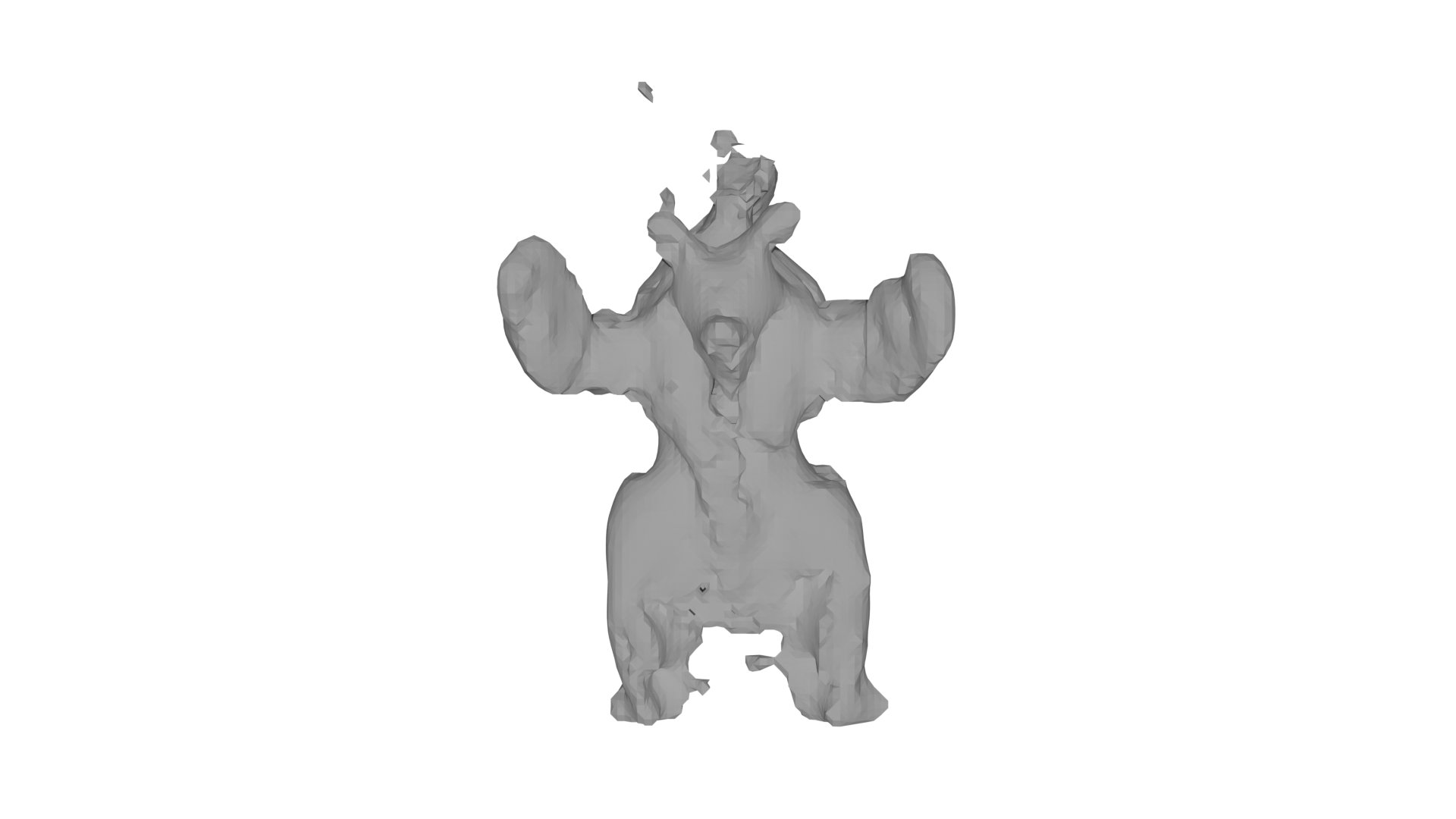} &
        \includegraphics[width=0.33\textwidth]{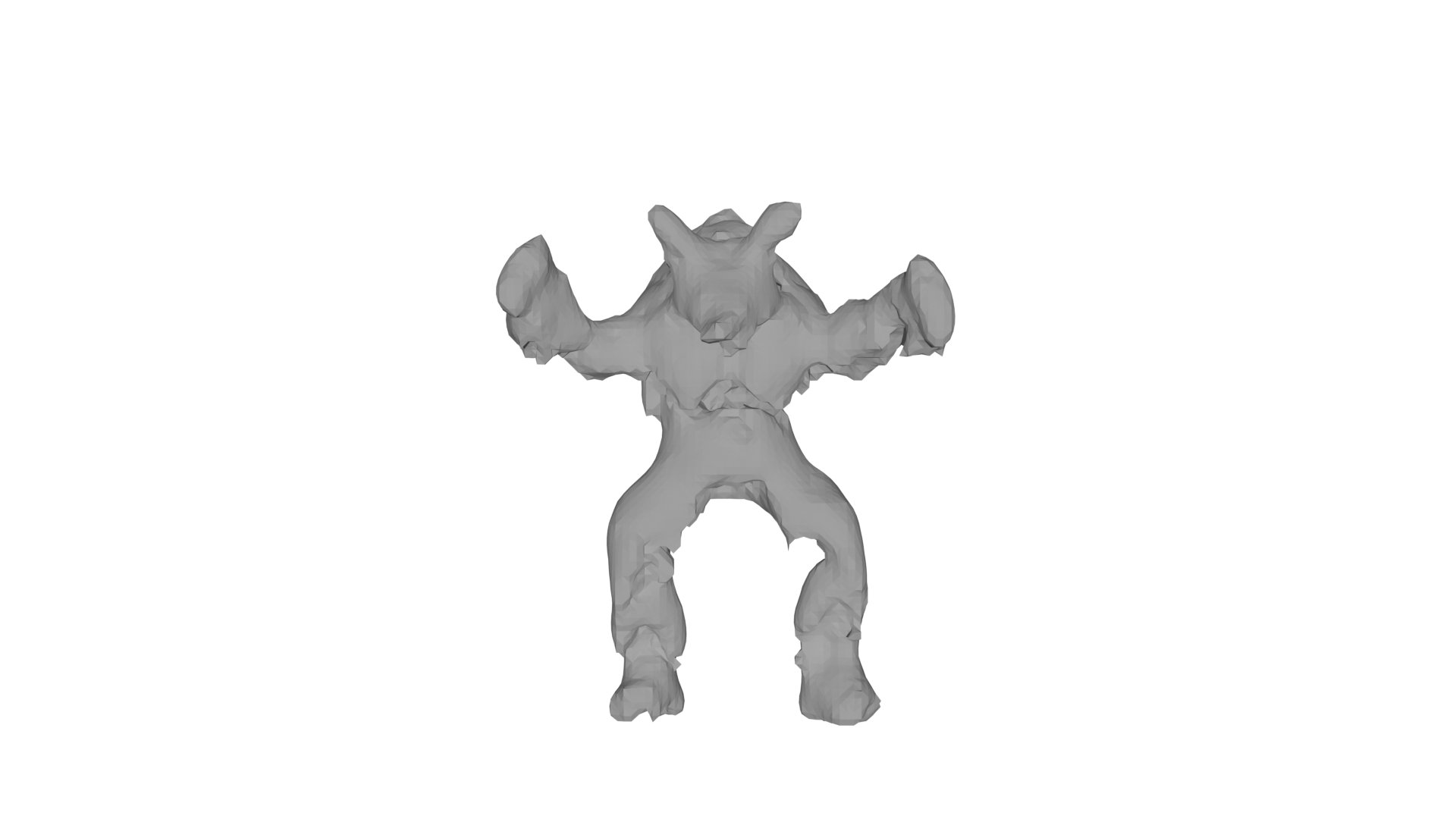} \\
    \end{tabular}
    \caption{Mesh consistency across threshold levels. Columns: methods. Rows: increasing threshold}
    \label{fig:threshold_comparison}
\end{figure*}

Figure~\ref{fig:threshold_comparison} examines how sensitive each method is to the marching-cubes threshold used to extract meshes from the learned field. As the threshold increases (top to bottom), Backprojection swings from an overfilled slab-like volume to severe erosion and fragmentation, indicating a poorly calibrated field whose level set is not tied to geometry. Reed et al.~\cite{reed2023neural} is more stable but still exhibits large changes: pedestal/floor artifacts at low thresholds. In contrast, Ours is markedly consistent across the changes in threshold and surfaces change mostly by a uniform thinning/thickening, with minimal new holes or floaters. This robustness suggests that our SH-based directional modeling learns a well-calibrated implicit field, making mesh visualization far less sensitive to the particular threshold choice.

\section{Signal fitting and novel-view transient synthesis}
\begin{figure*}[p]
  \centering
  % your existing 'novel view' content
  \setlength{\tabcolsep}{4pt}
  \begin{tabular}{@{}m{0.12\textwidth} m{0.28\textwidth} m{0.28\textwidth} m{0.28\textwidth}@{}}
    \centering\footnotesize Reed et al.~\cite{reed2023neural} &
    \includegraphics[width=\linewidth]{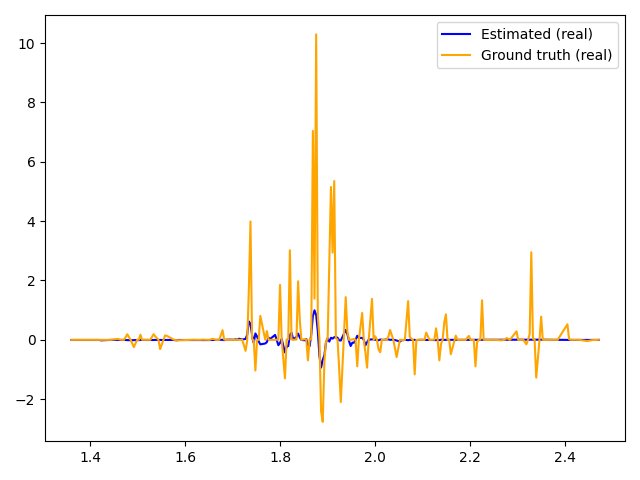} &
    \includegraphics[width=\linewidth]{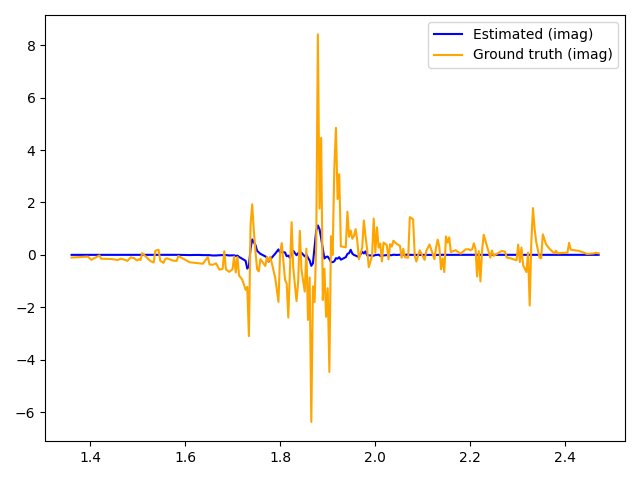} &
    \includegraphics[width=\linewidth]{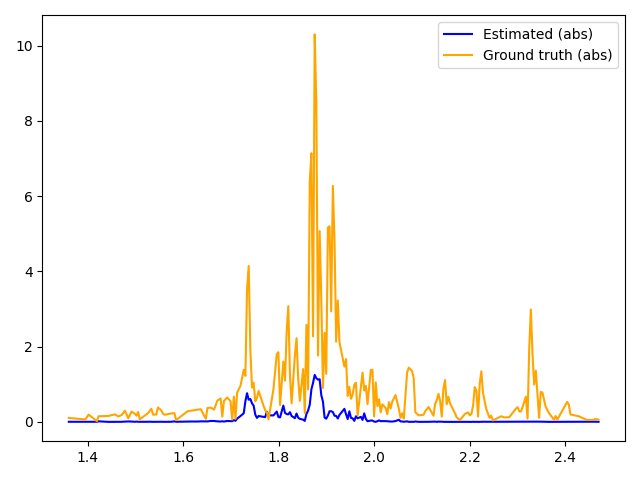} \\
    \centering\footnotesize Ours &
    \includegraphics[width=\linewidth]{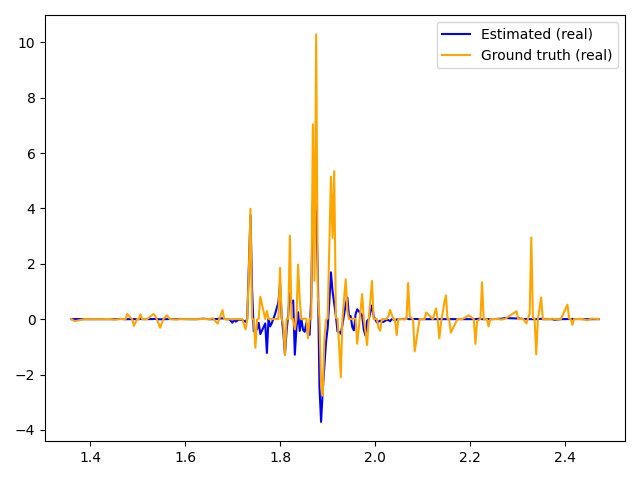} &
    \includegraphics[width=\linewidth]{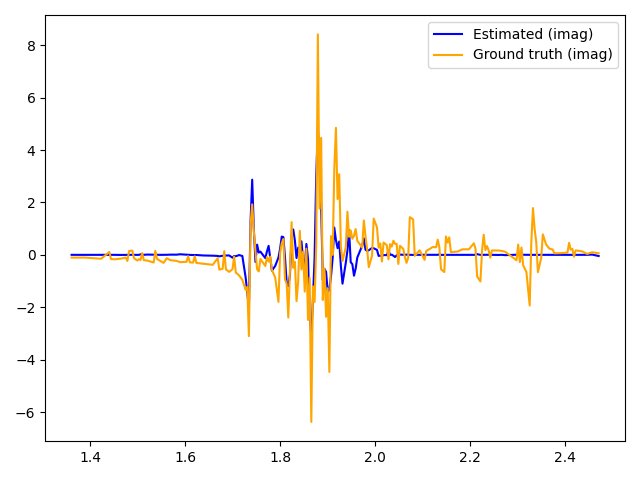} &
    \includegraphics[width=\linewidth]{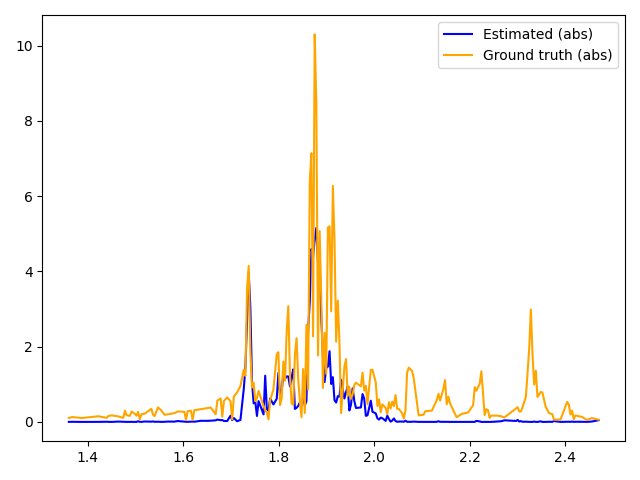}
  \end{tabular}
  \caption{Novel view signal synthesis results comparing Reed et al.~\cite{reed2023neural} (top row) and our method (bottom row). 
Each column shows a different component of the complex signal: real part (left), imaginary part (middle), and absolute magnitude (right). 
Within each plot, the estimated signal (blue) is compared against the ground-truth measurement (orange). 
Our method achieves closer alignment with ground-truth across all components, especially in regions with sharp variations.}

  \label{fig:novel_view}
\end{figure*}

We evaluate the network’s ability to fit measured signals and then use that fit to synthesize transients at unseen poses. We define \emph{novel-view transient synthesis} as generating the complex transient (analytic signal—real and imaginary parts) at a transmitter/receiver pose within the synthetic aperture where no physical measurement was taken; given a desired TX/RX location, our learned scene representation predicts the time-resolved response that would have been recorded at that pose. This capability enables simulation for sensor/trajectory design and target modeling, \emph{aperture densification} by interpolating missing pings to improve coherent processing, and data augmentation for downstream detection/classification. Related work in transient imaging motivates novel-view synthesis for virtual sensing and augmentation—TransientNeRF~\cite{malik2023transient} primarily operates in the LiDAR image domain, while TransientAngelo~\cite{Luo_2025_WACV} targets single-photon LiDAR and predicts normalized photon-count transients—but our focus is on \emph{analytic} (complex-valued) transients at arbitrary bistatic poses, consistent with the ellipsoidal ToF geometry. Operationally, for each target pose $(\mathbf{o}_T,\mathbf{o}_R)$ and orientation we form the constant-ToF ellipsoid per bin, compute ray–ellipsoid intersections to obtain geometry-aware samples, evaluate the learned SH-based directional scattering at those samples, and integrate through the forward model to yield the complex transient—identical to training-time analysis-by-synthesis, but evaluated at new poses. On the \emph{Armadillo} dataset, Fig.~\ref{fig:novel_view} shows that our predictions better preserve delay structure and phase, relative to~\cite{reed2023neural}; quantitatively, Table~4 reports consistent improvements across apertures for both the real and imaginary components—lower $\ell_1$ error and lower mean squared error (MSE).

\FloatBarrier
\clearpage

\begin{table*}[p]
\centering
\small
\begin{tabular}{llrrrr|rrrr}
\toprule
\multirow{2}{*}{Object} & \multirow{2}{*}{Method} &
\multicolumn{4}{c|}{Point cloud} &
\multicolumn{4}{c}{Mesh} \\
\cmidrule(lr){3-6}\cmidrule(lr){7-10}
& & Chamfer & IoU & Precision & F1 & Chamfer & IoU & Precision & F1 \\
\midrule
\multirow{3}{*}{Armadillo}
& Backprojection & \worst{8.204e-5} & \worst{0.243} & \worst{0.280} & \worst{0.391} & \second{1.081e-4} & \second{0.223} & \second{0.284} & \second{0.364} \\
& Reed et al.    & \second{7.871e-5} & \second{0.418} & \second{0.402} & \second{0.544} & \worst{1.209e-4} & \worst{0.154} & \worst{0.237} & \worst{0.266} \\
& Ours           & \best{7.009e-5} & \best{0.444} & \best{0.525} & \best{0.614} & \best{7.688e-5} & \best{0.249} & \best{0.393} & \best{0.398} \\
\midrule
\multirow{3}{*}{Buddha}
& Backprojection & \worst{9.546e-4} & \worst{0.131} & \worst{0.132} & \worst{0.232} & \worst{1.191e-3} & \worst{0.056} & \worst{0.074} & \second{0.107} \\
& Reed et al.    & \second{4.832e-4} & \second{0.601} & \second{0.439} & \second{0.608} & \second{1.184e-3} & \worst{0.056} & \second{0.101} & \worst{0.106} \\
& Ours           & \best{9.238e-5} & \best{0.611} & \best{0.736} & \best{0.756} & \best{7.211e-5} & \best{0.324} & \best{0.550} & \best{0.490} \\
\midrule
\multirow{3}{*}{Bunny}
& Backprojection & \best{1.233e-4} & \worst{0.227} & \worst{0.325} & \worst{0.371} & \best{1.551e-4} & \best{0.197} & \second{0.329} & \best{0.330} \\
& Reed et al.    & \worst{1.524e-4} & \best{0.412} & \second{0.480} & \best{0.584} & \worst{2.190e-4} & \worst{0.149} & \worst{0.101} & \worst{0.260} \\
& Ours           & \second{1.255e-4} & \second{0.398} & \best{0.518} & \second{0.569} & \second{1.798e-4} & \second{0.191} & \best{0.368} & \second{0.321} \\
\midrule
\multirow{3}{*}{Dragon}
& Backprojection & \best{5.679e-5} & \worst{0.202} & \worst{0.238} & \worst{0.336} & \worst{8.586e-5} & \worst{0.167} & \worst{0.212} & \worst{0.286} \\
& Reed et al.    & \worst{6.834e-5} & \best{0.360} & \second{0.390} & \best{0.529} & \second{6.306e-5} & \second{0.174} & \second{0.246} & \second{0.296} \\
& Ours           & \second{6.752e-5} & \second{0.357} & \best{0.402} & \second{0.526} & \best{5.925e-5} & \best{0.194} & \best{0.270} & \best{0.325} \\
\bottomrule
\end{tabular}
\caption{Core 3D metrics (point cloud vs. mesh) across objects. Chamfer uses scientific notation (3-dec mantissa); others rounded to 3 decimals. Colors indicate rank within each object/column (best=green, second=yellow, worst=red). Ties share the same color.}
\label{tab:objects-pc-vs-mesh-colored}
\end{table*}
\FloatBarrier
\clearpage

\FloatBarrier
\clearpage

\begin{figure*}[p]
    \centering

    \includegraphics[width=0.23\textwidth]{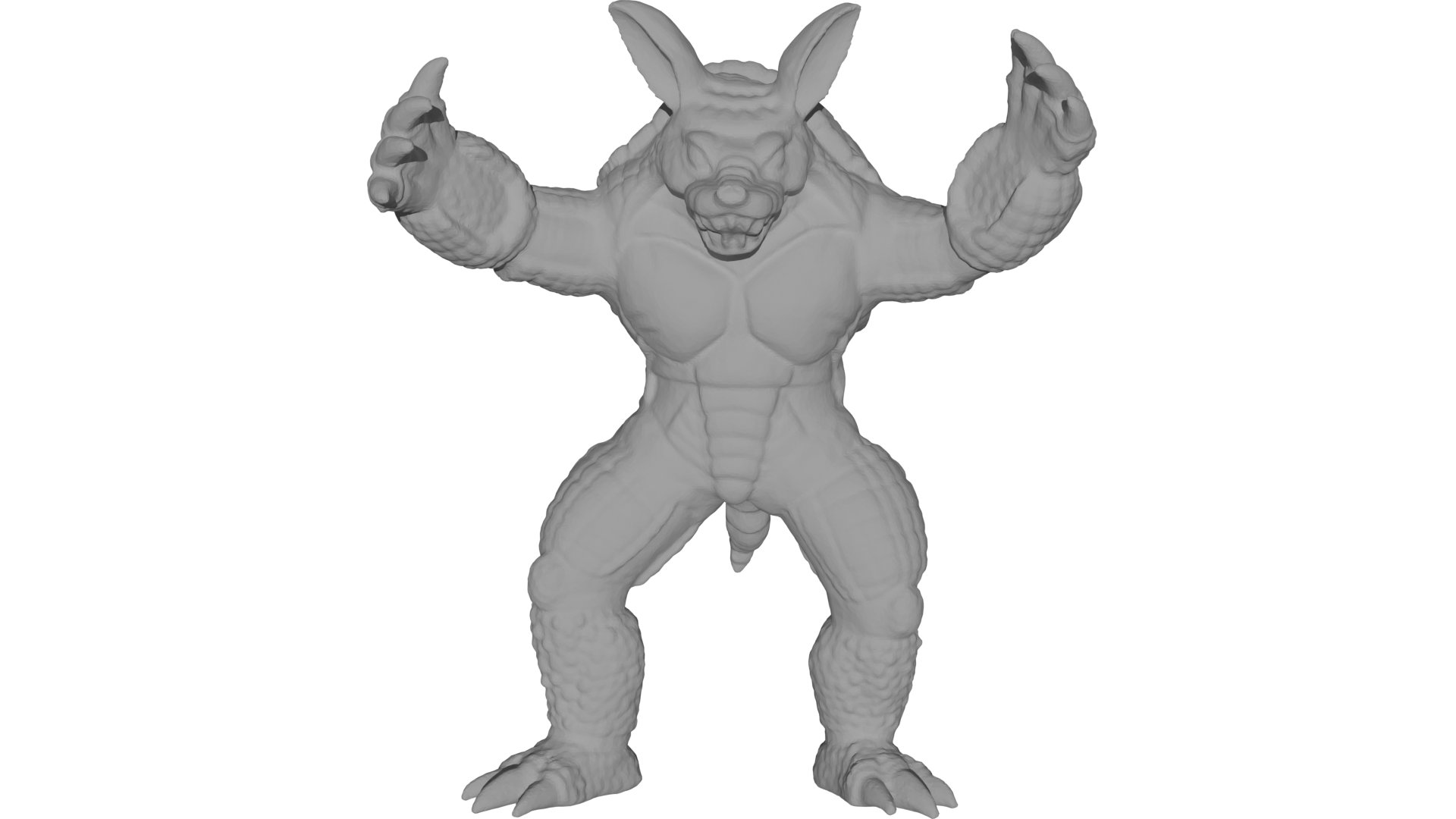}
    \includegraphics[width=0.23\textwidth]{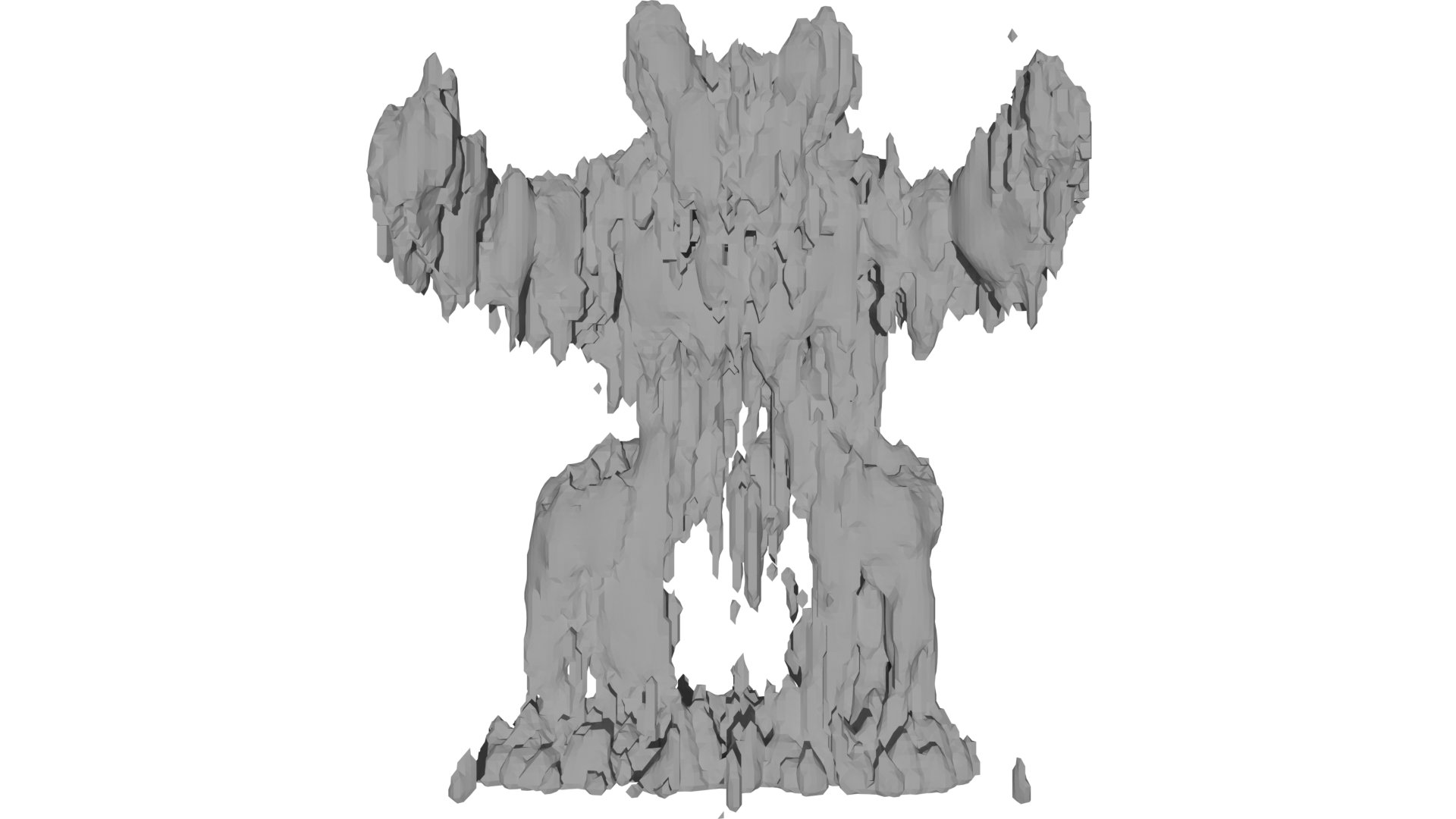}
    \includegraphics[width=0.23\textwidth]{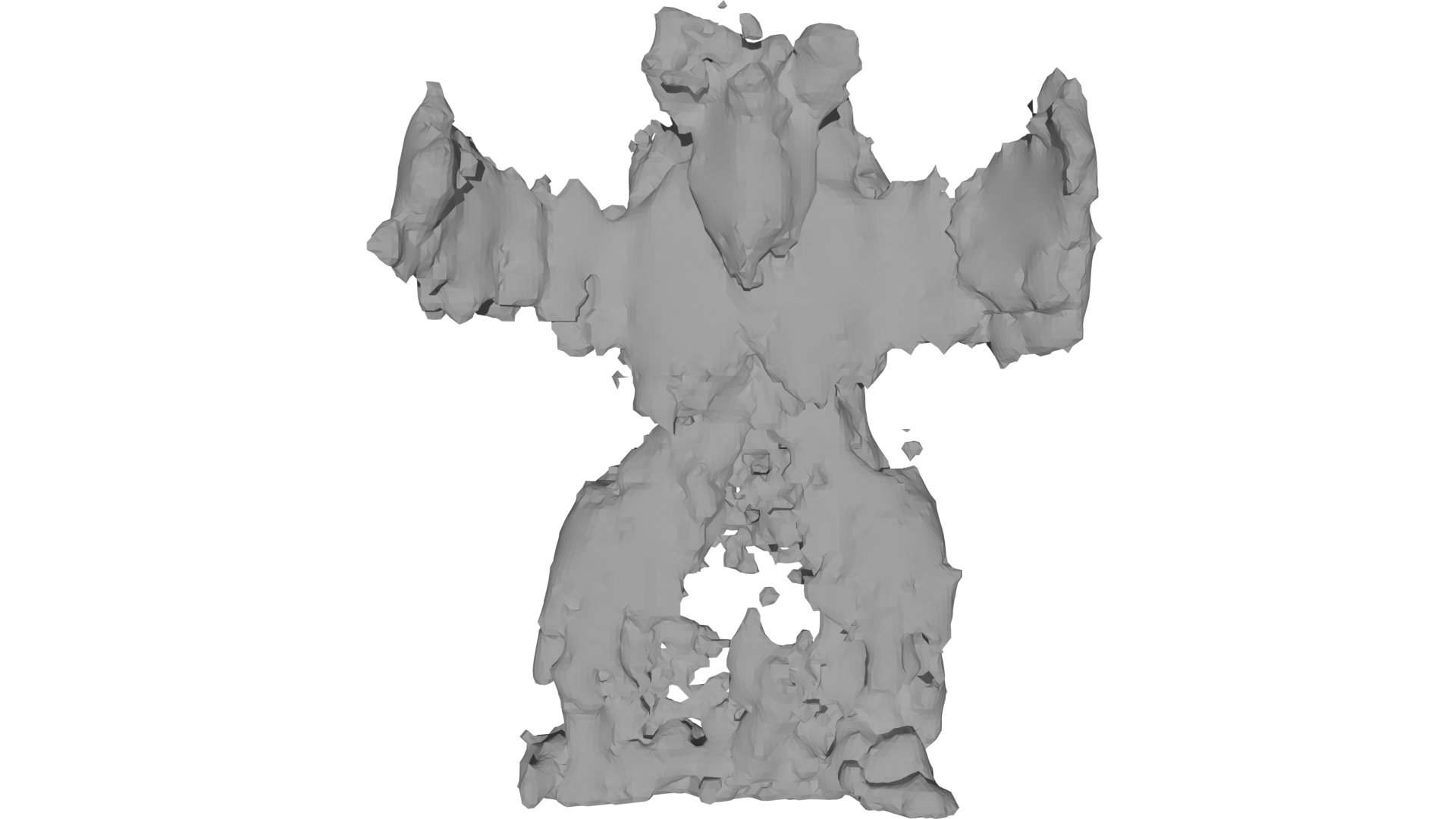}
    \includegraphics[width=0.23\textwidth]{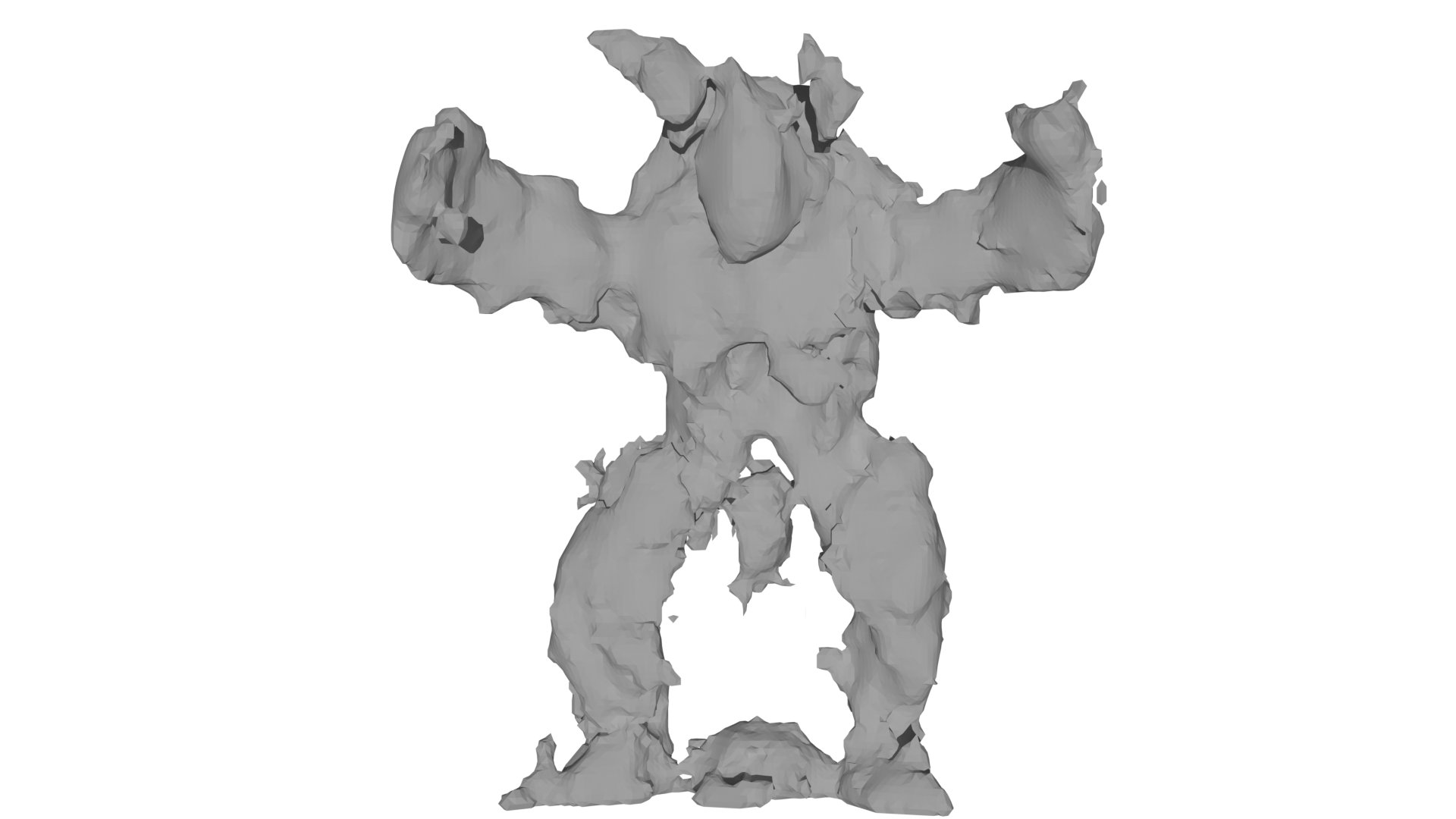}

    \vspace{5mm}
    
    \includegraphics[width=0.23\textwidth]{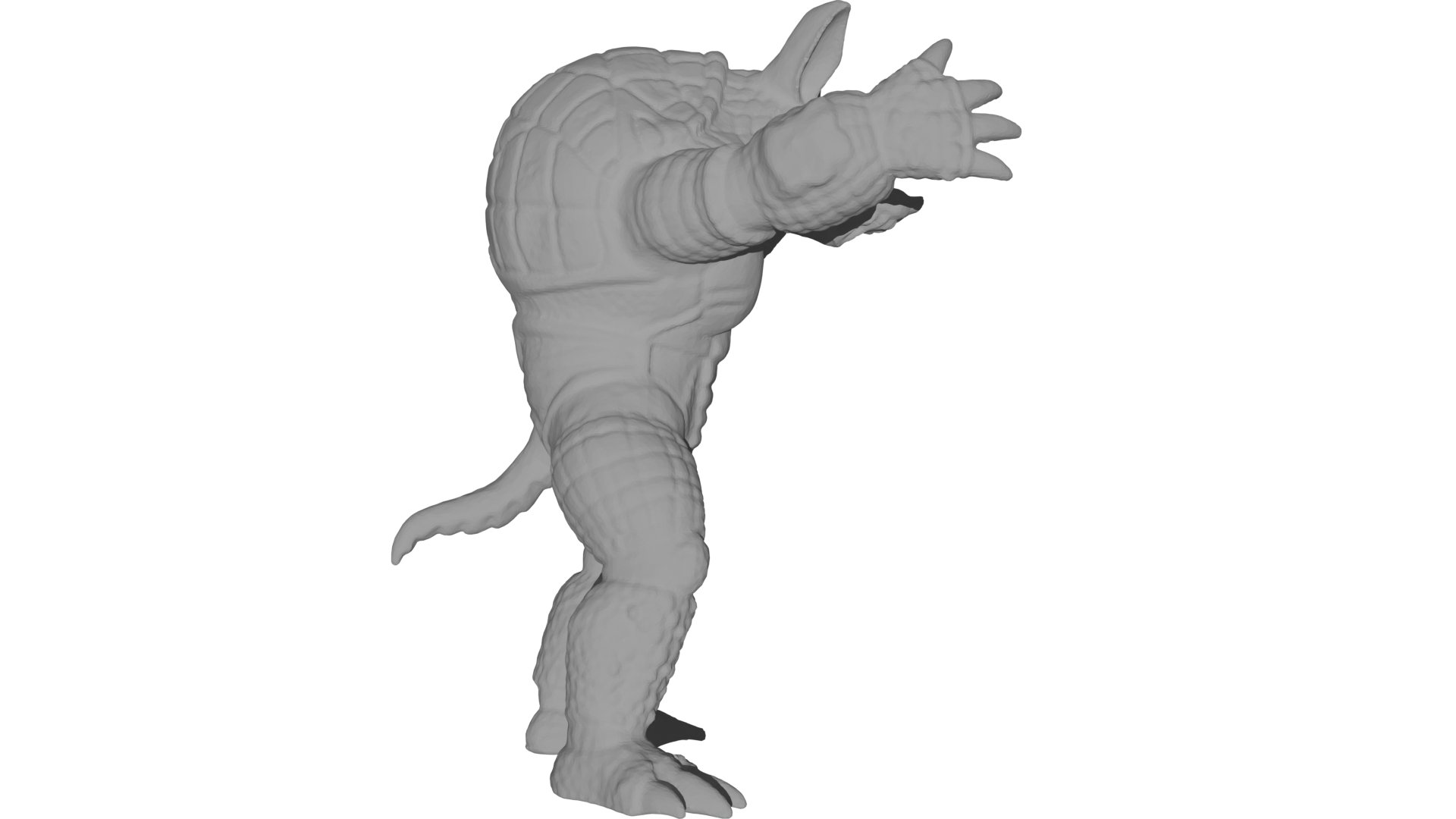}
    \includegraphics[width=0.23\textwidth]{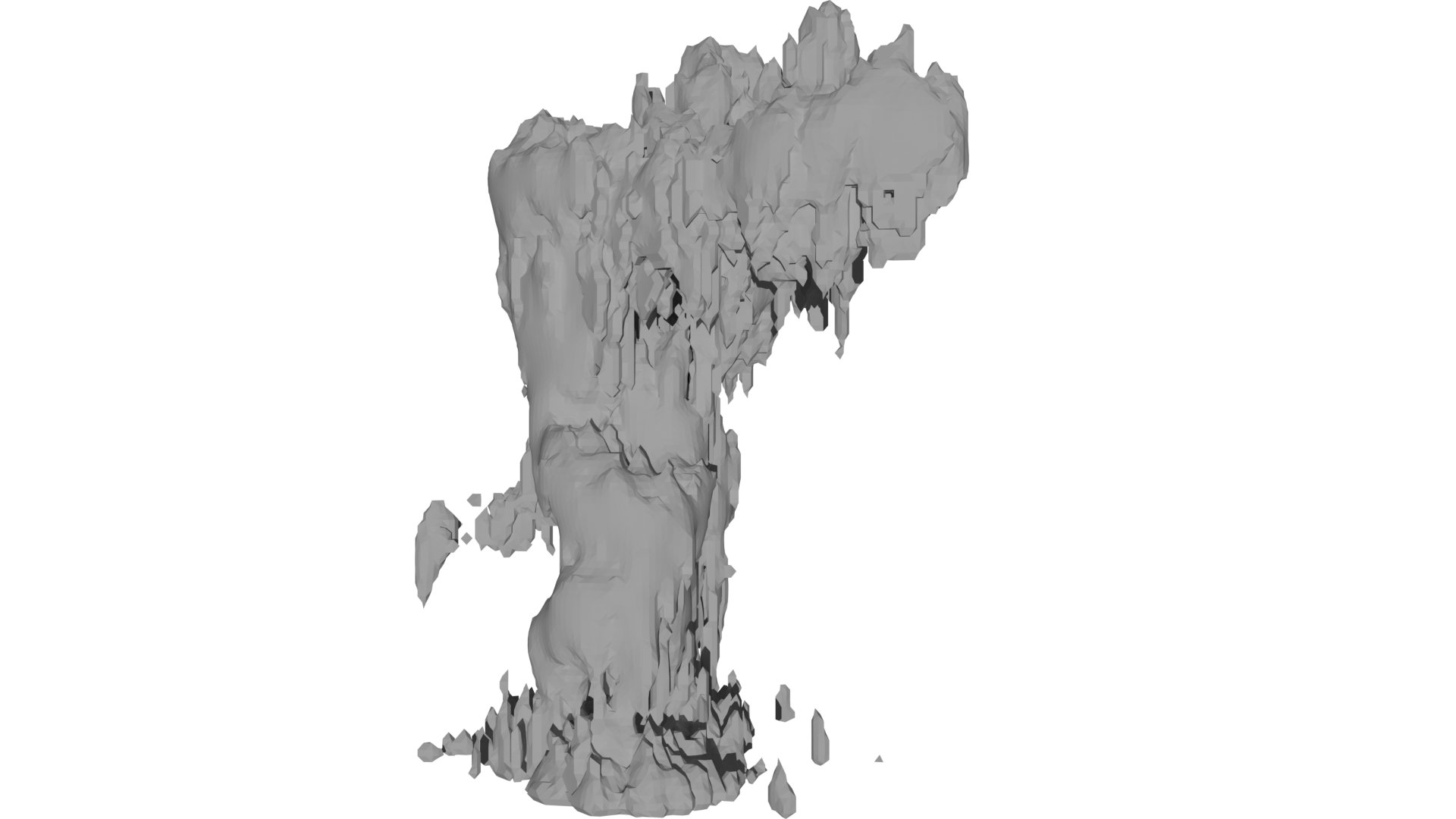}
    \includegraphics[width=0.23\textwidth]{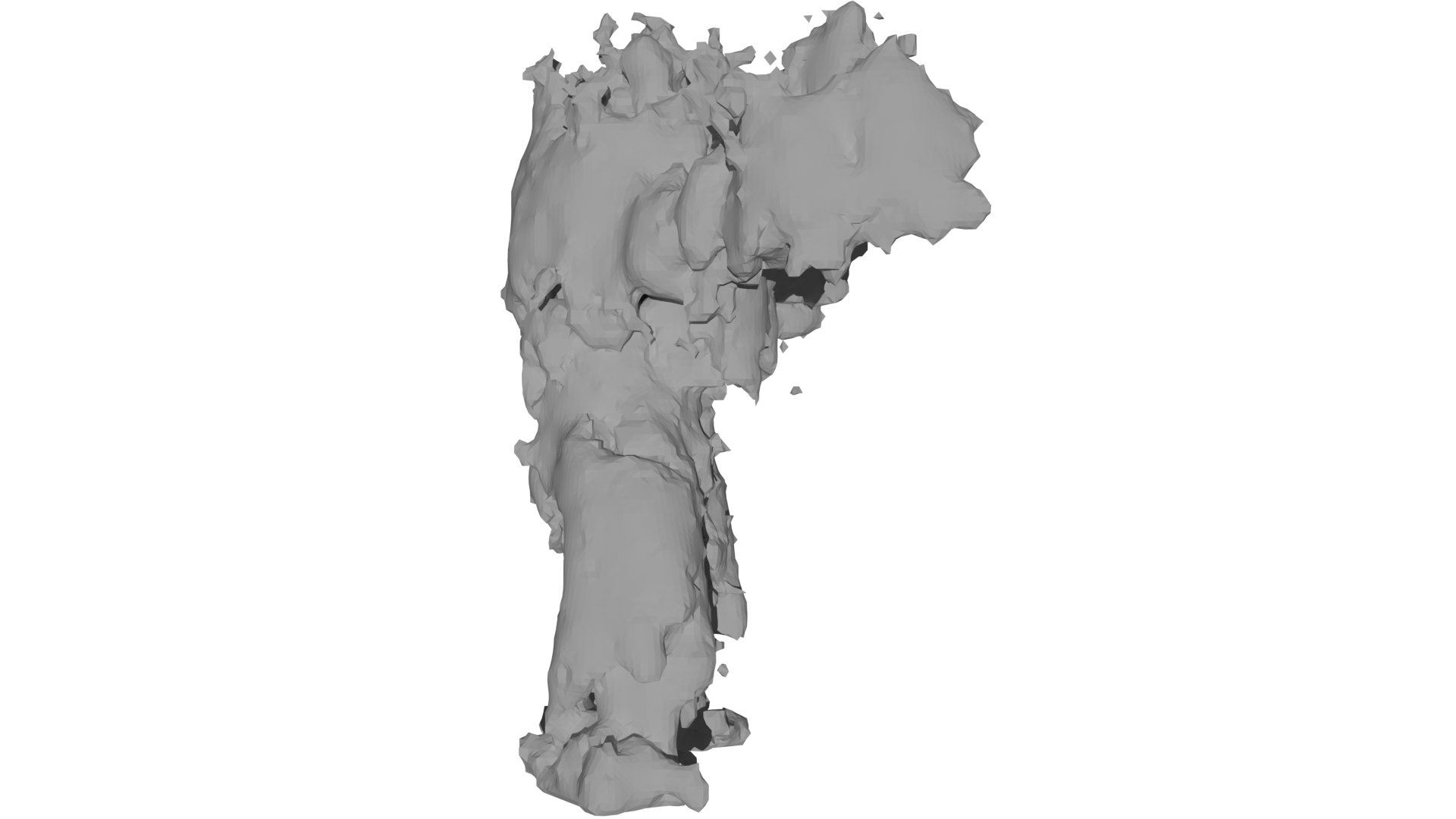}
    \includegraphics[width=0.23\textwidth]{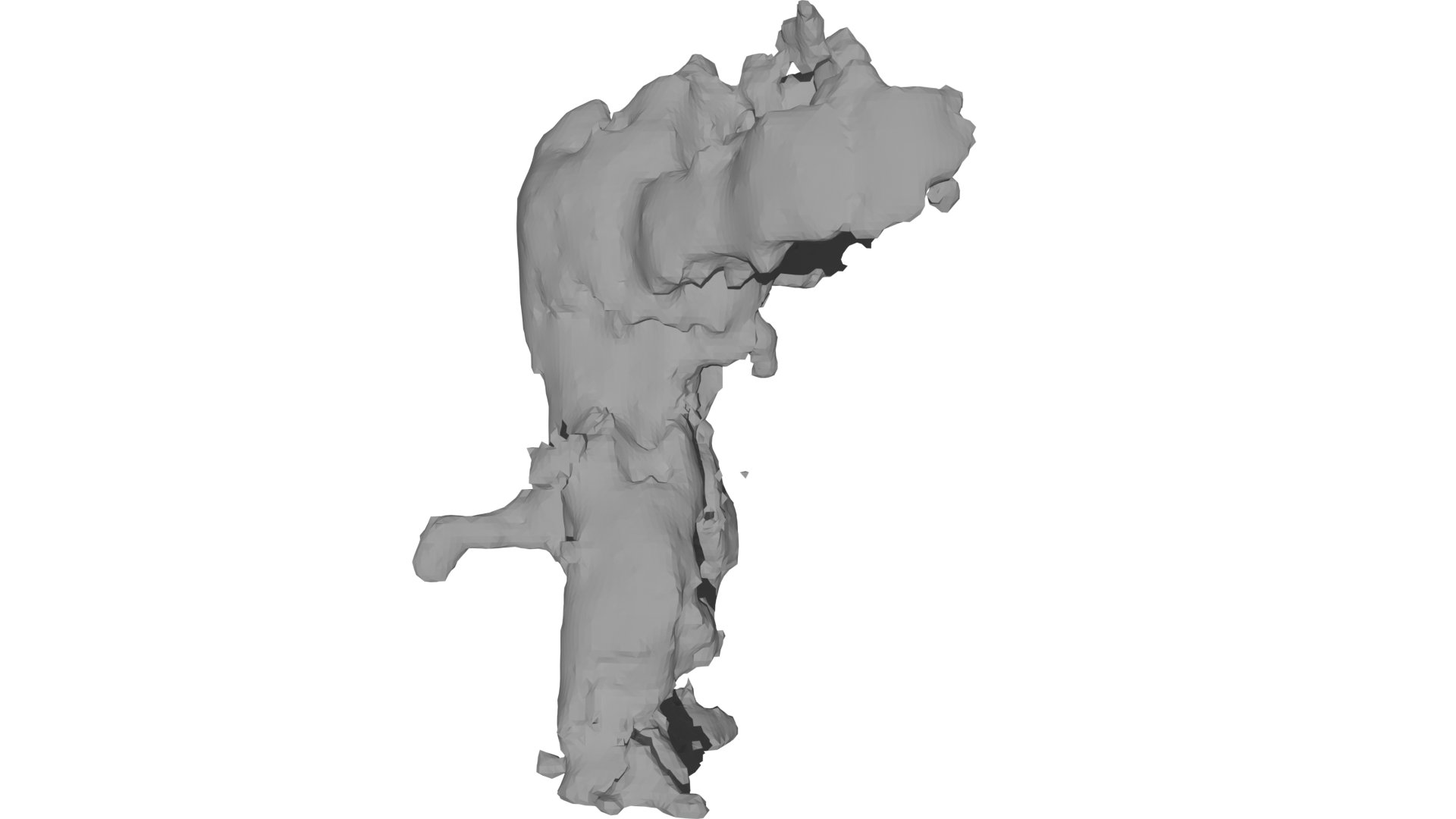}

    \makebox[0.23\textwidth]{\footnotesize Ground Truth}
    \makebox[0.23\textwidth]{\footnotesize Backprojection}
    \makebox[0.23\textwidth]{\footnotesize Reed et al.~\cite{reed2023neural}}
    \makebox[0.23\textwidth]{\footnotesize Ours}

    \caption{Real data: Armadillo. Top: frontal view. Bottom: side view.}
    \label{fig:real_arma_20k}
\end{figure*}

\begin{figure*}[p]
    \centering

    % \vspace{5mm}

    \includegraphics[width=0.23\textwidth]{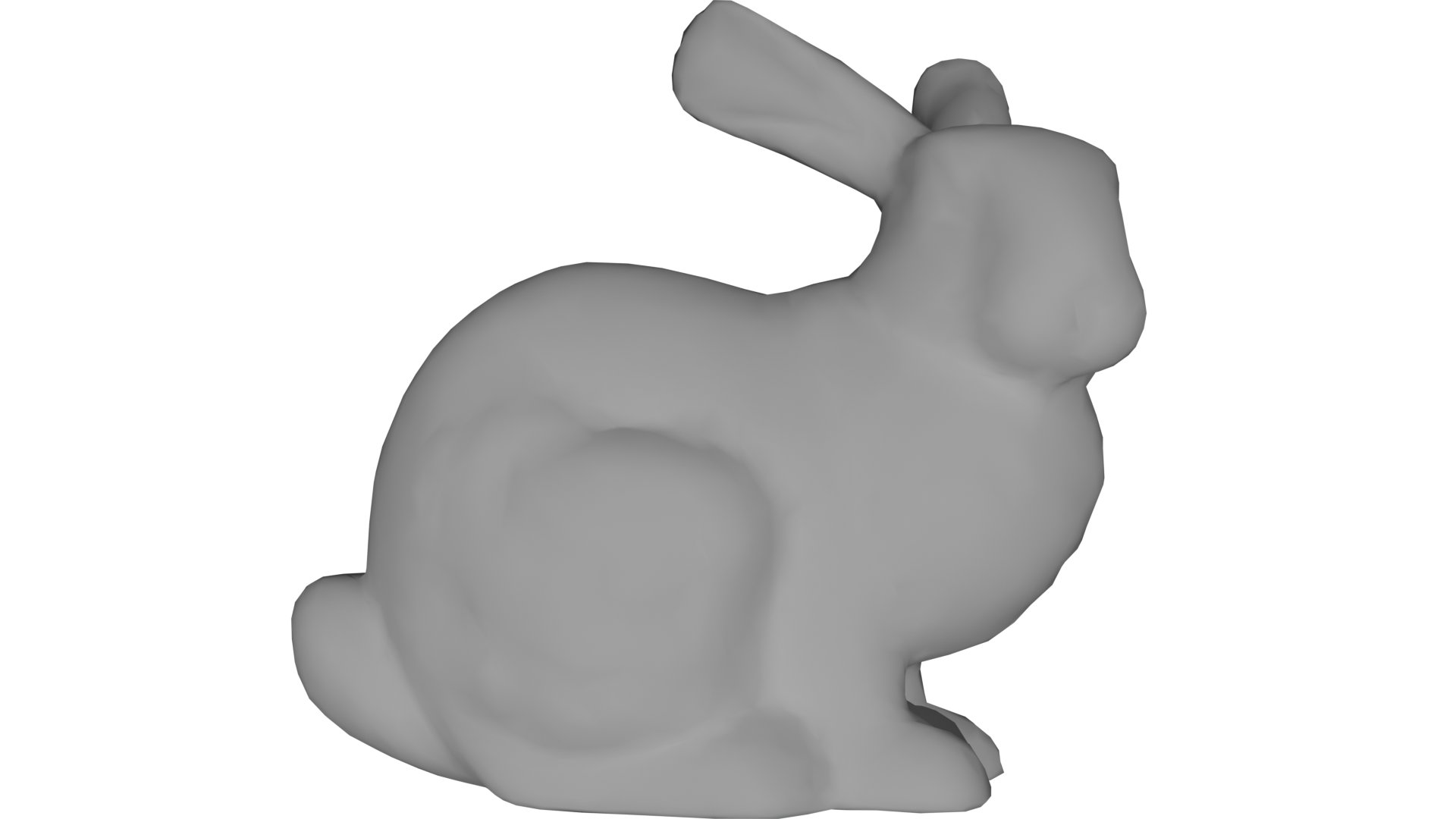}
    \includegraphics[width=0.23\textwidth]{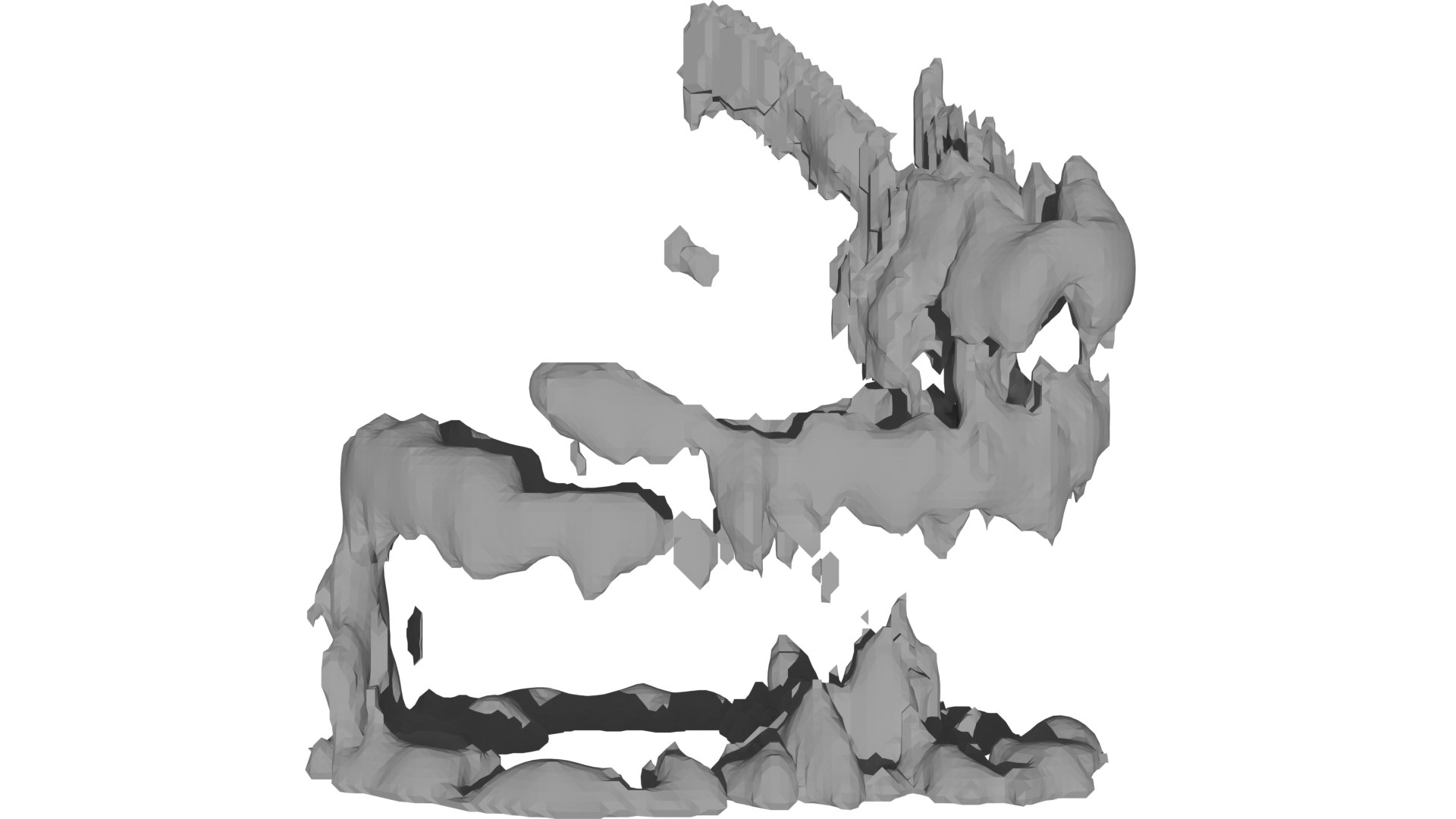}
    \includegraphics[width=0.23\textwidth]{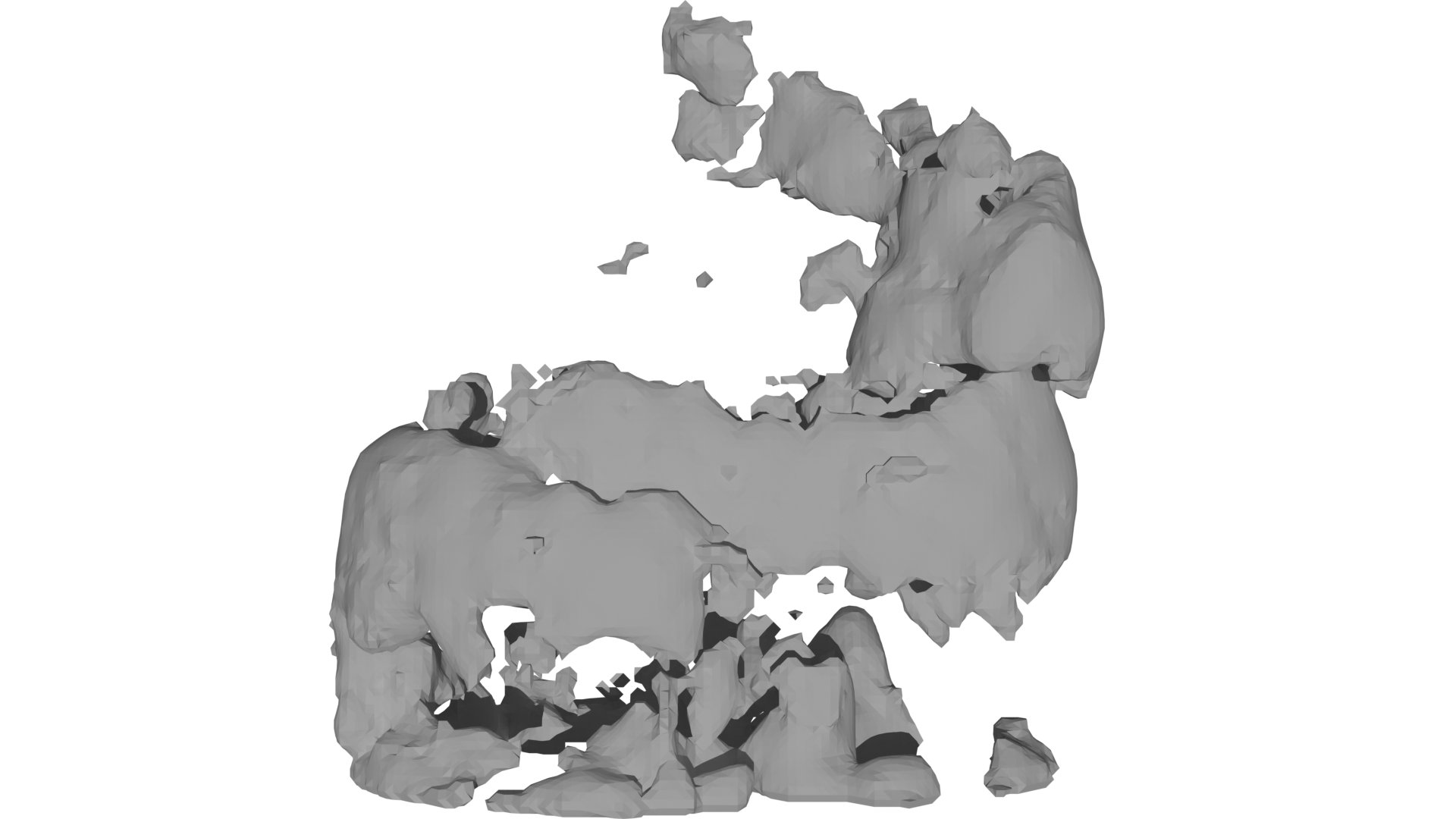}
    \includegraphics[width=0.23\textwidth]{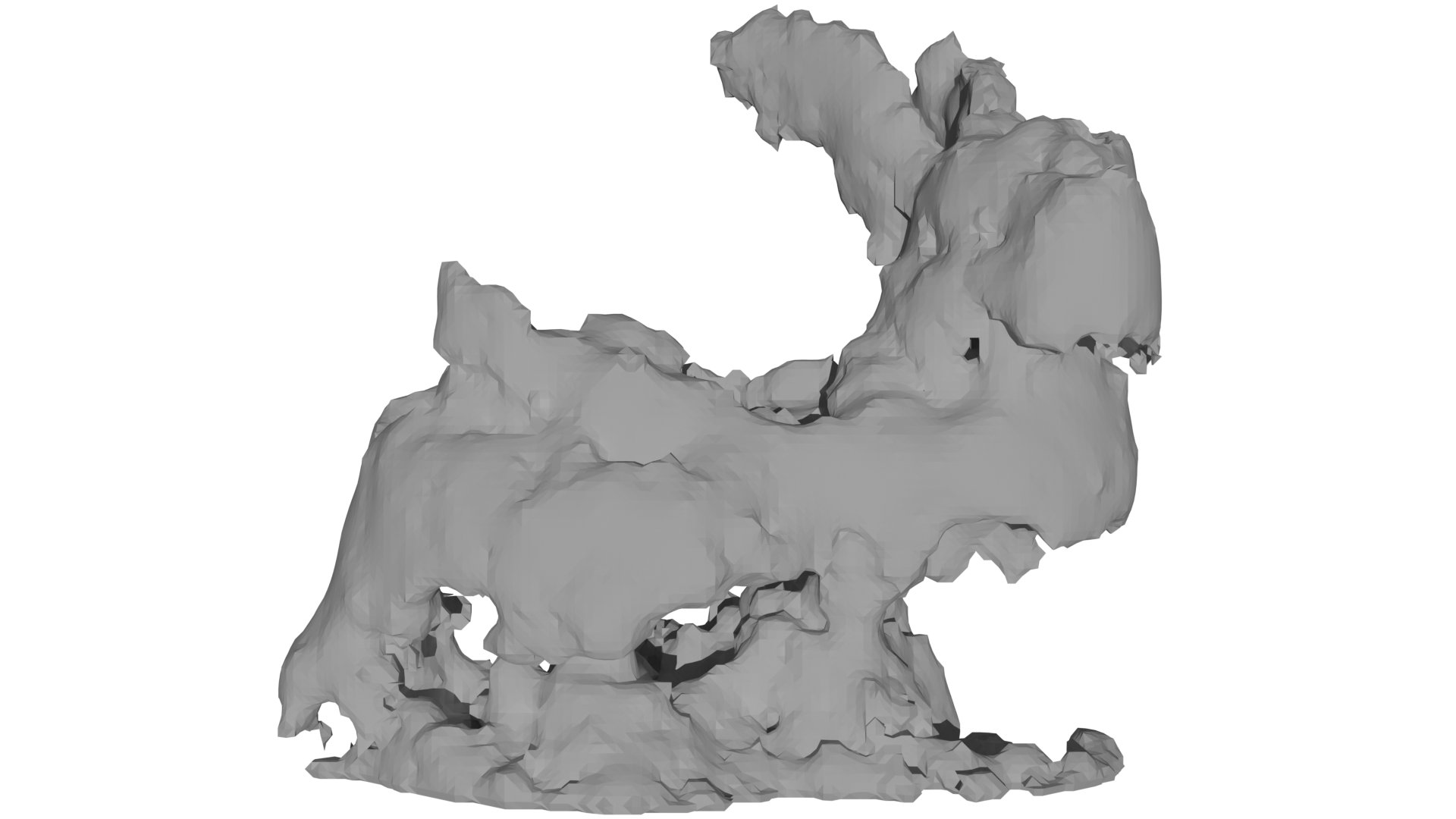}

    \vspace{5mm}
    
    \includegraphics[width=0.23\textwidth]{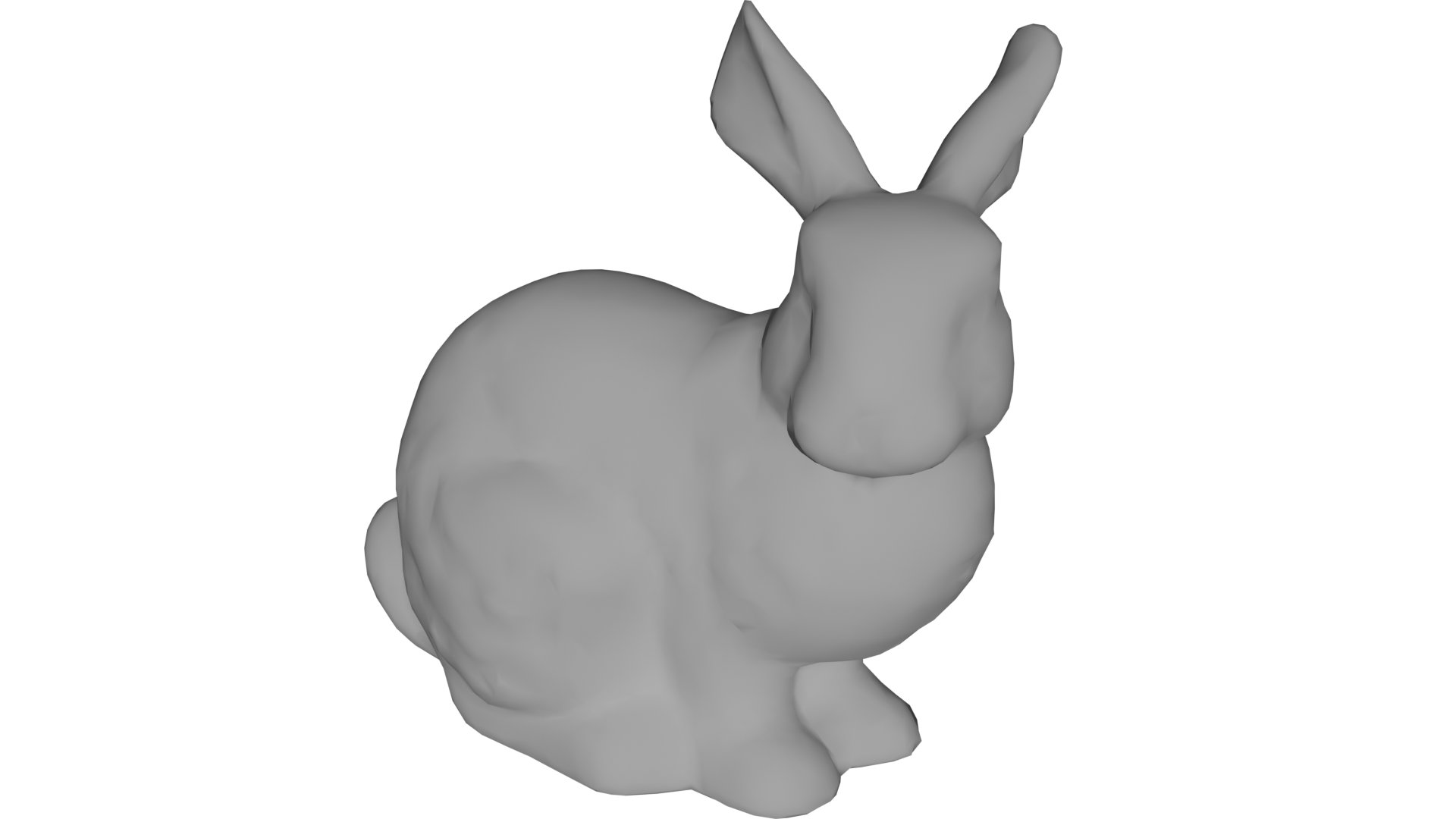}
    \includegraphics[width=0.23\textwidth]{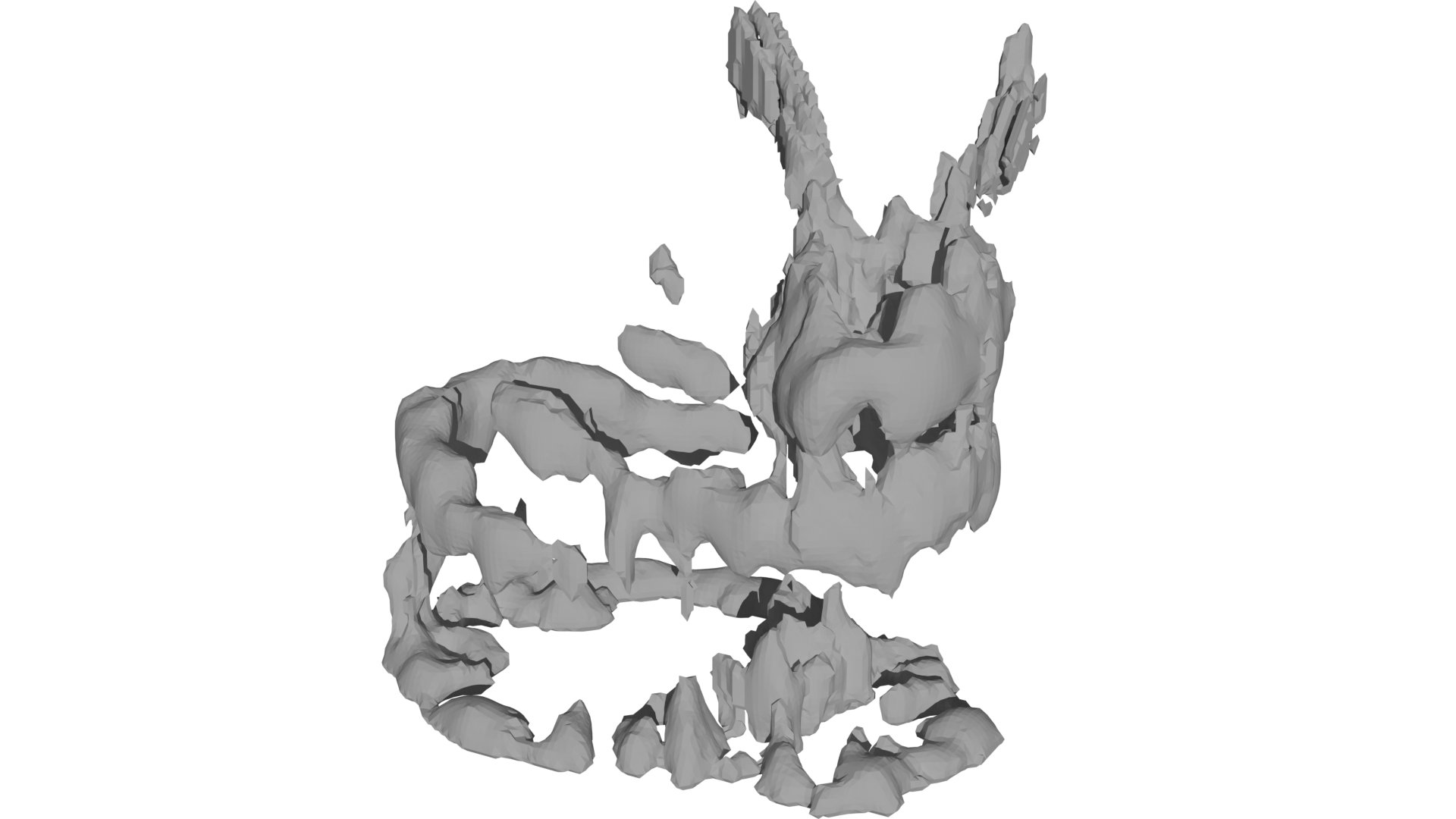}
    \includegraphics[width=0.23\textwidth]{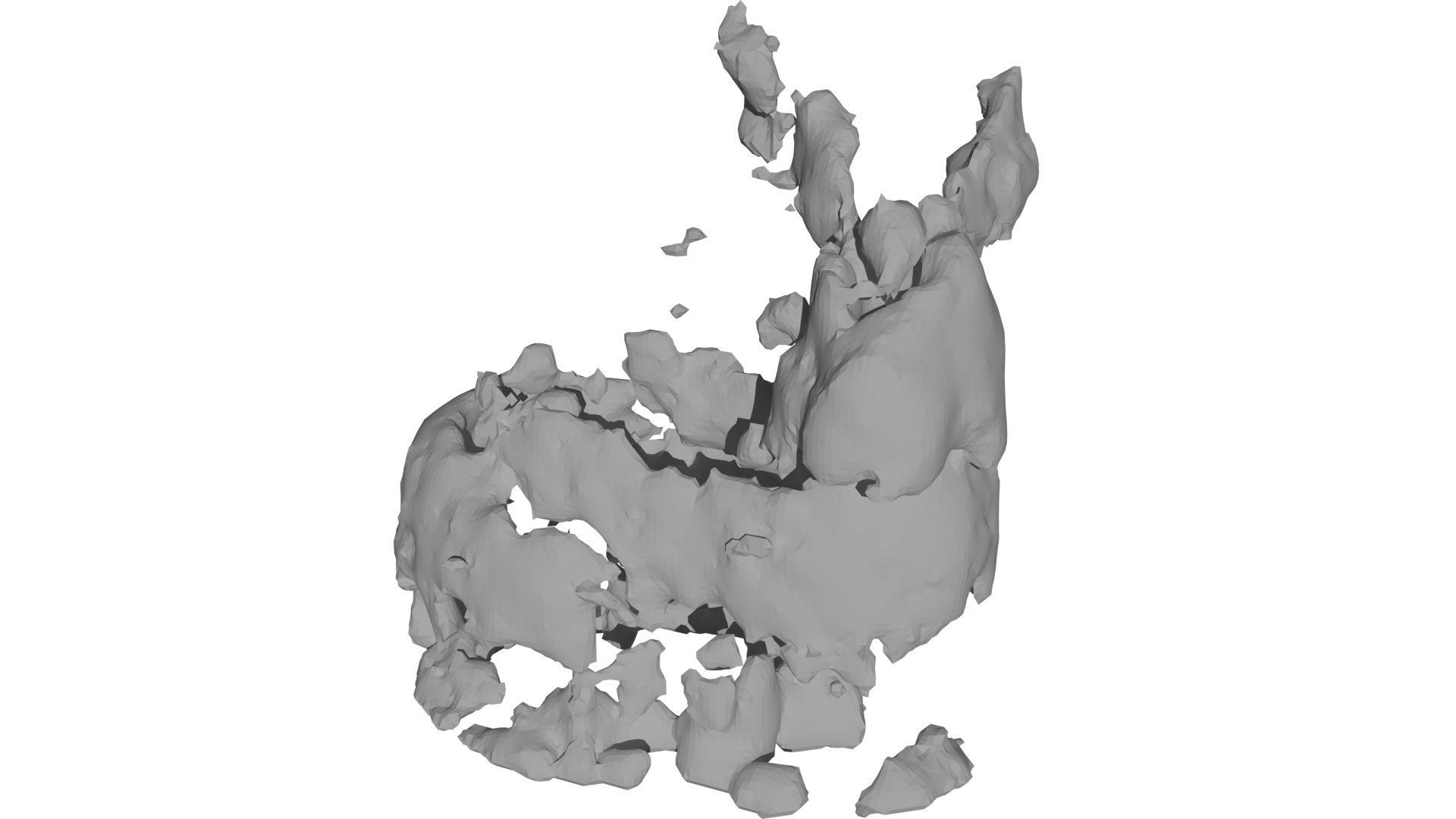}
    \includegraphics[width=0.23\textwidth]{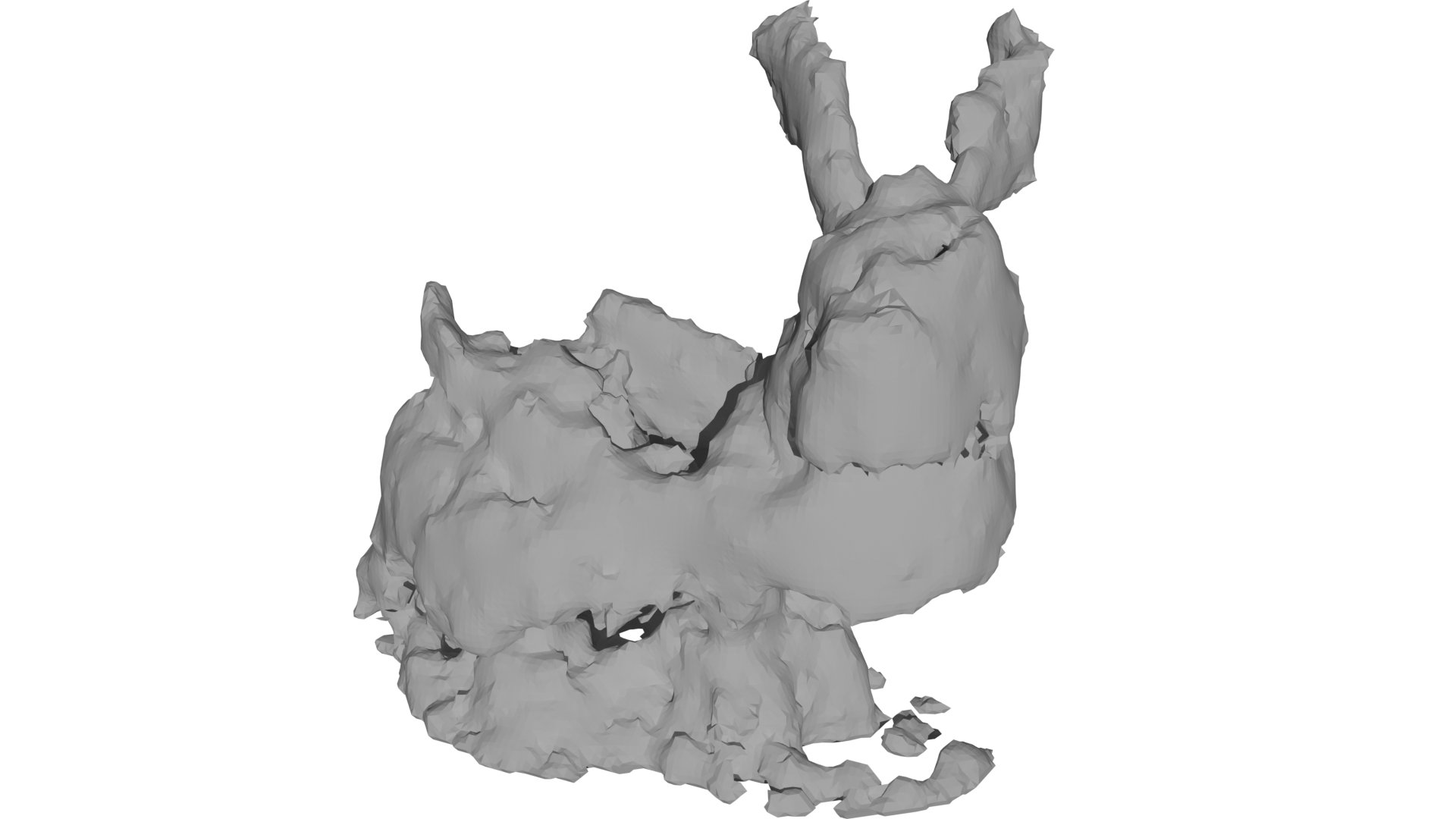}

    \makebox[0.23\textwidth]{\footnotesize Ground Truth}
    \makebox[0.23\textwidth]{\footnotesize Backprojection}
    \makebox[0.23\textwidth]{\footnotesize Reed et al.~\cite{reed2023neural}}
    \makebox[0.23\textwidth]{\footnotesize Ours}

    \caption{Real data: Stanford bunny. Top: frontal view. Bottom: top view.}
    \label{fig:real_bunny_20k}
\end{figure*}
\FloatBarrier
\clearpage

% --- Page 3: novel-view figure + table (same page) ---

\begin{table*}[p]
\centering
\small
\caption{Novel-view signal errors and GPU render time (lower is better).}
\label{tab:novel-view-metrics}
\begin{tabular}{l
                S[table-format=3.2]
                S[table-format=1.3]
                S[table-format=1.3]
                S[table-format=1.3]
                S[table-format=1.3]
                S[table-format=1.3]
                S[table-format=1.3]}
\toprule
Method & {Render (ms)} & {L1 (real)} & {L1 (imag)} & {L1 (abs)} & {MSE (real)} & {MSE (imag)} & {MSE (abs)} \\
\midrule
Reed et al. & 147.40          & 0.417          & 0.639          & 0.865          & 1.339          & 1.346          & 2.476          \\
Ours        & \textbf{126.11} & \textbf{0.402} & \textbf{0.576} & \textbf{0.665} & \textbf{0.797} & \textbf{0.810} & \textbf{1.171} \\
\bottomrule
\end{tabular}
\end{table*}
\FloatBarrier
\clearpage

\end{document}